\documentclass[prx, twocolumn, superscriptaddress]{revtex4-2}
\usepackage{bm, amsmath, amsfonts, amssymb, mathtools, braket}
\usepackage{times}
\usepackage{multirow}
\usepackage{graphicx}
\usepackage{float, color, xcolor}

\usepackage{bbm}
\usepackage{tabularx}

\usepackage[
pagebackref=false,
colorlinks=true,
linkcolor=blue,
urlcolor=blue,
filecolor=black,
citecolor=red,
pdfstartview=FitV,
pdftitle={},
pdfauthor={},
pdfsubject={},
pdfkeywords={},
pdfpagemode=None,
bookmarksopen=true
]{hyperref}

\usepackage[normalem]{ulem}

\newcommand{\tr}{\mathrm{Tr}}
\newcommand{\ra}{\rightarrow}

\newcommand{\dg}{\dagger}
\newcommand{\Wg}{\mathrm{Wg}}
\newcommand{\rU}{\mathrm{U}}
\newcommand{\rO}{\mathrm{O}}
\newcommand{\rSp}{\mathrm{Sp}}

\newcommand{\Ch}{\mathrm{Ch}}

\newcommand{\ii}{\text{i}}
\newcommand{\Tsign}{T^{*}T}
\newcommand{\Csign}{C^{*}C}

\begin{document}

\title{Symmetry classification of typical quantum entanglement}

\author{Yuhan Liu}
\email{yuhanliu@uchicago.edu}
\affiliation{Kadanoff Center for Theoretical Physics, University of Chicago, Chicago, Illinois 60637, USA}

\author{Jonah Kudler-Flam}
\email{jkudlerflam@ias.edu}
\affiliation{School of Natural Sciences, Institute for Advanced Study, Princeton, New Jersey 08540, USA}
\affiliation{Princeton Center for Theoretical Science, Princeton University, Princeton, New Jersey 08544, USA}

\author{Kohei Kawabata}
\email{kawabata@issp.u-tokyo.ac.jp}
\affiliation{Department of Physics, Princeton University, Princeton, New Jersey 08544, USA}
\affiliation{Institute for Solid State Physics, University of Tokyo, Kashiwa, Chiba 277-8581, Japan}

\date{\today}

\begin{abstract}
Entanglement entropy of typical quantum states, also known as the Page curve, plays an important role in quantum many-body systems and quantum gravity.
However, little has hitherto been understood about the role of symmetry in quantum entanglement.
Here, we establish the comprehensive classification of typical quantum entanglement for free fermions, or equivalently the quadratic Sachdev-Ye-Kitaev model with symmetry, on the basis of the tenfold fundamental symmetry classes of time reversal, charge conjugation, and chiral transformation.
Through both analytical and numerical calculations of random matrix theory, we show that the volume-law contribution to average entanglement entropy is robust and remains unaffected by symmetry.
Conversely, we uncover that the constant terms of the average and variance of entanglement entropy yield tenfold universal values unique to each symmetry class. 
These constant terms originate from the combination of a global scaling of the entanglement spectrum due to time-reversal symmetry and a singular peak at the center of the entanglement spectrum due to chiral or particle-hole symmetry.
Our work elucidates the interplay of symmetry and entanglement in quantum physics and provides characterization of symmetry-enriched quantum chaos.
\end{abstract}

\maketitle

\section{Introduction}

Quantum entanglement lies at the heart of quantum physics~\cite{Nielsen-textbook, Horodecki-review, Amico-review, Eisert-review}.
It provides fundamental characterizations of quantum phases of matter, such as critical phenomena~\cite{Osterloh-02, Osborne-02, Vidal-03, Calabrese-Cardy-04} and topological phases~\cite{Kitaev-Preskill-06, Levin-Wen-06, Ryu-06, Li-Haldane-08, Fidkowski-10, Pollmann-10}.
Moreover, quantum entanglement provides insight into thermalization of isolated quantum systems or lack thereof~\cite{Huse-review, Rigol-review, Abanin-review}.
In fact, entanglement entropy of typical states in quantum chaotic systems is maximal and proportional to the volume of the subsystem (i.e., volume law), resulting in thermalization and thereby validating thermodynamics and statistical physics~\cite{Goldstein-06, Popescu-06, Reimann-08, Garrison-Grover-18, Nakagawa-18, Vidmar-Rigol-17, Lu-Grover-19}.
Such typical quantum entanglement entropy also holds significant importance in black hole physics~\cite{Page-93a, *Page-93b, Hayden-Preskill-07}.
Recently, researchers have also studied quantum entanglement of typical Gaussian states in free fermions and found a signature of thermalization~\cite{Lai-15, Liu-18, Lydzba-20, *Lydzba-21, Bianchi-21, Bhattacharjee-21, Bianchi-22, Murciano-22, Yu-22}.
Correspondingly, the single-particle quantum chaos of free fermions has attracted growing interest~\cite{2016PhRvL.116c0401M, 2021PhRvB.104u4203L, 2022PhRvE.106c4118U, Lucas-22,  2022arXiv221000016L}.

An important signature of quantum chaos manifests in spectral statistics~\cite{Haake-textbook, Efetov-textbook}.
It is widely believed that the spectrum of a nonintegrable quantum system exhibits random-matrix statistics~\cite{BGS-84} whereas that of an integrable system obeys Poisson statistics~\cite{Berry-Tabor-77}.
The universality classes of random matrices are determined solely by the fundamental tenfold symmetry classes of time reversal, charge conjugation, and chiral transformation, known as the Altland-Zirnbauer (AZ) symmetry classes~\cite{AZ-97}.
As a prime example of quantum chaotic many-body systems, the Sachdev-Ye-Kitaev (SYK) model~\cite{Sachdev-Ye-93, Kitaev-KITP15, Rosenhaus-review, Sachdev-review} is classified by these tenfold symmetry classes and exhibits tenfold quantum chaotic behavior~\cite{You-17, Fu-16, GarciaGarcia-16, Cotler-17, Li-17, Kanazawa-17, Behrends-19, Sun-20}.
Furthermore, the AZ symmetry determines the universality classes of Anderson transitions~\cite{Beenakker-review-97, *Beenakker-review-15, Evers-review} and topological insulators and superconductors~\cite{Schnyder-08, *Ryu-10, Kitaev-09, HK-review, QZ-review, CTSR-review}.
However, perhaps surprisingly, little has been understood about the role of symmetry in entanglement theory.

\begin{figure}[b]
\centering
\includegraphics[width=\linewidth]{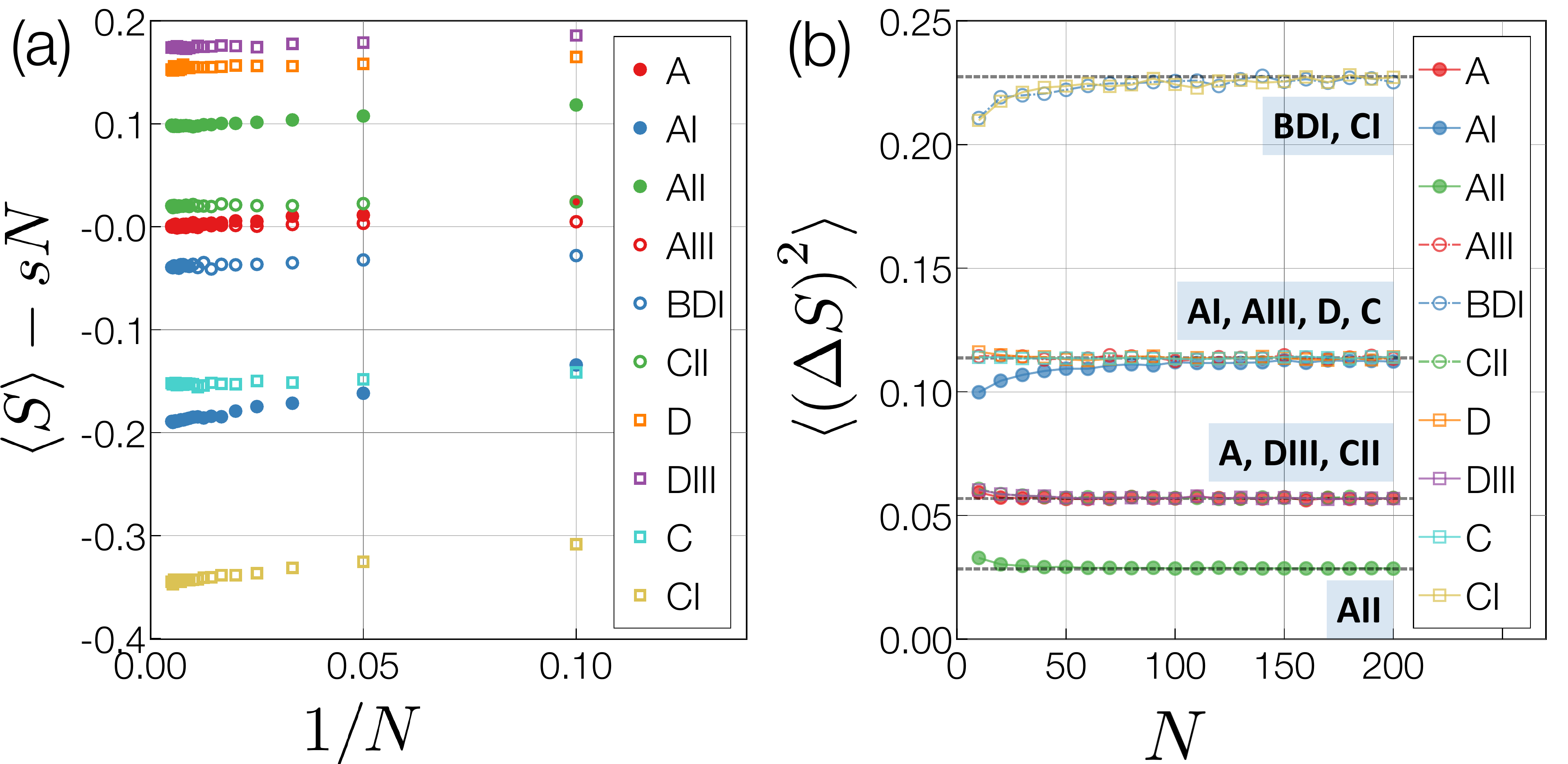} 
\caption{Typical entanglement entropy in the tenfold Altland-Zirnbauer symmetry classes. 
All the results are numerically calculated for particle-number-conserving free fermions at half filling and for half the degrees of freedom.
Each datum is averaged over $10^5$ disorder realizations.
(a)~Average entanglement entropy $\braket{S}$ with respect to the volume-law term $sN$ as functions of the inverse of the total system size $N$ without internal degrees of freedom [$s = \log 2 - 1/2$ in the standard classes and $s = 2 \left( \log 2 - 1/2 \right)$ in the chiral and Bogoliubov-de Gennes (BdG) classes]. 
(b)~Variance of entanglement entropy, $\braket{\left( \Delta S\right)^2}$, as functions of $N$.
The black dashed lines are the analytical results $\braket{\left( \Delta S \right)^2} = 2 \left( 3/4 - \log 2 \right)/\beta$ in the standard classes and $\braket{\left( \Delta S \right)^2} = 4 \left( 3/4 - \log 2 \right)/\beta$ in the chiral and BdG classes with the Dyson index $\beta = 1, 2, 4$.}
	\label{fig: Gaussian-Page-AZ}
\end{figure}

\begin{table*}[t]
	\centering
	\caption{Tenfold classification of typical entanglement entropy based on the Altland-Zirnbauer (AZ) symmetry classes. 
    The AZ symmetry classes consist of time-reversal symmetry (TRS), particle-hole symmetry (PHS), and chiral symmetry (CS). 
    For TRS and PHS, the entries ``$\pm 1$" mean the presence of symmetry and its sign, and the entries ``$0$" mean the absence of symmetry.
    For CS, the entries ``$1$" and ``$0$" mean the presence and absence of symmetry, respectively.
    Each class is characterized by the classifying space and the random-matrix indices $\left( \alpha, \beta \right)$.
    The constant terms of the average $\braket{S_0}$ and variance $\braket{\left( \Delta S \right)^2}$ of entanglement entropy are shown with $\sigma_0^2 \coloneqq 3/4 - \log 2$.
    All the results of entanglement entropy are calculated for particle-number-conserving free fermions with the half bipartition and half filling.
    For Bogoliubov-de Gennes Hamiltonians that do not conserve the particle number, the average is one half and the variance is one quarter in each symmetry class.
    }
     \begin{tabular}{cccccccccc} \hline \hline
     ~~AZ class~~ & ~~TRS~~ & ~~PHS~~ & ~~CS~~ & & ~Classifying space~ & ~~$\beta$~~ & ~~$\alpha$~~ & $\braket{S_0}$ & $\braket{\left( \Delta S \right)^2}$ \\ \hline
     A & $0$ & $0$ & $0$ & ~~${\cal C}_{0}$~~ & ${\rm U} \left( 2N\right) / {\rm U} \left( N \right) \times {\rm U} \left( N \right) $ & $2$ & ~~N/A~~ & $0$ & $\sigma_0^2$\\
     AIII & $0$ & $0$ & $1$ & ${\cal C}_{1}$ & ${\rm U} \left( N\right)$ & $2$ & $1$ & $0$ & $2 \sigma_0^2$\\ \hline
     AI & $+1$ & $0$ & $0$ & ${\cal R}_{0}$ & ${\rm O} \left( 2N\right) / {\rm O} \left( N \right) \times {\rm O} \left( N \right) $ & $1$ & N/A & $- \left( \log 2 - 1/2 \right)$ & $2 \sigma_0^2$\\
     BDI & $+1$ & $+1$ & $1$ & ${\cal R}_{1}$ & ${\rm O} \left( N\right)$ & $1$ & $0$ & $- \left( \left( 3/2 \right) \log 2 - 1 \right)$ & $4 \sigma_0^2$ \\
     D & $0$ & $+1$ & $0$ & ${\cal R}_{2}$ & ${\rm O} \left( 2N \right) / {\rm U} \left( N \right)$ & $2$ & $0$ & $\left( 1 - \log 2 \right)/2$ & $2\sigma_0^2$  \\
     DIII & $-1$ & $+1$ & $1$ & ${\cal R}_{3}$ & ${\rm U} \left( 2N \right) / {\rm Sp} \left( N \right)$ & $4$ & $1$ & $\left( \log 2 \right)/4$ & $\sigma_0^2$ \\
     AII & $-1$ & $0$ & $0$ & ${\cal R}_{4}$ & ~~${\rm Sp} \left( 2N\right) / {\rm Sp} \left( N \right) \times {\rm Sp} \left( N \right)$~~ & $4$ & N/A & $\left( \log 2 - 1/2 \right)/2$ & ~~$\sigma_0^2/2$~~ \\
     CII & $-1$ & $-1$ & $1$ & ${\cal R}_{5}$ & ${\rm Sp} \left( N\right)$ & $4$ & $3$ & ~~$\left( \left( 3/2 \right) \log 2 - 1 \right)/2$~~ & $\sigma_0^2$ \\
     C & $0$ & $-1$ & $0$ & ${\cal R}_{6}$ & ${\rm Sp} \left( N \right) / {\rm U} \left( N \right)$ & $2$ & $2$ & $-\left( 1 - \log 2 \right)/2$ & $2\sigma_0^2$ \\
     CI & $+1$ & $-1$ & $1$ & ${\cal R}_{7}$ & ${\rm U} \left( N \right) / {\rm O} \left( N \right)$ & $1$ & $1$ & $-\left( \log 2 \right)/2$ & $4\sigma_0^2$ \\ \hline \hline
     \end{tabular}
  	\label{tab: AZ}
\end{table*}

In this work, we establish the classification of typical quantum entanglement in free fermions on the basis of the tenfold fundamental symmetry classes. 
We show that the volume-law term of average entanglement entropy is invariant in all the classes.
Conversely, we find that the constant terms of the average and variance of entanglement entropy depend on symmetry and yield tenfold universal values unique to each symmetry class (Table~\ref{tab: AZ} and Fig.~\ref{fig: Gaussian-Page-AZ}).
In addition to numerical calculations, we analytically derive these tenfold universal values of typical entanglement, introducing the symmetry-enriched versions of Weingarten calculus~\cite{Collins-03, Collins-06, Collins-09, Matsumoto-13, Collins-review}.
Our findings elucidate the interplay of symmetry and entanglement in quantum physics.

The rest of this work is organized as follows.
In Sec.~\ref{msec: typical entanglement}, we develop the tenfold symmetry classification of typical quantum entanglement in free fermions (Table~\ref{tab: AZ} and Fig.~\ref{fig: Gaussian-Page-AZ}).
In Sec.~\ref{msec: Weingarten}, we analytically derive the typical quantum entanglement on the basis of the Weingarten calculus.
In Sec.~\ref{msec: Wigner}, we obtain typical quantum entanglement for small systems (Table~\ref{stab: Wigner surmise}), in the same spirit as the Wigner surmise.
In Sec.~\ref{msec: conclusion}, we conclude this work with several outlooks.
In Appendices~\ref{sec: numerics - AZ} and \ref{sec: numerics - AZ - BdG}, we explain details on typical entanglement entropy for particle-number-conserving free fermions and particle-number-nonconserving free fermions (i.e., BdG Hamiltonians), respectively.
In Appendix~\ref{sec: Weingarten}, we provide details on analytical derivations of typical quantum entanglement based on the Weingarten calculus.
In Appendix~\ref{sec: Wigner surmise}, we provide details on the Wigner surmise of typical quantum entanglement.

\section{Symmetry classification of typical quantum entanglement}
    \label{msec: typical entanglement}

We develop the tenfold classification of typical quantum entanglement in free fermions based on the AZ symmetry classes, as summarized in Table~\ref{tab: AZ} and Fig.~\ref{fig: Gaussian-Page-AZ}.
In Sec.~\ref{msubsec: AZ}, we begin with reviewing the AZ symmetry classification and explain calculations of typical entanglement entropy for free fermions.
Then, we provide typical entanglement entropy in the standard classes (Sec.~\ref{msubsec: standard}), chiral classes (Sec.~\ref{msubsec: chiral}), and Bogoliubov-de Gennes (BdG) classes (Sec.~\ref{msubsec: BdG}).
In Appendices~\ref{sec: numerics - AZ} and \ref{sec: numerics - AZ - BdG}, we explain details on calculations of typical entanglement entropy for particle-number-conserving free fermions and particle-number-nonconserving free fermions (i.e., BdG Hamiltonians), respectively.

\subsection{Altland-Zirnbauer (AZ) symmetry}
    \label{msubsec: AZ}

We consider a generic free fermionic system
\begin{align}
    \hat{H} = \sum_{ij} \hat{c}_{i}^{\dag} H_{ij} \hat{c}_{j},
\end{align} 
where $\hat{c}_{i}$'s ($i=1, 2, \cdots, N$) 
are complex fermion operators, and $H$ is an $N\times N$ single-particle Hamiltonian.
Since physical disorder breaks spatial symmetry, we focus on internal symmetry.
In general, $H$ is classified according to the fundamental internal symmetries of time reversal, charge conjugation, and chiral transformation:
\begin{gather}
    T^{-1} H^{*} T = H,\quad \Tsign = \pm 1, \label{eq: TRS} \\
    C^{-1} H^{*} C = - H,\quad \Csign = \pm 1, \label{eq: PHS} \\
    S^{-1} H S = - H,\quad S^2 = +1, \label{eq: CS}
\end{gather}
where, $T$, $C$, and $S$ are unitary operators.
Fermions include the spin (particle-hole) degree of freedom in the presence of time-reversal symmetry with sign $\Tsign = -1$ (chiral or particle-hole symmetry). 
Time-reversal symmetry $T$ gives the threefold Wigner-Dyson symmetry classes~\cite{Wigner-51, *Wigner-58, Dyson-62}, and its combination with particle-hole symmetry $C$ and chiral symmetry $S$ gives the tenfold AZ symmetry classes~\cite{AZ-97}.
The tenfold classifying spaces in Table~\ref{tab: AZ} provide all the possible symmetric spaces for free fermionic systems.
Spectral properties of Hermitian random matrices are universally determined by their symmetry~\cite{Mehta-textbook, Forrester-textbook, Haake-textbook}, characterized by the random-matrix indices $\left( \alpha, \beta \right)$.

Depending on the symmetry classes, we randomly choose single-particle eigenstates by the Haar measure from different classifying spaces
and calculate entanglement entropy of a subsystem with half the size for half-filled many-body eigenstates 
(see Appendix~\ref{sec: numerics - AZ} for details).
Thereby, we study the quantum chaotic behavior of typical thermal eigenstates, instead of special eigenstates such as ground states at zero temperature.
Notably, our random free fermionic Hamiltonians are equivalent to the two-body SYK model with complex fermions~\cite{Sachdev-Ye-93, Kitaev-KITP15, Rosenhaus-review, Sachdev-review}, whose entanglement entropy was studied in the absence of symmetry (i.e., class A)~\cite{Liu-18, Bianchi-22}.
Below, we demonstrate that additional symmetry changes the constant terms of the average and variance of entanglement entropy and yields tenfold universal values unique to each symmetry class.
The AZ symmetry classification does not include unitary symmetry that commutes with Hamiltonians, the effect of which can be studied in subspaces of fixed conserved charge~\cite{Murciano-22, Lau-22}.
By contrast, we show that the AZ symmetries cannot be captured in such a manner and play a more fundamental role in typical quantum entanglement.

\subsection{Standard (Wigner-Dyson) classes}
    \label{msubsec: standard}

The threefold standard classes are concerned only with time-reversal symmetry in Eq.~(\ref{eq: TRS})~\cite{Wigner-51, *Wigner-58, Dyson-62}.
The symmetry class without any symmetry is called class A, and the symmetry class with time-reversal symmetry having sign $+1$ ($-1$) is called class AI (AII).
In class AII, time-reversal symmetry leads to the Kramers degeneracy, and we calculate entanglement entropy only from half of the entanglement spectrum.
While the single-particle eigenstates generally form the unitary group $\mathrm{U} \left( N \right)$ in class A, time-reversal symmetry makes them belong to the orthogonal group $\mathrm{O} \left( N \right)$ and the symplectic group $\mathrm{Sp} \left( N \right)$ in classes AI and AII, respectively.
Taking the eigenstates Haar-randomly from these classifying spaces, we obtain the average entanglement entropy
\begin{align}
    \braket{S} = \left( \log 2 - \frac{1}{2} \right) \left( N+1-\frac{2}{\beta} \right) + o \left( 1 \right)
        \label{eq: average-WD}
\end{align}
with the Dyson index $\beta = 1$ (class AI), $\beta = 2$ (class A), and $\beta = 4$ (class AII).
In class A, the constant term 
\begin{align}
    \braket{S_0} = \left( 1 - \frac{2}{\beta} \right) \left( \log 2 - \frac{1}{2} \right) 
\end{align}
of the average entanglement entropy vanishes, consistent with Refs.~\cite{Liu-18, Bianchi-22}.
Even in the presence of time-reversal symmetry, the leading term proportional to $N$ does not change.
However, time-reversal symmetry gives rise to the nonzero constant term $\braket{S_0}$, which is negative (positive) for class AI (AII).
Equation~(\ref{eq: average-WD}) may be understood as one particle being effectively removed in class AI and half of a particle being effectively added in class AII.

We also obtain the variance of entanglement entropy as
\begin{align}
    \braket{\left( \Delta S \right)^2} = \frac{2}{\beta} \left( \frac{3}{4} - \log 2\right) + o \left( 1 \right).
        \label{eq: variance-WD}
\end{align}
The nonvanishing variance $\braket{\left( \Delta S \right)^2}$ is a feature unique to free fermions~\cite{Bianchi-21, Bianchi-22}. 
For many-body chaotic systems, the entanglement entropy is highly self-averaging, with variance suppressed by a power of the Hilbert space dimension.
Notably, $\braket{\left( \Delta S \right)^2}$ is twice larger in class AI ($\Tsign = +1$, $\beta = 1$) than in class A and reduces by half in class AII ($\Tsign = -1$, $\beta = 4$).
This is consistent with the universal spectral statistics of random matrices and quantum chaotic systems where level repulsion is suppressed in class AI and enhanced in class AII~\cite{Mehta-textbook, Forrester-textbook, Haake-textbook}, which results in the larger and smaller variances.
In contrast to the average, the variance of entanglement entropy is universally determined solely by time-reversal symmetry and the Dyson index $\beta$ even in the presence of chiral and particle-hole symmetries.

\subsection{Chiral classes}
    \label{msubsec: chiral}

In the presence of chiral symmetry in Eq.~(\ref{eq: CS}), single-particle Hamiltonians $H$ generally take the form
\begin{align}
    H = \begin{pmatrix}
        0 & h \\
        h^{\dag} & 0
    \end{pmatrix},
        \label{eq: chiral structure}
\end{align}
with $h \in \mathrm{U} \left( N \right)$, $ \mathrm{O} \left( N \right)$, and $ \mathrm{Sp} \left( N \right)$ in classes AIII, BDI, and CII, respectively.
Even in the presence of chiral symmetry, the volume-law term of the average entanglement entropy is invariant.
By contrast, we find that the chiral structure leads to the constant terms 
\begin{align}
    \braket{S_0} = \left( 1 - \frac{2}{\beta} \right) \left( \frac{3}{2} \log 2 - 1 \right), 
        \label{eq: average-chiral}
\end{align}
with the Dyson index $\beta = 1$ (class BDI), $\beta = 2$ (class AIII), and $\beta = 4$ (class CII). 
Similarly to class A, $\braket{S_0}$ vanishes in the absence of time-reversal symmetry.
Additional time-reversal symmetry with sign $\Tsign = +1$ ($\Tsign = -1$) gives rise to negative (positive) $\braket{S_0}$, which is roughly one fifth of $\braket{S_0}$ in the standard classes.
Furthermore, because of the chiral structure in Eq.~(\ref{eq: chiral structure}), the variance of entanglement entropy is twice larger than Eq.~(\ref{eq: variance-WD}):
\begin{align}
    \braket{\left( \Delta S \right)^2} = \frac{4}{\beta} \left( \frac{3}{4} - \log 2\right) + o \left( 1 \right).
        \label{eq: variance-chiral&BdG}
\end{align}
Still, it is determined solely by the Dyson index $\beta$, which signals the underlying universality.

\subsection{Bogoliubov-de Gennes (BdG) classes}
    \label{msubsec: BdG}

In the BdG classes, where particle-hole symmetry in Eq.~(\ref{eq: PHS}) is respected, the volume-law term of the average entanglement entropy is invariant. 
However, the constant term $\braket{S_0}$ exhibits distinctive values unique to the BdG classes.
In classes D and C, single-particle eigenstates are respectively characterized by $\mathrm{O} \left( 2N \right)/\mathrm{U} \left( N \right)$ and $\mathrm{Sp} \left( N \right)/\mathrm{U} \left( N \right)$,
and the average entanglement entropy is obtained as
\begin{align}
    \braket{S_0} = \frac{1}{2} \left( 1-\alpha \right) \left( 1-\log 2 \right).  
        \label{eq: average-BdG1}
\end{align}
Here, the random-matrix index $\alpha = 0$ (class D) and $\alpha = 2$ (class C) controls the spectral statistics around the zero eigenvalue in contrast with the Dyson index $\beta$ that controls the spectral statistics for generic eigenvalues~\cite{AZ-97, Haake-textbook}.
Notably, $\braket{S_0}$ is different despite the same Dyson index $\beta = 2$ in classes D and C.
On the other hand, in classes DIII and CI, both time-reversal and particle-hole symmetries are relevant, and single-particle Hamiltonians $H$ are given by Eq.~(\ref{eq: chiral structure}) with $h$ in the circular symplectic and orthogonal ensembles~\cite{Mehta-textbook, Forrester-textbook, Haake-textbook}, respectively, leading to
\begin{align}
    \braket{S_0} = \frac{1}{2} \left( 1-\frac{2}{\beta}\right) \log 2 
        \label{eq: average-BdG2}
\end{align}
with the Dyson index $\beta = 1$ (class CI) and $\beta = 4$ (class DIII).
Similarly to the chiral classes, the variance of entanglement entropy is twice larger than Eq.~(\ref{eq: variance-WD}) [i.e., Eq.~(\ref{eq: variance-chiral&BdG})].

While we have hitherto focused on free fermions that conserve the particle number, we also investigate typical entanglement entropy in particle-number-nonconserving BdG Hamiltonians
\begin{align}
    \hat{H} = \hat{\Gamma}^{\dag} H \hat{\Gamma} 
\end{align}
with the Nambu spinor $\hat{\Gamma} \coloneqq ( \hat{c}_{1}~\cdots~\hat{c}_{N}~\hat{c}_{1}^{\dag}~\cdots~\hat{c}_{N}^{\dag} )^{T}$ that consists of both annihilation and creation operators 
(see Appendix~\ref{sec: numerics - AZ - BdG} for details).
We find that the average $\braket{S_0}$ is one half and the variance $\braket{\left( \Delta S \right)^2}$ is one quarter in BdG Hamiltonians in comparison with their particle-number-conserving cousins in the same symmetry classes.
This is consistent with the analytical results of BdG Hamiltonians in class D~\cite{Bianchi-21}.
While the fundamental constituents of particle-number-conserving free fermions are complex fermions, those of BdG Hamiltonians are Majorana fermions.
Majorana fermions effectively have half the degree of freedom compared with complex fermions, which results in the half average and quarter variance of entanglement entropy.
While the particle-number conservation was previously considered important~\cite{Liu-18, Bianchi-21, Bianchi-22}, our results show that the AZ symmetries play a more fundamental role in the free-fermion Page curve.

\section{Weingarten calculus}
    \label{msec: Weingarten}

Now, we analytically derive the typical entanglement entropy.
Since it is not straightforward to generalize the analytical approaches in Refs.~\cite{Liu-18, Bianchi-21, Bianchi-22} in the presence of symmetry, we introduce the symmetry-enriched versions of Weingarten calculus~\cite{Collins-03, Collins-06, Collins-09, Matsumoto-13, Collins-review}.
As a special feature of free fermions, the entanglement entropy is obtained from the single-particle correlation matrix $C$ constrained on the subsystem~\cite{Peschel-03, Peschel-review}, defined as 
\begin{align}
    C_{ij} \coloneqq \braket{ \Psi | \hat{c}_{i}^{\dag} \hat{c}_{j} | \Psi} 
\end{align}
with a many-body eigenstate $\ket{\Psi}$.
Let $\lambda_i$'s ($0 \leq \lambda_i \leq 1$) be the eigenspectrum of $C$ (i.e., single-particle entanglement spectrum).
Then, the average entanglement entropy reads
\begin{align}
    \braket{S} = \int_{0}^{1} d\lambda~s \left( \lambda \right) \braket{D \left( \lambda \right)},
        \label{eq: average entanglement entropy}
\end{align}
with 
\begin{align}
s \left( \lambda \right) \coloneqq - \lambda \log \lambda - \left( 1-\lambda \right) \log \left( 1- \lambda \right)
\end{align}
and the density of the entanglement spectrum,
\begin{align}
D \left( \lambda \right) \coloneqq \sum_{i} \delta \left( \lambda - \lambda_i \right).
\end{align}
The bracket denotes the average over the Haar measure on each classifying space.
From the standard procedure of the resolvent method, the average density is obtained as 
\begin{align}
\braket{D \left( \lambda \right)} = - \frac{1}{\pi}\,\mathrm{Im} \lim_{\varepsilon \to 0^{+}} \braket{R \left( \lambda + \ii \varepsilon \right)} 
\end{align}
with the resolvent
\begin{align}
    R \left( z \right) \coloneqq \tr \left( \frac{I}{zI - C} \right) = \tr \left( \frac{I}{z} \right) + \sum_{n=1}^{\infty} \frac{\tr\,C^n}{z^{n+1}}.
\end{align}
Thus, the calculations of the typical entanglement entropy in Eq.~(\ref{eq: average entanglement entropy}) reduce to $\braket{\tr\,C^n}$,
which can be constructed from random unitary matrices and hence
systematically carried out by the Weingarten calculus~\cite{Collins-03, Collins-06, Collins-09, Matsumoto-13, Collins-review} (see Appendix~\ref{sec: Weingarten} for details).
Depending on different classifying spaces in different symmetry classes, different types of the Weingarten functions are relevant, which leads to the tenfold typical entanglement entropy.

\begin{figure}[t]
\centering
\includegraphics[width=\linewidth]{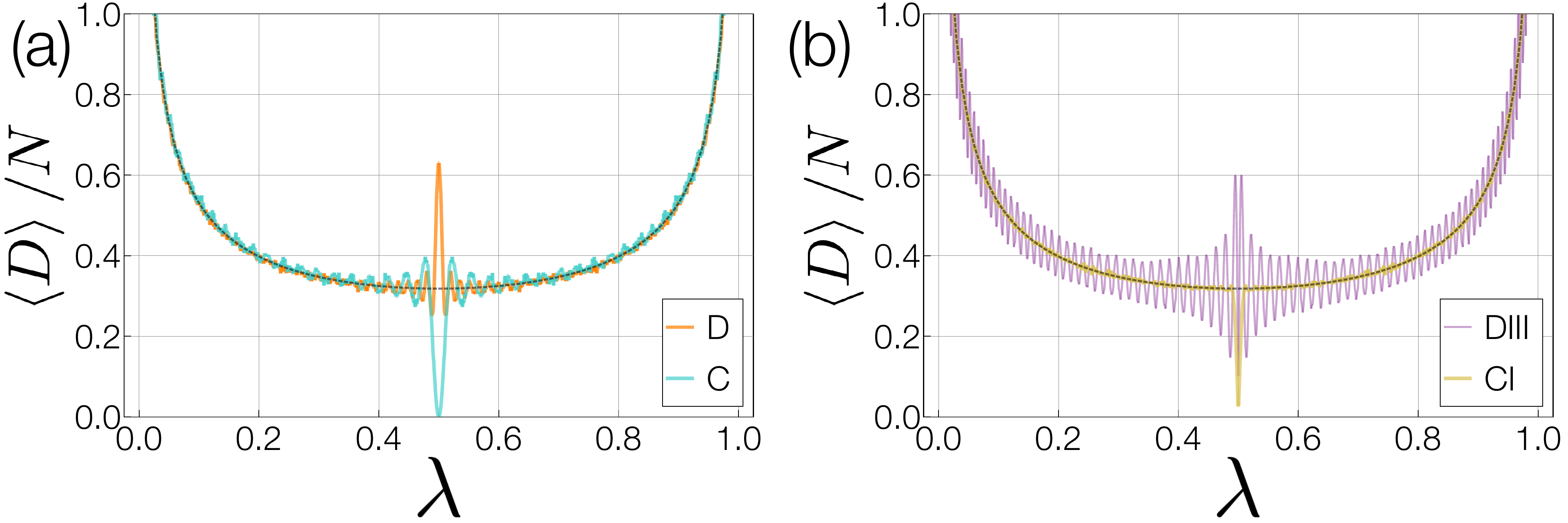} 
\caption{Average density of the single-particle entanglement spectrum for the quadratic Sachdev-Ye-Kitaev model enriched by symmetry, or equivalently, random Bogoliubov-de Gennes Hamiltonians in (a)~classes D (orange curve), C (blue curve), (b)~DIII (purple curve), and CI (yellow curve).
Each datum is averaged over $10^5$ disorder realizations for $N=100$.
The black dashed curves are the volume-law term $\braket{D}/N = 1/2\pi\sqrt{\lambda \left( 1- \lambda \right)}$.
The singular peaks or dips appear at $\lambda = 1/2$, consistent with the analytical results $\left( 1-\alpha/2 - 1/\beta \right) \delta \left( \lambda - 1/2 \right)$.}
	\label{fig: SYK-ES}
\end{figure}

\begin{table*}[t]
	\centering
	\caption{Wigner surmise of typical quantum entanglement.
    The tenfold Altland-Zirnbauer (AZ) symmetry classes consist of time-reversal symmetry (TRS), particle-hole symmetry (PHS), and chiral symmetry (CS). 
    For TRS and PHS, the entries ``$\pm 1$" mean the presence of symmetry and its sign, and the entries ``$0$" mean the absence of symmetry.
    For CS, the entries ``$1$" and ``$0$" mean the presence and absence of symmetry, respectively.
    The average and variance of entanglement entropy are calculated both analytically and numerically for $N=2$.
    In the numerical calculations, each datum is averaged over $10^8$ ensembles.
    All the results of entanglement entropy are calculated for particle-number-conserving free fermions with the half bipartition and the half filling.
    }
     \footnotesize
     \begin{tabular}{cccccccc} \hline \hline
     ~~AZ class~~ & ~~TRS~~ & ~~PHS~~ & ~~CS~~ & $\braket{S}_{\rm analytical}$ & $\braket{S}_{\rm numerical}$ & $\braket{\left( \Delta S\right)^2}_{\rm analytical}$ & $\braket{\left( \Delta S\right)^2}_{\rm numerical}$  \\ \hline
     A & $0$ & $0$ & $0$ & $1/2$ & $0.5000$ & $\left( 21-2\pi^2 \right)/36$ & $0.0350$ \\
     AIII & $0$ & $0$ & $1$ & $\left( 2 \log 2 + 1\right)/3$ & $0.7954$ & $( 3 - 8 \log 2 + 8 \left( \log 2\right)^2 )/9$ & $0.1443$ \\ \hline
     AI & $+1$ & $0$ & $0$ & $2\log 2 - 1$ & $0.3863$ & $5\pi^2/24 - 2$ & $0.0562$ \\
     BDI & $+1$ & $+1$ & $1$ & $2 \left( 2 \log 2 - 1\right)$ & $0.7726$ & $5\pi^2/6 - 8$ & $0.2247$ \\
     D & $0$ & $+1$ & $0$ & $1$ & $1.0000$ & $\left( 21-2\pi^2 \right)/9$ & $0.1401$\\
     DIII & $-1$ & $+1$ & $1$ & $1.013604 \cdots$ [Eq.~(\ref{seq: Wigner-DIII-average})] & $1.0136$ & $0.0818924 \cdots$ [Eq.~(\ref{seq: Wigner-DIII-variance})] & $0.0819$ \\
     AII & $-1$ & $0$ & $0$ & $7/12$ & $0.5833$ & $97/144 - \pi^2/15$ &  $0.0156$\\
     CII & $-1$ & $-1$ & $1$ & $\left( 79 + 132 \log 2\right)/210$ & $0.8119$ & $( 16021 - 44160 \log 2 + 38016 \left( \log 2\right)^2 )/44100$ & $0.0834$ \\
     C & $0$ & $-1$ & $0$ & $2/3$ & $0.6666$ & $0.128497 \cdots$ [Eq.~(\ref{seq: Wigner-C-variance})] & $0.1285$ \\
     CI & $+1$ & $-1$ & $1$ & $0.554363 \cdots$ [Eq.~(\ref{seq: Wigner-CI-average})] & $0.5543$ & $0.177858 \cdots$ [Eq.~(\ref{seq: Wigner-CI-variance})] & $0.1779$ \\ \hline \hline
     \end{tabular}
  	\label{stab: Wigner surmise}
\end{table*}

From the aforementioned Weingarten calculus in the standard classes, we obtain (see Appendix~\ref{sec:standard-analy} for derivations)
\begin{align}
    \braket{D \left( \lambda \right)} = \frac{N + 1 - 2/\beta}{2\pi \sqrt{\lambda \left( 1- \lambda \right)}} 1_{[0, 1]} + O \left( 1/N \right) ,
\end{align}
which leads to Eq.~(\ref{eq: average-WD}).
Here, we define $1_{[0,1]}$ to be $1$ ($0$) for $0 \leq \lambda \leq 1$ (otherwise) and neglect the additional delta functions at $\lambda = 0, 1$ irrelevant to entanglement entropy. 
The overall difference in the entanglement spectrum, which originates from the different Weingarten functions for $\mathrm{U} \left( N \right)$, $\mathrm{O} \left( N \right)$, and $\mathrm{Sp} \left( N \right)$, results in the different constant terms of the typical entanglement entropy in Eq.~(\ref{eq: average-WD}).
In the chiral and BdG classes, by contrast, an additional delta function appears at the center $\lambda = 1/2$ of the entanglement spectrum.
In the chiral classes, we have (see Appendix~\ref{sec:chiral-analy} for derivations)
\begin{align}
    &\braket{D \left( \lambda \right)} = \frac{N + 1 - 2/\beta}{\pi \sqrt{\lambda \left( 1- \lambda \right)}} 1_{[0, 1]} \nonumber \\
    &\qquad\qquad\quad +\frac{1}{2} \left( 1- \frac{2}{\beta} \right) \delta \left( \lambda - \frac{1}{2} \right) + O \left( 1/N \right).
\end{align}
In the BdG classes, we have (see Appendix~\ref{sec:BdG-analy} for derivations)
\begin{align}
    &\braket{D \left( \lambda \right)} = \frac{N - \left( 1-\alpha \right)/2}{\pi \sqrt{\lambda \left( 1- \lambda \right)}} 1_{[0, 1]} \nonumber \\
    &\qquad\qquad - \left( 1- \frac{\alpha}{2} - \frac{1}{\beta} \right) \delta \left( \lambda - \frac{1}{2} \right) + O \left( 1/N \right).
\end{align}
These additional delta functions also contribute to the constant terms of the average entanglement entropy, leading to Eqs.~(\ref{eq: average-chiral}), (\ref{eq: average-BdG1}), and (\ref{eq: average-BdG2}). 
Notably, the two types of constant terms originate from different origins and behave in a different manner.
While the density of states is generally less universal than the higher-order correlation functions (e.g., level-spacing statistics), that around the chiral-symmetric or particle-hole-symmetric point is known to be universal~\cite{AZ-97, Haake-textbook}.
Consequently, the delta functions at the center of the entanglement spectrum should be more universal than the other constant contributions.

We also numerically calculate the average density $\braket{D \left( \lambda \right)}$ of the single-particle entanglement spectrum for the quadratic SYK model with Majorana fermions (Fig.~\ref{fig: SYK-ES}).
The delta-function peaks or dips are resolved for finite $N$, their smooth signatures appearing at the center $\lambda = 1/2$ of the entanglement spectrum, consistent with the analytical results for infinite $N\to \infty$.
Different symmetry leads to distinctive behavior at $\lambda = 1/2$, characterizing symmetry-enriched quantum chaos of the SYK model.

The variance in entanglement entropy is evaluated as 
\begin{align}
    \braket{\left( \Delta S \right)^2} = \int_0^1 d\lambda_1 \int_0^1 d\lambda_2\,s \left(\lambda_1 \right) s \left( \lambda_2 \right) D_2 \left( \lambda_1 ,\lambda_2 \right)
        \label{eq: variance entanglement entropy}
\end{align}
with the two-point correlation function 
\begin{align}
    D_2 \left( \lambda_1, \lambda_2 \right) &\coloneqq \Braket{\sum_{ij} \delta \left( \lambda_1 - \lambda_i \right) \delta \left( \lambda_2 - \lambda_j \right)} \nonumber \\
    &\qquad\qquad\qquad\quad - \Braket{D \left( \lambda_1 \right)} \Braket{D \left( \lambda_2 \right)}.
\end{align}
It is known that $D_2 \left( \lambda_1, \lambda_2 \right)$ for large $N$ has the universal $1/\beta$ dependence for general matrix potentials~\cite{Beenakker-review-97, Mehta-textbook, Forrester-textbook, Haake-textbook}, which explains the universal $1/\beta$ dependence in Eq.~(\ref{eq: variance-WD}). 
Furthermore, we can fix the multiplicative factor $\sigma_0^2 \coloneqq 3/4 - \log 2$ in the standard classes, using the known analytical result for class A~\cite{Bianchi-22}. 
We have strong numerical evidence that this holds for all the AZ symmetry classes [Fig.~\ref{fig: Gaussian-Page-AZ}\,(b)].  


\section{Wigner surmise}
    \label{msec: Wigner}

In addition to the large-$N$ results, we analytically derive the average $\braket{S}$ and variance $\braket{\left( \Delta S \right)^2}$ of the entanglement entropy for small systems $N=2$, in the same spirit as the Wigner surmise~\cite{Wigner-51} (Table~\ref{stab: Wigner surmise}; see Appendix~\ref{sec: Wigner surmise} for details).
In contrast with the large-$N$ results, there is no way to differentiate between the volume-law and constant terms.
Still, the qualitative differences between different symmetry classes are found even for $N=2$.
Notably, ten different values of $\braket{\left( \Delta S \right)^2}$ appear, which contrast with the large-$N$ results that depend solely on the Dyson index $\beta$.
This implies that the quantized variance of entanglement entropy in Eq.~(\ref{eq: variance-WD}) originates from the many-level effect.
Similarly, the singular peak of $\braket{D \left( \lambda \right)}$ at the symmetric point $\lambda = 1/2$, universal feature in the chiral and BdG classes for large $N$, does not appear for $N=2$.

\section{Discussions}
    \label{msec: conclusion}

Symmetry plays a pivotal role throughout physics.
The fundamental tenfold internal symmetries, AZ symmetries, determine the universality classes of quantum chaos and the physics of free fermions.
However, the role of symmetry in quantum entanglement has remained largely unclear.
In this work, we developed the symmetry classification of typical quantum entanglement in free fermions.
We demonstrated that while the volume-law term of average entanglement entropy is unaffected by symmetry, the constant terms of the average and variance of entanglement entropy yield the tenfold universal values unique to each symmetry class.

Importantly, typical quantum entanglement underlies thermalization of isolated quantum systems and validates thermodynamics and statistical physics~\cite{Huse-review, Rigol-review, Abanin-review}.
Our findings should be useful in understanding the role of symmetry in quantum chaos and thermalization.
While we focused on random-matrix models in this work, our results should be relevant to the symmetry-enriched quantum chaos even in finite-dimensional systems without disorder.
On the other hand, different entanglement properties appear at Anderson transitions~\cite{Evers-review}, which merit further study.
It is also noteworthy that additional topological contributions can manifest themselves in the typical entanglement spectra.

Remarkably, our findings of typical quantum entanglement have a similarity to mesoscopic transport phenomena~\cite{Datta-textbook, Imry-textbook}.
For example, the sign of the quantum corrections to the conductance in disordered electronic systems depends on the sign of time-reversal symmetry~\cite{Abrahams-79, Gorkov-79, Altshuler-80, Hikami-80, Evers-review} in a similar manner to the average entanglement entropy in Eq.~(\ref{eq: average-WD}).
Furthermore, the conductance fluctuations in the diffusive regime are universally given as $\braket{\left( \Delta G \right)^2} \propto 1/\beta$~\cite{Washburn-86, Beenakker-review-97}, akin to the fluctuations of entanglement entropy in Eq.~(\ref{eq: variance-WD}).
Here, the transmission probability holds a parallel role to the single-particle entanglement spectrum.
These mesoscopic transport phenomena are universally described by field theory of nonlinear sigma model, with target manifolds classified by the AZ symmetries~\cite{Efetov-textbook}.
Similarly, it would be significant to develop field theory for typical quantum entanglement.

\begin{table*}[t]
	\centering
	\caption{Tenfold classification of typical quantum entanglement based on the Altland-Zirnbauer (AZ) symmetry classes. 
    The AZ symmetry classes consist of time-reversal symmetry (TRS), particle-hole symmetry (PHS), and chiral symmetry (CS). 
    For TRS and PHS, the entries ``$\pm 1$" mean the presence of symmetry and its sign, and the entries ``$0$" mean the absence of symmetry.
    For CS, the entries ``$1$" and ``$0$" mean the presence and absence of symmetry, respectively.
    The ten AZ classes are divided into two complex classes that only involve unitary symmetry (i.e., CS) and the eight real classes that involve antiunitary symmetry (i.e., TRS and PHS).
    Each class is characterized by the classifying space and the random-matrix indices $\left( \alpha, \beta \right)$.
    The numerical fitting results of the average entanglement entropy by $\braket{S} = S_1 N + S_0 + S_{-1}/N$, as well as those of the variance of entanglement entropy by $\braket{\left( \Delta S \right)^2} = \sigma_0^2 + \sigma_{-}^{2}/N$, are shown [$S_1 = \log 2 - 1/2$ for the standard classes (classes A, AI, and AII) and $S_1 = 2 \left( \log 2 - 1/2 \right)$ for the chiral classes (classes AIII, BDI, and CII) and Bogoliubov-de Gennes classes (classes D, DIII, C, and CI)].
    All the results of entanglement entropy are calculated for particle-number-conserving free fermions with the half bipartition and the half filling.
    }
     \begin{tabular}{cccccccccccc} \hline \hline
     ~~AZ class~~ & ~~TRS~~ & ~~PHS~~ & ~~CS~~ & & ~Classifying space~ & ~$\beta$~ & ~$\alpha$~ & $S_0$ & $S_{-1}$ & $\sigma_0^2$ & $\sigma_{-1}^2$  \\ \hline
     A & $0$ & $0$ & $0$ & ~~${\cal C}_{0}$~~ & ${\rm U} \left( 2N \right) / {\rm U} \left( N \right) \times {\rm U} \left( N \right)$ & $2$ & ~N/A~ & ~~$2 \times 10^{-5}$~~ & ~~$0.245$~~ & $0.057$ & $0.023$\\
     AIII & $0$ & $0$ & $1$ & ${\cal C}_{1}$ & ${\rm U} \left( N\right)$ & $2$ & $1$ & $-1 \times 10^{-4}$ & $0.056$ & $0.114$ & $0.0082$ \\ \hline
     AI & $+1$ & $0$ & $0$ & ${\cal R}_{0}$ & ${\rm O} \left( 2N\right) / {\rm O} \left( N \right) \times {\rm O} \left( N \right) $ & $1$ & N/A & $-0.192$ & $0.587$ & ~~$0.113$~~ & ~~$-0.142$~~ \\
     BDI & $+1$ & $+1$ & $1$ & ${\cal R}_{1}$ & ${\rm O} \left( N\right)$ & $1$ & $0$ & $-0.0392$ & $0.118$ & $0.227$ & $-0.168$ \\
     D & $0$ & $+1$ & $0$ & ${\cal R}_{2}$ & ${\rm O} \left( 2N \right) / {\rm U} \left( N \right)$ & $2$ & $0$ & $0.154$ & $0.107$ & $0.113$ & $0.028$ \\
     DIII & $-1$ & $+1$ & $1$ & ${\cal R}_{3}$ & ${\rm U} \left( 2N \right) / {\rm Sp} \left( N \right)$ & $4$ & $1$ & $0.173$ & $0.123$ & $0.057$ & $0.037$ \\
     AII & $-1$ & $0$ & $0$ & ${\cal R}_{4}$ & ~${\rm Sp} \left( 2N\right) / {\rm Sp} \left( N \right) \times {\rm Sp} \left( N \right)$~ & $4$ & N/A & ~~$0.0964$~~ & $0.219$ & $0.028$ & $0.044$ \\
     CII & $-1$ & $-1$ & $1$ & ${\cal R}_{5}$ & ${\rm Sp} \left( N\right)$ & $4$ & $3$ & $0.0198$ & $0.045$ & $0.057$ & $0.039$ \\
     C & $0$ & $-1$ & $0$ & ${\cal R}_{6}$ & ${\rm Sp} \left( N \right) / {\rm U} \left( N \right)$ & $2$ & $2$ & $-0.154$ & $0.118$ & $0.114$ & $0.0027$ \\
     CI & $+1$ & $-1$ & $1$ & ${\cal R}_{7}$ & ${\rm U} \left( N \right) / {\rm O} \left( N \right)$ & $1$ & $1$ & $-0.346$ & $0.389$ & $0.227$ & $-0.177$ \\ \hline \hline
     \end{tabular}
  	\label{stab: classifying space}
\end{table*}

\begin{acknowledgments}
We thank Shinsei Ryu for helpful discussions.
Y.L. is supported in part by the National Science Foundation under Grant No.~NSF PHY-1748958, the Heising-Simons Foundation, and the Simons Foundation (216179, LB) at the Kavli Institute for Theoretical Physics, and in part by the National Science Foundation under Award No.~DMR-2001181 and Simons Investigator Grant from the Simons Foundation (Award
No.~566116) through Shinsei Ryu.
J.K.F. is supported by the Institute for Advanced Study and the National Science Foundation under Grant No.~PHY-2207584.
K.K. is supported by the Japan Society for the Promotion of Science (JSPS) through the Overseas Research Fellowship, and by the Gordon and Betty Moore Foundation through Grant No.~GBMF8685 toward the Princeton theory program. 
\end{acknowledgments}

\appendix

\section{Typical entanglement entropy in the Altland-Zirnbauer (AZ) symmetry classification}
    \label{sec: numerics - AZ}

We obtain the typical quantum entanglement entropy for all the ten 
AZ symmetry classes~\cite{AZ-97, Beenakker-review-97,*Beenakker-review-15, Evers-review, CTSR-review}.
We consider particle-number-conserving free fermions
\begin{align}
    \hat{H} = \sum_{mn} \hat{c}_{m}^{\dag} H_{mn} \hat{c}_{n},
\end{align}
where $\hat{c}_n$ ($\hat{c}_n^{\dag}$) annihilates (creates) a fermion, and $H = \left( H_{mn} \right)_{m, n}$ is a single-particle Hamiltonian subject to certain symmetry.
Each symmetry class is characterized by the classifying space, summarized in Table~\ref{stab: classifying space}.
To obtain the typical quantum entanglement, we consider eigenstates constructed from the Haar-random matrices in these classifying spaces.
For clarity, we focus on the entanglement entropy for the half bipartition and the half filling.
The numerical results are also summarized in Table~\ref{stab: classifying space}.

\subsection{Standard (Wigner-Dyson) class (classes A, AI, and AII)}

In the standard (Wigner-Dyson) class (classes A, AI, and AII), Hamiltonians are only concerned with time-reversal symmetry.
In the many-body Hilbert (Fock) space, time-reversal symmetry is described by the antiunitary operation
\begin{align}
    \hat{\cal T} \hat{c}_{m} \hat{\cal T}^{-1}
    = \sum_{n} T_{mn} \hat{c}_{n},\quad
    \forall\,z\in\mathbb{C}~~~\hat{\cal T}z\hat{\cal T}^{-1} = z^{*}.
\end{align}
Here, $\hat{\cal T}$ is an antiunitary operator that acts on the many-body fermionic Fock space while $T = \left( T_{mn} \right)_{m, n}$ is a unitary matrix in the single-particle Hilbert space.
A system respects time-reversal invariance if the Hamiltonian $\hat{H}$ satisfies
\begin{align}
    \hat{\cal T} \hat{H} \hat{\cal T}^{-1}
    = \hat{H}.
\end{align}
In fact, if this relation is satisfied, we have 
$\hat{\cal T} \hat{O} \left( t \right) \hat{\cal T}^{-1} = \hat{O} \left( -t \right)$,
where $\hat{O} \left( t \right) = e^{\ii \hat{H} t} \hat{O} e^{-\ii \hat{H} t}$ is the time-evolved operator of an arbitrary operator $\hat{O}$. 
In terms of the single-particle Hamiltonian $H$, time-reversal invariance is equivalent to
\begin{align}
    T^{-1} H^{*} T = H.
\end{align}
Because of antiunitarity of time-reversal symmetry, the symmetry operators are required to satisfy
\begin{align}
    \hat{\cal T}^{2} = \left( \pm 1 \right)^{\hat{N}},\quad
    T^{*} T = \pm 1
\end{align}
with the number operator $\hat{N} \coloneqq \sum_{n} \hat{c}_{n}^{\dag} \hat{c}_{n}$.
Then, the standard (Wigner-Dyson) classes---classes A, AI, and AII---are defined as follows:

\begin{itemize}
\item In the absence of time-reversal symmetry (and any other internal symmetry), Hamiltonians are defined to belong to the unitary class (class A).

\item In the presence of time-reversal symmetry with the sign $T^{*}T=+1$, Hamiltonians are defined to belong to the orthogonal class (class AI).

\item In the presence of time-reversal symmetry with the sign $T^{*}T=-1$, Hamiltonians are defined to belong to the symplectic class (class AII).
An important feature of symplectic time-reversal symmetry is the Kramers degeneracy.
\end{itemize}

In the following, suppose the total system size is $N$, the subsystem size is $N_{A}$, and the particle number is $M$ for classes A and AI.
For class AII, on the other hand, the degree of freedom is double because of the Kramers degeneracy; 
the total system size is $2N$, the subsystem size is $2N_{A}$, and the particle number is $2M$.
Importantly, depending on time-reversal symmetry, generic single-particle Hamiltonians $H$ are diagonalized by the matrix $U$ that belongs to the unitary group $\mathrm{U} \left( N \right)$, orthogonal group $\mathrm{O} \left( N \right)$, and symplectic group $\mathrm{Sp} \left( N \right)$:
\begin{align}
    U \in \begin{cases}
        \mathrm{U} \left( N \right) & \left( \text{class A}\right); \\
        \mathrm{O} \left( N \right) & \left( \text{class AI}\right); \\
        \mathrm{Sp} \left( N \right) & \left( \text{class AII}\right).
    \end{cases}
        \label{seq: U-group}
\end{align}
These groups are called classifying spaces and characterize each symmetry class (Table~\ref{stab: classifying space}).

Using the eigenstate matrix $U$ introduced above, we calculate entanglement entropy~\cite{Peschel-03, Peschel-review}.
The truncated correlation matrix is given as
\begin{align}
    C_A = V^{\dag} V,
        \label{seq: CA}
\end{align}
where $V$ is an $M \times N_A$ matrix satisfying
\begin{align}
    U = \begin{pmatrix}
        V & V' \\
        W & W' 
    \end{pmatrix}^{T}.
        \label{seq: submatrix}
\end{align}
The entanglement entropy is obtained as
\begin{align}
    S = - \sum_{i} \left[ \lambda_i \log \lambda_i + \left( 1- \lambda_i \right) \log \left( 1-\lambda_i \right) \right],
        \label{seq: entanglement sum}
\end{align}
where $\lambda_i$'s ($i=1, 2, \cdots, N_A$) are the eigenvalues of $C_A$.
In class AII, $V$ is a $2M \times 2N_A$ matrix, and the number of eigenvalues of $C_A$ is $2N_A$.
The spectrum of the correlation matrix $C_A$ is two-fold degenerate because of time-reversal symmetry with the sign $-1$, and we calculate the entanglement entropy only from half of the entanglement spectrum.

\begin{figure*}[t]
\centering
\includegraphics[width=\linewidth]{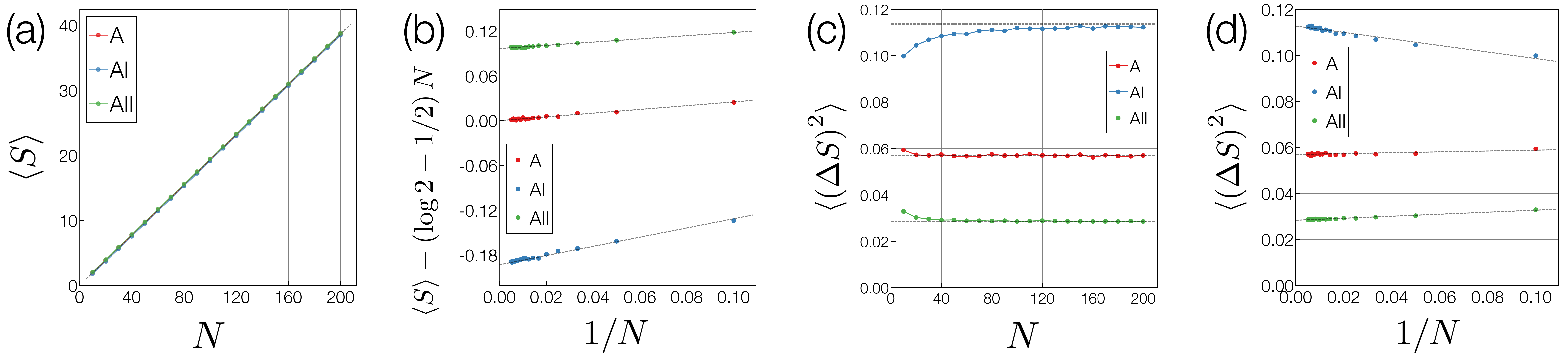} 
\caption{Typical quantum entanglement entropy in the standard (Wigner-Dyson) classes with the half bipartition and half filling for classes A (red dots), AI (blue dots), and AII (green dots). 
Each datum is averaged over $10^5$ ensembles. 
(a)~Average $\braket{S}$ of entanglement entropy as functions of the system size $N$. 
The black dashed line is the analytical result $\braket{S} \simeq \left( \log 2-1/2 \right) N$. 
(b)~Deviation of $\braket{S}$ from the volume-law term $\left( \log 2 - 1/2 \right) N$ as functions of $1/N$. 
(c)~Variance $\braket{\left( \Delta S\right)^2}$ of entanglement entropy as functions of $N$. 
The black dashed lines are the analytical results $\braket{\left( \Delta S\right)^2} = 2\left( 3/4 - \log 2\right)/\beta$ with the Dyson index $\beta = 1$ (class AI), $\beta = 2$ (class A), and $\beta = 4$ (class AII). 
(d)~Variance $\braket{\left( \Delta S\right)^2}$ as functions of $1/N$.}
	\label{sfig: standard}
\end{figure*}

To obtain the typical entanglement entropy for the standard classes, we numerically calculate the entanglement entropy for $U$ Haar-randomly distributed in the classifying spaces in Eq.~(\ref{seq: U-group}).
In Ref.~\cite{Bianchi-22}, the average and variance of such typical entanglement entropy without symmetry (i.e., class A) were analytically derived as
\begin{align}
        \braket{S} &= \left( \log 2 - \frac{1}{2} \right) N + \frac{1}{4N} + O \left( 1/N^3 \right) \nonumber \\
        &= \left( 0.193147 \cdots \right) N + \frac{1}{4N} + O \left( 1/N^3 \right),
            \label{seq: Bianchi-A-average}
        \\
        \braket{\left( \Delta S \right)^2} &= \frac{3}{4} - \log 2 + o \left( 1 \right) \nonumber \\
        &= 0.0568528 \cdots + o \left( 1 \right)
            \label{seq: Bianchi-A-variance}
\end{align}
for the half bipartition and the half filling.
However, no analytical or numerical results have 
been
obtained in the presence of time-reversal symmetry (i.e., classes AI and AII).
Below, we numerically calculate the average and variance of typical entanglement entropy and show that the $O \left( 1 \right)$ constant terms of the average and the variance crucially depend on time-reversal symmetry.
In Appendix~\ref{sec: Weingarten}, we also derive these results analytically.

As shown in Fig.~\ref{sfig: standard}\,(a), the average $\braket{S}$ of entanglement entropy grows almost linearly with respect to the system size $N$ for all the three symmetry classes.
It is also consistent with the analytical result for class A in Eq.~(\ref{seq: Bianchi-A-average}).
Then, we fit the numerical results by [Fig.~\ref{sfig: standard}\,(b)]
\begin{align}
    \braket{S} = \left( \log 2 - \frac{1}{2} \right) N + S_0 + \frac{S_{-1}}{N} + o \left( 1/N \right).
        \label{seq: fit-av-standard}
\end{align}
The fitting results for all the three symmetry classes are summarized in Table~\ref{stab: classifying space}.
The tiny $O \left( 1 \right)$ term $S_0 \simeq 2 \times 10^{-5}$ and $O \left( 1/N \right)$ term $S_{-1} \simeq 0.245$ for class A are compatible with the analytical result in Eq.~(\ref{seq: Bianchi-A-average}).
In contrast to class A, the $O \left( 1 \right)$ constant terms are present in classes AI and AII.
They are negative for class AI and positive for class AII.
On the basis of $\log 2 - 1/2 = 0.193147 \cdots$, the average entanglement entropy is supposed to be
\begin{align}
    \braket{S} = \left( \log 2 - \frac{1}{2} \right) \left( N+1-\frac{2}{\beta} \right) + o \left( 1 \right)
    \label{eqn:ent-standard}
\end{align}
with the Dyson index $\beta =1$ (class AI), $\beta = 2$ (class A), and $\beta = 4$ (class AII).
We analytically demonstrate this result in Appendix~\ref{sec: Weingarten}.
We speculate that
this result implies that one particle is effectively removed in class AI and one half of particle is added in class AII in comparison with class A.

Figure~\ref{sfig: standard}\,(c) shows the variance $\braket{\left( \Delta S \right)^2}$ of entanglement entropy for the three symmetry classes.
Clearly, each symmetry class is characterized by the different values of $\braket{\left( \Delta S \right)^2}$;
in comparison with class A, $\braket{\left( \Delta S \right)^2}$ increases in class AI and decreases in class AII.
In general, level repulsion of random matrices is suppressed (enhanced) in class AI (AII), which is also compatible with the larger (smaller) value of $\braket{\left( \Delta S \right)^2}$ in class AI (AII)~\cite{Mehta-textbook, Forrester-textbook, Haake-textbook}.
We fit the numerical results of $\braket{\left( \Delta S \right)^2}$ by [Fig.~\ref{sfig: standard}\,(d)]
\begin{align}
    \braket{\left( \Delta S \right)^2} = \sigma_0^2 + \frac{\sigma_{-1}^2}{N} + o \left( 1/N \right),
        \label{seq: fit-variance}
\end{align}
which are summarized in Table~\ref{stab: classifying space}.
Here, the numerically obtained variance $\sigma_0^2 \simeq 0.057$ in class A is compatible with the analytical result in Eq.~(\ref{seq: Bianchi-A-variance}).
The variance $\sigma_0^2$ in class AI is almost twice larger than that in class A while that in class AII is almost half of that in class A.
These numerical results are 
\begin{align}
    \braket{\left( \Delta S\right)^2} = \frac{2}{\beta} \left( \frac{3}{4} - \log 2 \right) + o \left( 1 \right)
        \label{seq: variance-standard}
\end{align}
with the Dyson index $\beta =1$ (class AI), $\beta = 2$ (class A), and $\beta = 4$ (class AII).
This behavior of the variance of entanglement entropy reminds us of the universal conductance fluctuations, which are one of the important applications of random matrix theory in condensed matter physics~\cite{Washburn-86, Beenakker-review-97, *Beenakker-review-15}.
In fact, in the diffusive regime of mesoscopic wires, the variance of conductance universally behaves as $\propto 1/\beta$, similar to the variance of entanglement entropy in our work.

\subsection{Chiral class (classes AIII, BDI, and CII)}

In the chiral classes (classes AIII, BDI, and CII), Hamiltonians are concerned with chiral symmetry.
In the many-body Fock space, chiral symmetry (or equivalently, sublattice symmetry) is defined by the antiunitary operation
\begin{align}
    \hat{\cal S} \hat{c}_{m} \hat{\cal S}^{-1}
    = \sum_{n} S_{mn} \hat{c}_{n}^{\dag},
\end{align}
where $\hat{\cal S}$ is an antiunitary operator on the many-body fermionic Fock space, and $S = \left( S_{mn} \right)_{m, n}$ is a unitary matrix on the single-particle Hilbert space.
In contrast to time-reversal symmetry, this operation mixes fermion annihilation and creation operators.
In the simultaneous presence of time-reversal symmetry and particle-hole symmetry, chiral symmetry appears as a combination of the two symmetries.
Even in the absence of time-reversal symmetry and particle-hole symmetry, chiral symmetry can be respected, for example, in bipartite hopping models.
The system respects chiral symmetry if the many-body Hamiltonian satisfies 
\begin{align}
    \hat{\cal S} \hat{H} \hat{\cal S}^{-1}
    = \hat{H},
\end{align}
which leads to $\tr\,H = 0$ and
\begin{align}
    S^{-1} H S = - H.
\end{align}
The matrix $S$ can be chosen to be Hermitian and satisfy $S^2 = 1$ without loss of generality.
In the presence of chiral symmetry, single-particle eigenenergies appear in opposite-sign pairs $\left( E, -E \right)$, and zero-energy modes are subject to a special constraint.
According to the combination of chiral symmetry and time-reversal symmetry, the chiral classes---classes AIII, BDI, and CII---are defined as follows:

\begin{itemize}
\item In the absence of time-reversal symmetry, chiral-symmetric Hamiltonians are defined to belong to the chiral unitary class (class AIII).

\item In the presence of time-reversal symmetry with the sign $T^{*}T=+1$, chiral-symmetric Hamiltonians are defined to belong to the chiral orthogonal class (class BDI).

\item In the presence of time-reversal symmetry with the sign $T^{*}T=-1$, chiral-symmetric Hamiltonians are defined to belong to the chiral symplectic class (class CII).
Because of symplectic time-reversal symmetry, Hamiltonians in class CII generally exhibit the Kramers degeneracy.
\end{itemize}
In classes BDI and CII, time-reversal symmetry is imposed so that it will commute with chiral symmetry.

\begin{figure*}[t]
\centering
\includegraphics[width=\linewidth]{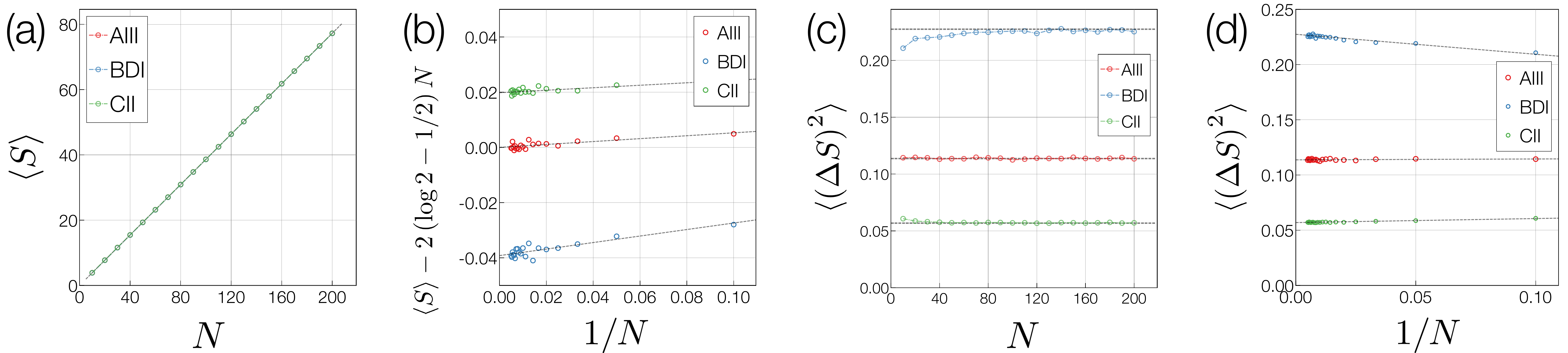} 
\caption{Typical quantum entanglement entropy in the chiral classes with the half bipartition and half filling for classes AIII (red dots), BDI (blue dots), and CII (green dots). 
Each datum is averaged over $10^5$ ensembles. 
(a)~Average $\braket{S}$ of entanglement entropy as functions of the system size $N$. 
The black dashed line is the analytical result $\braket{S} \simeq 2 \left( \log 2-1/2 \right) N$. 
(b)~Deviation of $\braket{S}$ from the volume-law term $2 \left( \log 2 - 1/2 \right) N$ as functions of $1/N$. 
(c)~Variance $\braket{\left( \Delta S\right)^2}$ of entanglement entropy as functions of $N$. 
The black dashed lines are the analytical results $\braket{\left( \Delta S\right)^2} = 4\left( 3/4 - \log 2\right)/\beta$ with the Dyson index $\beta = 1$ (class BDI), $\beta = 2$ (class AIII), and $\beta = 4$ (class CII). 
(d)~Variance $\braket{\left( \Delta S\right)^2}$ as functions of $1/N$.}
	\label{sfig: chiral}
\end{figure*}

We obtain the typical quantum entanglement entropy in the chiral classes.
Suppose the total system size is $2N$ ($4N$), the subsystem size is $N_A$ ($2N_A$), and the particle number is $M$ ($2M$) in classes AIII and BDI (class CII).
Here, $N$ denotes the number of the unit cell.
Because of chiral symmetry, the single-particle Hamiltonian can be expressed as
\begin{align}
    H = \begin{pmatrix}
        0 & h \\
        h^{\dag} & 0
    \end{pmatrix},
        \label{seq: H - chiral - h}
\end{align}
where $h$ is an $N \times N$ ($2N \times 2N$) matrix in classes AIII and BDI (class CII).
In this representation, the chiral-symmetry operator is chosen as $S = \sigma_z \otimes I_{N (2N)}$, where $I_{N (2N)}$ is the $N \times N$ ($2N \times 2N$) identity matrix.
In the following, we assume the absence of zero modes.
The same assumption was imposed also in Refs.~\cite{Liu-18, Bianchi-21, Bianchi-22} for BdG Hamiltonians in class D.
We are interested only in the entanglement properties of eigenstates;
we can flatten the spectrum of the Hamiltonian to $\pm 1$ and deform the matrix $h$ so that it will be unitary.
Depending on additional time-reversal symmetry, the matrix $h$ belongs to the unitary group $\mathrm{U} \left( N \right)$, orthogonal group $\mathrm{O} \left( N \right)$, and symplectic group $\mathrm{Sp} \left( N \right)$:
\begin{align}
    h \in \begin{cases}
        \mathrm{U} \left( N \right) & \left( \text{class AIII}\right); \\
        \mathrm{O} \left( N \right) & \left( \text{class BDI}\right); \\
        \mathrm{Sp} \left( N \right) & \left( \text{class CII}\right).
    \end{cases}
        \label{seq: h-group}
\end{align}
These groups give the classifying spaces in the chiral classes and characterize eigenstates and their entanglement (Table~\ref{stab: classifying space}).

A generic normalized single-particle eigenstates of $H$ in Eq.~(\ref{seq: H - chiral - h}) is given as
\begin{align}
    \ket{u_i} = \frac{1}{\sqrt{2}} \begin{pmatrix}
        \ket{n_i} \\ \epsilon h^{\dag} \ket{n_i}
    \end{pmatrix},
\end{align}
where $\epsilon \in \{ \pm 1\}$ is the flattened single-particle eigenenergy (i.e., $H \ket{u_i} = \epsilon \ket{u_i}$), and $\ket{n_i}$ is the $N$-dimensional ($2N$-dimensional) orthonormal basis in classes AIII and BDI (class CII) that satisfies $\ket{n_i}_j = \delta_{ij}$.
Then, the $2N \times 2N$ ($4N \times 4N$) unitary matrix $U$ collecting all the $2N$ ($4N$) eigenstates in classes AIII and BDI (class CII) is given as
\begin{align}
    U = \frac{1}{\sqrt{2}} \begin{pmatrix}
    I_{N (2N)} & I_{N (2N)} \\
    h^{\dag} & -h^{\dag}
    \end{pmatrix}.
    \label{eqn:U-chiral}
\end{align}
From this eigenstate matrix $U$, the truncated correlation matrix $C_A$ is given as Eq.~(\ref{seq: CA}), and the entanglement entropy is obtained as Eq.~(\ref{seq: entanglement sum}).
Because of chiral symmetry, the eigenvalues of $C_A$ come in $\left( \lambda, 1-\lambda \right)$ pairs.
We numerically calculate the average and variance of entanglement entropy in the chiral classes by Haar-randomly choosing $h$ in each classifying space in Eq.~(\ref{seq: h-group}).
Here, we focus on the half-filled many-body eigenstates constructed from all the single-particle eigenstates with negative eigenenergies.
Similarly to class AII, the spectrum of the correlation matrix $C_A$ is two-fold degenerate in class CII, and we calculate the entanglement entropy only from half of the entanglement spectrum.

In a similar manner to the standard classes, the average $\braket{S}$ of entanglement entropy grows almost linearly with respect to the system size $N$ for all the three symmetry classes [Fig.~\ref{sfig: chiral}\,(a)].
Then, we fit the numerical results by [Fig.~\ref{sfig: chiral}\,(b)]
\begin{align}
    \braket{S} = 2 \left( \log 2 - \frac{1}{2} \right) N + S_0 + \frac{S_{-1}}{N} + o \left( 1/N \right).
        \label{seq: fit-av-chiral-BdG}
\end{align}
We note that the total number of sites is chosen to be $2N$ in the chiral classes. 
The fitting results for all the three symmetry classes are summarized in Table~\ref{stab: classifying space}.
In class AIII, the $O \left( 1 \right)$ term $S_0 \simeq - 1 \times 10^{-4}$ is tiny, similar to class A.
On the other hand, the $O \left( 1/N \right)$ term $S_1 \simeq 0.056$ in class AIII is much smaller than that in class A.
Time-reversal symmetry with the sign $+1$ gives rise to the negative $O \left( 1 \right)$ term $S_0 \simeq -0.039$ in class BDI while time-reversal symmetry with the sign $-1$ gives rise to the positive $O \left( 1 \right)$ term $S_0 \simeq 0.020$ in class CII, both of which are much smaller than the $O \left( 1 \right)$ terms in classes AI and AII.
In Appendix~\ref{sec: Weingarten}, we analytically show
\begin{align}
    \braket{S} = 2\left( \log 2 - \frac{1}{2} \right) N + \left( 1- \frac{2}{\beta} \right) \left(  \frac{3}{2} \log 2 - 1\right) + o \left( 1 \right)
        \label{eqn:ent-chiral}
\end{align}
with the Dyson index $\beta =1$ (class BDI), $\beta = 2$ (class AIII), and $\beta = 4$ (class CII), which are consistent with the numerical results. 

Figure~\ref{sfig: chiral}\,(c) shows the variance $\braket{\left( \Delta S \right)^2}$ of entanglement entropy for the chiral classes.
Each symmetry class is characterized by the different values of $\braket{\left( \Delta S \right)^2}$.
In comparison with class AIII, $\braket{\left( \Delta S \right)^2}$ increases in class BDI and decreases in class CII, which is similar to the standard classes.
We fit the numerical results of $\braket{\left( \Delta S \right)^2}$ by Eq.~(\ref{seq: fit-variance}), as summarized in Fig.~\ref{sfig: chiral}\,(d) and Table~\ref{stab: classifying space}.
Notably, these numerical results are 
\begin{align}
    \braket{\left( \Delta S\right)^2} = \frac{4}{\beta} \left( \frac{3}{4} - \log 2 \right) + o \left( 1 \right)
        \label{seq: variance-formula-chiral-BdG}
\end{align}
with the Dyson index $\beta =1$ (class BDI), $\beta = 2$ (class AIII), and $\beta = 4$ (class CII).
This is twice larger than the variance in the standard classes [i.e., Eq.~(\ref{seq: variance-standard})].

\subsection{Bogoliubov-de Gennes (BdG) class (classes D, DIII, C, and CI)}

The BdG classes (classes, D, C, DIII, and CI) are concerned with particle-hole symmetry.
Particle-hole symmetry (or equivalently, charge-conjugation symmetry) is described by the unitary operation defined by
\begin{align}
    \hat{\cal C} \hat{c}_{m} \hat{\cal C}^{-1}
    = \sum_{n} C^{*}_{mn} \hat{c}_{n}^{\dag},
\end{align}
where $\hat{\cal C}$ and $C = \left( C_{mn} \right)_{m, n}$ are unitary many-body operators and single-particle matrices, respectively.
It describes the transformation between particles and holes, and flips the sign of the electron charge with respect to the charge neutral point 
$\hat{\cal C} \hat{Q} \hat{\cal C}^{-1} = - \hat{Q}$
with $\hat{Q} \coloneqq \hat{N} - N/2$.
The many-body Hamiltonian is particle-hole symmetric if it satisfies
\begin{align}
    \hat{\cal C} \hat{H} \hat{\cal C}^{-1}
    = \hat{H},
\end{align}
which leads to $\tr\,H = 0$ and
\begin{align}
    C^{-1} H^{T} C = - H.
\end{align}
Thus, particle-hole symmetry acts as unitary symmetry on the many-body fermionic Fock space but acts as antiunitary symmetry on the single-particle Hilbert space.
Similarly to time-reversal symmetry, the symmetry operator and matrix are required to satisfy
\begin{align}
    \hat{\cal C}^{2} = \left( \pm 1 \right)^{\hat{N}},\quad
    C^{*} C = \pm 1.
\end{align}
In the presence of particle-hole symmetry, single-particle eigenenergies appear in opposite-sign pairs $\left( E, -E \right)$, and zero-energy modes are subject to a special constraint.
According to the combination of particle-hole symmetry and time-reversal symmetry, the BdG classes---classes D, C, DIII, and CI---are defined as follows:

\begin{itemize}
\item In the sole presence of particle-hole symmetry with the sign $C^{*} C = +1$, particle-hole-symmetric Hamiltonians are defined to belong to class D.
In the additional presence of time-reversal symmetry with the sign $T^{*} T = -1$, Hamiltonians are defined to belong to class DIII.

\item In the sole presence of particle-hole symmetry with the sign $C^{*} C = -1$, particle-hole-symmetric Hamiltonians are defined to belong to class C.
In the additional presence of time-reversal symmetry with the sign $T^{*} T = +1$, Hamiltonians are defined to belong to class CI.
\end{itemize}
In classes DIII and CI, the combination of time-reversal symmetry and particle-hole symmetry gives rise to chiral symmetry, which anticommutes with time-reversal symmetry and particle-hole symmetry.

We obtain the typical quantum entanglement entropy in the BdG classes.
Similarly to the chiral classes, we assume the absence of zero modes and flatten the spectrum to be $\{ \pm 1 \}$.
First, $2N \times 2N$ single-particle Hamiltonians $H$ in class D respect particle-hole symmetry
\begin{align}
    H^{*} = - H,
\end{align}
where the particle-hole-symmetry operator is chosen as $C = I_{2N}$.
If we define $A$ by $H \eqqcolon \ii A$, $A$ is a real antisymmetric matrix.
Then, we diagonalize $A$ with a proper basis as
\begin{align}
    A = O \begin{pmatrix}
        0 & I_N \\
        -I_N & 0
    \end{pmatrix} O^{-1},
        \label{seq: class D A-O}
\end{align}
where $O$ is a $2N \times 2N$ orthogonal matrix
\begin{align}  
    O \in \mathrm{O} \left( 2N \right).
        \label{seq: classifying space D}
\end{align}
This orthogonal matrix $O$ contains all information about single-particle eigenstates, and the orthogonal group $O \in \mathrm{O} \left( 2N \right)$ gives the classifying space in class D (Table~\ref{stab: classifying space}).
In fact, from this orthogonal matrix $O$, a generic normalized eigenstate of $H$ is given as
\begin{align}
    \ket{u_i} = \frac{O}{\sqrt{2}} \begin{pmatrix}
        \ket{n_i} \\ - \ii \epsilon \ket{n_i}
    \end{pmatrix},
\end{align}
where $\epsilon \in \{ \pm 1\}$ is the flattened eigenenergy (i.e., $H \ket{u_i} = \epsilon \ket{u_i}$), and $\ket{n_i}$ is the $N$-dimensional orthonormal basis that satisfies $\ket{n_i}_j = \delta_{ij}$.
Then, the $2N \times 2N$ unitary matrix $U$ collecting all the $2N$ eigenstates is given as
\begin{align}
    U = \frac{O}{\sqrt{2}} \begin{pmatrix}
        I_N & I_N \\
        -\ii \times I_N & \ii \times I_N
    \end{pmatrix},
    \label{eqn:U-class-D}
\end{align}
where $I_N$ is the $N \times N$ identify matrix.
From this unitary matrix $U$, the truncated correlation matrix $C_A$ is given as Eq.~(\ref{seq: CA}), and the entanglement entropy is obtained as Eq.~(\ref{seq: entanglement sum}).
Because of particle-hole symmetry, the eigenvalues of $C_A$ come in $\left( \lambda, 1-\lambda \right)$ pairs.
We numerically calculate the average and variance of entanglement entropy in class D by Haar-randomly choosing $O$ in the classifying space in Eq.~(\ref{seq: classifying space D}).
Similarly to the chiral classes, we focus on the half-filled many-body eigenstates constructed from all the single-particle eigenstates with negative eigenenergies.

\begin{figure*}[t]
\centering
\includegraphics[width=\linewidth]{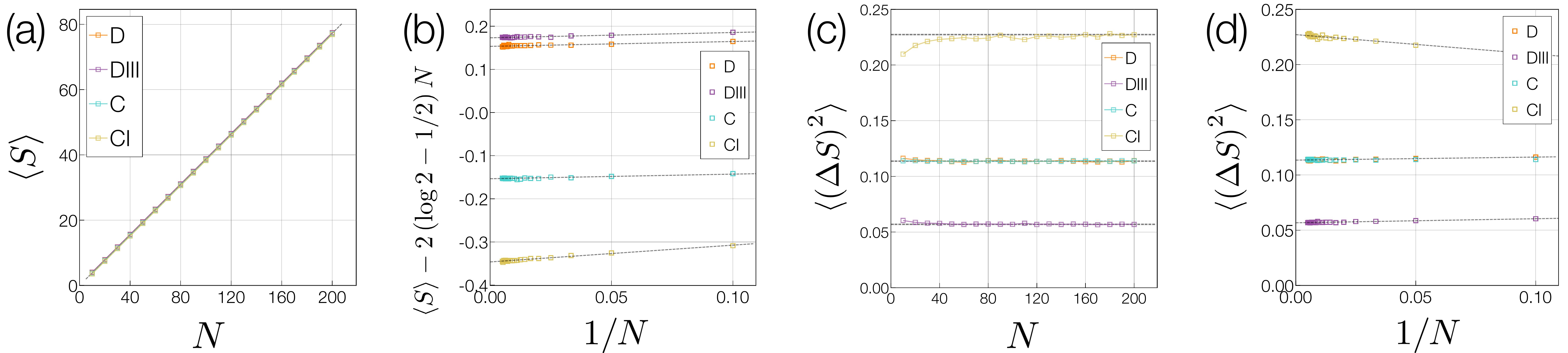} 
\caption{Typical quantum entanglement entropy in the Bogoliubov-de Gennes (BdG) classes with the half bipartition and half filling for classes D (orange dots), DIII (purple dots), C (light-blue dots), and CI (yellow dots). 
Each datum is averaged over $10^5$ ensembles. 
(a)~Average $\braket{S}$ of entanglement entropy as functions of the system size $N$. 
The black dashed line is the analytical result $\braket{S} \simeq 2 \left( \log 2-1/2 \right) N$. 
(b)~Deviation of $\braket{S}$ from the volume-law term $2 \left( \log 2 - 1/2 \right) N$ as functions of $1/N$. 
(c)~Variance $\braket{\left( \Delta S\right)^2}$ of entanglement entropy as functions of $N$. 
The black dashed lines are the analytical results $\braket{\left( \Delta S\right)^2} = 4\left( 3/4 - \log 2\right)/\beta$ with the Dyson index $\beta = 1$ (class CI), $\beta = 2$ (classes D and C), and $\beta = 4$ (class DIII). 
(d)~Variance $\braket{\left( \Delta S\right)^2}$ as functions of $1/N$.}
	\label{sfig: BdG}
\end{figure*}

Next, $2N \times 2N$ single-particle Hamiltonians $H$ in class C respect particle-hole symmetry
\begin{align}
    \sigma_y H^{*} \sigma_y = - H,
\end{align}
where the particle-hole-symmetry operator is chosen as $C = \sigma_y \otimes I_{N}$.
If we define $A$ by $H \eqqcolon \ii A$, $A$ satisfies $A^{\dag} = - A$ and $\sigma_y A^{*} \sigma_y = A$.
Then, we diagonalize $A$ with a proper basis as
\begin{align}
    A = U \begin{pmatrix}
        0 & I_N \\
        -I_N & 0
    \end{pmatrix} U^{-1},
        \label{seq: class C A-U}
\end{align}
where $U$ is a $2N \times 2N$ symplectic matrix
\begin{align}  
    U \in \mathrm{Sp} \left( N \right).
        \label{seq: classifying space C}
\end{align}
Using this symplectic matrix $U$ instead of the orthogonal matrix $O$ in Eq.~(\ref{seq: classifying space D}), we obtain entanglement entropy in a similar manner to class D.

In class DIII, chiral symmetry is present as a combination of time-reversal symmetry and particle-hole symmetry.
Owing to chiral symmetry, $4N \times 4N$ single-particle Hamiltonians in class DIII are generally written as Eq.~(\ref{seq: H - chiral - h}) with $h \in \mathrm{U} \left( 2N \right)$, where the chiral-symmetry operator is chosen as $S \coloneqq \sigma_z \otimes I_{2N}$.
In addition, we choose the time-reversal-symmetry operator as $T = \sigma_x \otimes \ii \sigma_y \otimes I_{N}$.
Then, time-reversal symmetry $T^{-1} H^{*} T = H$ imposes 
\begin{align}
    \left( \sigma_y \otimes I_{N} \right) h^{T} \left( \sigma_y \otimes I_{N} \right) = h,
        \label{seq: DIII-TRS}
\end{align}
leading to the general representation 
\begin{align}
    h = f^{T} \left( \sigma_y \otimes I_{N} \right) f \left( \sigma_y \otimes I_{N} \right),\quad f \in \mathrm{U} \left( 2N \right).
        \label{seq: classifying space DIII}
\end{align}
Thus, $f \in \mathrm{U} \left( 2N \right)$ contains all information about the single-particle eigenstates, and the unitary group $\mathrm{U} \left( 2N \right)$ gives the classifying space in class DIII (Table~\ref{stab: classifying space}).
In other words, $h$ belongs to the circular symplectic ensemble.
Similarly to classes AII and CII, the spectrum of the correlation matrix $C_A$ is two-fold degenerate in class DIII, and we calculate the entanglement entropy only from half of the entanglement spectrum.

In class CI, $2N \times 2N$ single-particle Hamiltonians also respect chiral symmetry and are generally written as Eq.~(\ref{seq: H - chiral - h}) with $h \in \mathrm{U} \left( N \right)$, where the chiral-symmetry operator is chosen as $S \coloneqq \sigma_z \otimes I_{N}$.
Then, we choose the time-reversal-symmetry operator as $T = \sigma_x \otimes I_N$.
Time-reversal symmetry imposes 
\begin{align}
    h^{T} = h,
        \label{seq: CI-TRS}
\end{align}
leading to the general representation
\begin{align}
    h = f^{T} f,\quad f \in \mathrm{U} \left( N \right).
        \label{seq: classifying space CI}
\end{align}
Using this general representation of $h$, we calculate entanglement entropy in a similar manner to class DIII.
Here, $h$ belongs to the circular orthogonal ensemble.

In passing, we note that the random matrices in Eqs.~(\ref{seq: classifying space D}), (\ref{seq: classifying space C}), (\ref{seq: classifying space DIII}), and (\ref{seq: classifying space CI}) have gauge ambiguity.
For example, in class D, the orthogonal matrix $O$ in Eq.~(\ref{seq: classifying space D}) obeys the gauge transformation 
$O \to O \tilde{O}$ satisfying
\begin{align}
    \tilde{O} \begin{pmatrix}
        0 & I_N \\
        -I_{N} & 0
    \end{pmatrix} \tilde{O}^{-1} = \begin{pmatrix}
        0 & I_N \\
        -I_{N} & 0
    \end{pmatrix},\quad
    \tilde{O} \in \mathrm{O} \left( 2N \right).
\end{align}
If we introduce a matrix $G$ that rotates $\sigma_z \otimes I_N$ to $\sigma_y \otimes I_N$, i.e.,
\begin{gather}
    \sigma_y \otimes I_N = G \left( \sigma_z \otimes I_N \right) G^{-1}, \\
    G \coloneqq \frac{1}{\sqrt{2}} \begin{pmatrix} 
        I_N & - \ii \times I_N \\
        \ii \times I_N & I_N
    \end{pmatrix},
\end{gather}
the above gauge transformation reduces to
\begin{align}
    ( G^{-1} \tilde{O} G ) \begin{pmatrix}
        I_N & 0 \\
        0 & - I_N
    \end{pmatrix} ( G^{-1} \tilde{O} G )^{-1} = \begin{pmatrix}
        I_N & 0 \\
        0 & - I_N
    \end{pmatrix}.
\end{align}
Hence, the allowed gauge transformation is generally given by
\begin{align}
    \tilde{O} = G \begin{pmatrix}
        W_n & 0 \\
        0 & W_{n}^{*}
    \end{pmatrix} G^{-1},\quad W_{n} \in \mathrm{U} \left( N \right).
\end{align}
Thus, $O$ in Eq.~(\ref{seq: class D A-O}) belongs to $\mathrm{O} \left( 2N \right)/\mathrm{U} \left( N \right)$, which precisely gives the classifying space in class D.
Such gauge ambiguity is also summarized in Table~\ref{stab: classifying space} for all the symmetry classes.
In our calculations of typical quantum entanglement entropy, the gauge ambiguity of the classifying spaces is irrelevant, although it is relevant, for example, to Anderson localization~\cite{Evers-review} and topological insulators and superconductors~\cite{CTSR-review}.

From the above representations, we calculate the average and variance of entanglement entropy in the BdG classes (classes D, DIII, C, and CI), as summarized in Fig.~\ref{sfig: BdG}.
Similarly to the other symmetry classes, the average $\braket{S}$ of entanglement entropy grows almost linearly with respect to the system size $N$ for all the four symmetry classes [Fig.~\ref{sfig: BdG}\,(a)].
Then, we fit the numerical results by Eq.~(\ref{seq: fit-av-chiral-BdG}) [Fig.~\ref{sfig: BdG}\,(b)].
The fitting results for all the four symmetry classes are summarized in Table~\ref{stab: classifying space}.
The $O \left( 1 \right)$ term of $\braket{S}$ in class D is positive, and that in class C is negative, both of which have almost the same absolute value.
In Appendix~\ref{sec: Weingarten}, we analytically show
\begin{align}
    \braket{S} = 2\left( \log 2 - \frac{1}{2} \right) N + \frac{1}{2} \left( 1-\alpha \right) \left( 1 - \log 2 \right) + o \left( 1 \right)
        \label{eqn:ent-D&C}
\end{align}
with the index $\alpha = 0$ (class D) and $\alpha = 2$ (class C)~\cite{AZ-97, Haake-textbook}. 
On the other hand, the $O \left( 1 \right)$ term of $\braket{S}$ in class CI is twice larger than that in class DIII, the signs of which are opposite to each other.
This behavior is similar to the average entanglement in classes AI and AII, as well as that in classes BDI and CII.
In Appendix~\ref{sec: Weingarten}, we analytically show
\begin{align}
    \braket{S} = 2\left( \log 2 - \frac{1}{2} \right) N + \frac{1}{2} \left( 1 - \frac{2}{\beta} \right) \log 2 + o \left( 1 \right)
        \label{eqn:ent-DIII&CI}
\end{align}
with the Dyson index $\beta = 1$ (class CI) and $\beta = 4$ (class DIII), which are consistent with the numerical results. 

Figure~\ref{sfig: BdG}\,(c) shows the variance $\braket{\left( \Delta S \right)^2}$ of entanglement entropy for the BdG classes.
The variance $\braket{\left( \Delta S \right)^2}$ of entanglement entropy coincides with each other in classes D and C,
and $\braket{\left( \Delta S \right)^2}$ in classes DIII and CI is twice larger and smaller than that, respectively.
We fit the numerical results of $\braket{\left( \Delta S \right)^2}$ by Eq.~(\ref{seq: fit-variance}), as summarized in Fig.~\ref{sfig: BdG}\,(d) and Table~\ref{stab: classifying space}.
Notably, these numerical results follow Eq.~(\ref{seq: variance-formula-chiral-BdG}) with the Dyson index $\beta =1$ (class CI), $\beta = 2$ (classes D and C), and $\beta = 4$ (class DIII).
This is twice larger than the variances in the standard classes [i.e., Eq.~(\ref{seq: variance-standard})] and the same as those in the chiral classes.

\section{Typical entanglement entropy of Bogoliubov-de Gennes (BdG) Hamiltonians}
    \label{sec: numerics - AZ - BdG}

\begin{table*}[t]
	\centering
	\caption{Typical quantum entanglement entropy of particle-number-nonconserving Bogoliubov-de Gennes (BdG) Hamiltonians in classes D, DIII, C, and CI.
    The Altland-Zirnbauer (AZ) symmetry classes consist of time-reversal symmetry (TRS), particle-hole symmetry (PHS), and chiral symmetry (CS). 
    For TRS and PHS, the entries ``$\pm 1$" mean the presence of symmetry and its sign, and the entries ``$0$" mean the absence of symmetry.
    For CS, the entries ``$1$" and ``$0$" mean the presence and absence of symmetry, respectively.
    Each class is characterized by the classifying space and the random-matrix indices $\left( \alpha, \beta \right)$.
    The numerical fitting results of the average entanglement entropy by $\braket{S} = \left( \log 2 - 1/2 \right) N + S_0 + S_{-1}/N$, as well as those of the variance of entanglement entropy by $\braket{\left( \Delta S \right)^2} = \sigma_0^2 + \sigma_{-}^{2}/N$, are shown.
    All the results of entanglement entropy are calculated for particle-number-nonconserving BdG Hamiltonians with the half bipartition.
    }
     \begin{tabular}{cccccccccccc} \hline \hline
     ~~AZ class~~ & ~~TRS~~ & ~~PHS~~ & ~~CS~~ & & ~Classifying space~ & ~$\beta$~ & ~$\alpha$~ & $S_0$ & $S_{-1}$ & $\sigma_0^2$ & $\sigma_{-1}^2$  \\ \hline
     D & $0$ & $+1$ & $0$ & ${\cal R}_{2}$ & ${\rm O} \left( 2N \right) / {\rm U} \left( N \right)$ & $2$ & $0$ & ~~$0.077$~~ & ~~$0.064$~~ & ~~$0.028$~~ & ~~$0.004$~~\\
     DIII & $-1$ & $+1$ & $1$ & ${\cal R}_{3}$ & ~~${\rm U} \left( 2N \right) / {\rm Sp} \left( N \right)$~~ & $4$ & $1$ & $0.087$ & $0.059$ & $0.014$ & $0.009$ \\
     C & $0$ & $-1$ & $0$ & ${\cal R}_{6}$ & ${\rm Sp} \left( N \right) / {\rm U} \left( N \right)$ & $2$ & $2$ & ~~$-0.077$~~ & $0.070$ & $0.028$ &  ~~$0.003$~~ \\
     CI & $+1$ & $-1$ & $1$ & ${\cal R}_{7}$ & ${\rm U} \left( N \right) / {\rm O} \left( N \right)$ & $1$ & $1$ & $-0.173$ & $0.182$ & $0.057$ & ~~$-0.042$~~ \\ \hline \hline
     \end{tabular}
  	\label{stab: BdG}
\end{table*}

\begin{figure*}[t]
\centering
\includegraphics[width=\linewidth]{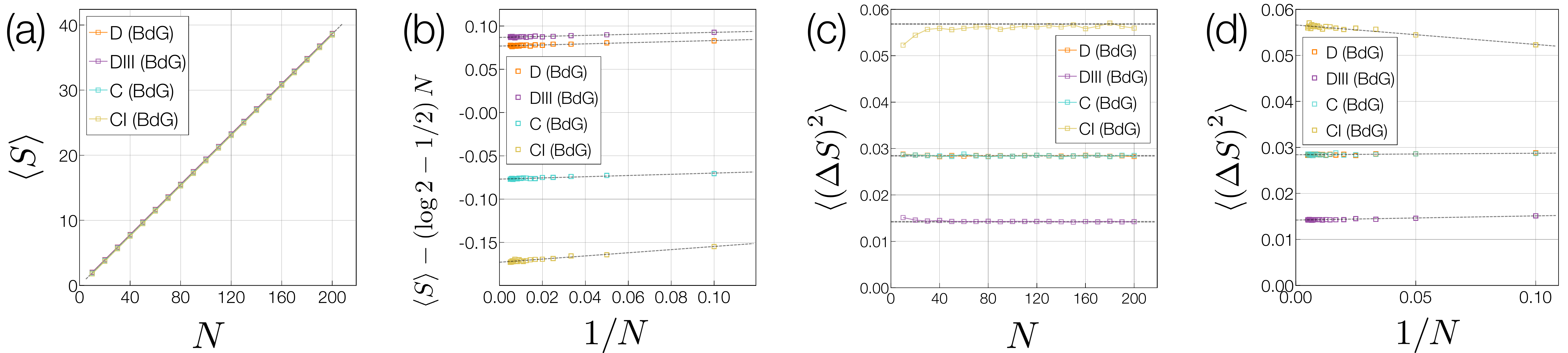} 
\caption{Typical quantum entanglement entropy of particle-number-nonconserving Bogoliubov-de Gennes (BdG) Hamiltonians with the half bipartition for classes D (orange dots), DIII (purple dots), C (light-blue dots), and CI (yellow dots). 
Each datum is averaged over $10^5$ ensembles. 
(a)~Average $\braket{S}$ of entanglement entropy as functions of the system size $N$. 
The black dashed line is the analytical result $\braket{S} \simeq \left( \log 2-1/2 \right) N$. 
(b)~Deviation of $\braket{S}$ from the volume-law term $\left( \log 2 - 1/2 \right) N$ as functions of $1/N$. 
(c)~Variance $\braket{\left( \Delta S\right)^2}$ of entanglement entropy as functions of $N$. 
The black dashed lines are the analytical results $\braket{\left( \Delta S\right)^2} = \left( 3/4 - \log 2\right)/\beta$ with the Dyson index $\beta = 1$ (class CI), $\beta = 2$ (classes D and C), and $\beta = 4$ (class DIII). 
(d)~Variance $\braket{\left( \Delta S\right)^2}$ as functions of $1/N$.}
	\label{sfig: BdG-Majorana}
\end{figure*}

We obtain typical quantum entanglement entropy of BdG Hamiltonians that violate the conservation of the particle number.
In general, BdG Hamiltonians read
\begin{align}
    \hat{H} = \hat{\Gamma}^{\dag} H \hat{\Gamma},
        \label{seq: BdG}
\end{align}
where $\hat{\Gamma}$ is the Nambu spinor that consists of both annihilation and creation operators, and $H$ is the single-particle BdG Hamiltonian.
For the BdG classes (i.e., classes D, DIII, C, and CI), we calculate the average and variance of entanglement entropy for eigenstates Haar-randomly chosen from the corresponding classifying spaces (see below for details).
The results are summarized in Table~\ref{stab: BdG} and Fig.~\ref{sfig: BdG-Majorana}.
Notably, the average entanglement entropy is half as large as that of particle-number-conserving free fermions in the corresponding symmetry classes, and the variance of entanglement entropy is quarter as large as that of particle-number-conserving free fermions in the corresponding symmetry classes (compare Table~\ref{stab: classifying space} with Table~\ref{stab: BdG}).
In particular, the variance of entanglement entropy is 
\begin{align}
    \braket{\left( \Delta S\right)^2} = \frac{1}{\beta} \left( \frac{3}{4} - \log 2 \right) + o \left( 1 \right)
        \label{seq: BdG-variance-Majorana}
\end{align}
with the Dyson index $\beta =1$ (class CI), $\beta = 2$ (classes D and C), and $\beta = 4$ (class DIII).
While the fundamental constituents of particle-number-conserving free fermions are complex fermions, those of BdG Hamiltonians are Majorana fermions.
Majorana fermions effectively have half the degree of freedom in comparison with complex fermions, which results in the half average and quarter variance of entanglement entropy.
In the rest of this section, we describe how to calculate entanglement entropy of BdG Hamiltonians in each symmetry class.

\subsection{Class D}

In class D, generic BdG Hamiltonians read Eq.~(\ref{seq: BdG}) with the spinless Nambu spinor $\hat{\Gamma} \coloneqq ( \hat{c}_1~\cdots~\hat{c}_{N}~\hat{c}_1^{\dag}~\cdots~\hat{c}_{N}^{\dag} )^{T}$.
Here, the number of fermions is denoted by $N$, and $H$ is a $2N \times 2N$ matrix.
Notably, $\hat{\Gamma}$ and $\hat{\Gamma}^{\dag}$ are not independent but are related to each other by 
\begin{align}
    \left[ \left( \tau_x \otimes I_N \right) \hat{\Gamma} \right]^{T} = \hat{\Gamma}^{\dag},\quad
    \left[ \hat{\Gamma}^{\dag} \left( \tau_x \otimes I_N \right) \right]^{T} = \hat{\Gamma}.
\end{align}
Because of this constraint, the BdG Hamiltonian satisfies
\begin{align}
    \hat{H} 
    &= \left[ \left( \tau_x \otimes I_N \right) \hat{\Gamma} \right]^{T} H \left[ \hat{\Gamma}^{\dag} \left( \tau_x \otimes I_N \right) \right]^{T} \nonumber \\
    &= - \hat{\Gamma}^{\dag} \left[ \left( \tau_x \otimes I_{N} \right) H^{T} \left( \tau_x \otimes I_{N} \right) \right] \hat{\Gamma} + \tr\,H,
\end{align}
leading to the particle-hole constraints $\tr\,H = 0$ and
\begin{align}
    \left( \tau_x \otimes I_{N} \right) H^{T} \left( \tau_x \otimes I_{N} \right) = - H.
        \label{seq: PHS constraint BdG-D}
\end{align}
Hence, the single-particle BdG Hamiltonian $H$ belongs to class D.

In class D, it is useful to introduce the Majorana basis.
We define the Majorana fermion operators $\hat{a}_i$ and $\hat{b}_i$ ($i=1, 2, \cdots, N$) by
\begin{align}
    \hat{a}_i \coloneqq \hat{c}_{i} + \hat{c}_{i}^{\dag},\quad
    \hat{b}_i \coloneqq -\ii\,( \hat{c}_{i} - \hat{c}_{i}^{\dag} ).
\end{align}
Then, the BdG Hamiltonian is rewritten as
\begin{align}
    \hat{H} = \frac{1}{2} \begin{pmatrix}
        \hat{a}_1 & \hat{b}_1 & \cdots & \hat{a}_N & \hat{b}_N
    \end{pmatrix}~\tilde{H}~\begin{pmatrix}
        \hat{a}_1 \\ \hat{b}_1 \\ \vdots \\ \hat{a}_N \\ \hat{b}_N
    \end{pmatrix},
\end{align}
with the single-particle Hamiltonian 
\begin{align}
    \tilde{H} = V H V^{\dag},\quad
    V \coloneqq \frac{1}{\sqrt{2}} \begin{pmatrix}
        1 & 1 \\
        -\ii & \ii 
    \end{pmatrix} \otimes I_N.
\end{align}
This is equivalent to the two-body ($q=2$) SYK Hamiltonian~\cite{Liu-18}.
In the Majorana basis, the particle-hole constraint in Eq.~(\ref{seq: PHS constraint BdG-D}) reduces to $\tilde{H}^{T} = - \tilde{H}$.
Then, similarly to Eq.~(\ref{seq: class D A-O}), $\tilde{H}$ is diagonalized by 
\begin{align}
    \tilde{H} = \ii O \begin{pmatrix}
        0 & I_N \\
        -I_N & 0
    \end{pmatrix} O^{T},\quad O \in \mathrm{O} \left( 2N \right).
\end{align}
Here, we assume the absence of zero-energy modes and flatten the single-particle spectrum. 
Introducing 
\begin{align}
    \hat{\tilde{\Gamma}} \coloneqq
    \begin{pmatrix}
        \hat{\tilde{c}}_1 \\ \vdots \\ \hat{\tilde{c}}_N \\ \hat{\tilde{c}}_1^{\dag} \\ \vdots \\ \hat{\tilde{c}}_{N}^{\dag} \\
    \end{pmatrix} \coloneqq \frac{V^{\dag}}{\sqrt{2}}
    \begin{pmatrix}
        \hat{\tilde{a}}_1 \\ \hat{\tilde{b}}_1 \\ \vdots \\ \hat{\tilde{a}}_N \\ \hat{\tilde{b}}_N
    \end{pmatrix} 
    &\coloneqq \frac{V^{\dag} O^{T}}{\sqrt{2}} \begin{pmatrix}
        \hat{a}_1 \\ \hat{b}_1 \\ \vdots \\ \hat{a}_N \\ \hat{b}_N
    \end{pmatrix} \nonumber \\
    &= \left( V^{\dag} O^{T} V \right) \hat{\Gamma},
\end{align}
we have
\begin{align}
    \hat{H} = \sum_{i=1}^{N} \ii \hat{\tilde{a}}_i \hat{\tilde{b}}_i
    = 2 \sum_{i=1}^{N} \left( \tilde{c}_i^{\dag} \tilde{c}_i - \frac{1}{2} \right).
\end{align}
Therefore, the ground state $\ket{\Omega}$ is given as the state annihilated by all $\hat{\tilde{c}}_i$'s (i.e., $\forall\,i~~\hat{\tilde{c}}_i \ket{\Omega} = 0$).

We obtain typical quantum entanglement entropy for the BdG Hamiltonians in class D by calculating the entanglement entropy of $\ket{\Omega}$ for $O$ Haar-randomly chosen from the classifying space $\mathrm{O} \left( 2N \right)$.
In a similar manner to the particle-number-conserving case, the entanglement entropy of BdG Hamiltonians is obtained from the correlation matrix, which reads
\begin{align}
    C &\coloneqq \braket{\Omega\,|\,\hat{\Gamma} \hat{\Gamma}^{\dag}\,|\,\Omega} \nonumber \\
    &= (V^{\dag} O V) \braket{\Omega\,|\,\hat{\tilde{\Gamma}} \hat{\tilde{\Gamma}}^{\dag} \,|\,\Omega} (V^{\dag} O^{T} V) \nonumber \\
    &= (V^{\dag} O V) \begin{pmatrix}
    I_{N} & 0 \\
    0 & 0 
    \end{pmatrix} (V^{\dag} O^{T} V).
\end{align}
Then, the entanglement entropy is obtained as~\cite{Peschel-03, Peschel-review}
\begin{align}
    S = - \frac{1}{2}\sum_{i} \left[ \lambda_i \log \lambda_i + \left( 1- \lambda_i \right) \log \left( 1-\lambda_i \right) \right],
        \label{seq: entanglement sum - BdG}
\end{align}
where $\lambda_i$'s ($i=1, 2, \cdots, 2N_A$) are the eigenvalues of the correlation matrix $C$ constrained to the subsystem.
Here, the eigenvalues of $C$ come in $\left( \lambda, 1-\lambda \right)$ pairs because of particle-hole symmetry.
The coefficient $1/2$ comes from the particle-number-nonconserving nature of BdG Hamiltonians.

According to the numerical results in Table~\ref{stab: BdG} and Fig.~\ref{sfig: BdG-Majorana}, the average entanglement entropy of BdG Hamiltonians in class D is 
\begin{align}
    \braket{S} = \left( \log 2 - \frac{1}{2} \right) N + \frac{1}{4} \left( 1-\alpha \right) \left( 1 - \log 2 \right) + o \left( 1 \right)
\end{align}
with the index $\alpha = 0$ (class D), and the variance of entanglement entropy is obtained as Eq.~(\ref{seq: BdG-variance-Majorana}).
In Ref.~\cite{Bianchi-21}, the average and the variance of the subsystem entanglement entropy were analytically obtained as
\begin{align}
    \braket{S} &= \left( \log 2 - \frac{1}{2} \right) N + \frac{1-\log 2}{4} + O \left( 1/N \right) \nonumber \\
    &= \left( 0.193147 \cdots \right) N + 0.0767132 \cdots + O \left( 1/N \right),  \\
    \braket{\left( \Delta S \right)^2} &= \frac{1}{2} \left( \frac{3}{4} - \log 2 \right) + o \left( 1 \right) \nonumber \\
    &= 0.0284264 \cdots + o \left( 1 \right),
\end{align}
for the half bipartition, which are consistent with our results.

\subsection{Class DIII}

While class D is concerned with spinless BdG Hamiltonians, class DIII is concerned with spinful BdG Hamiltonians respecting time-reversal symmetry.
Generic BdG Hamiltonians in class DIII read Eq.~(\ref{seq: BdG}), where the Nambu spinor is given as $\hat{\Gamma} \coloneqq ( \hat{c}_{1\uparrow}~\cdots~\hat{c}_{N\uparrow}~\hat{c}_{1\downarrow}~\cdots~\hat{c}_{N\downarrow}~\hat{c}_{1\uparrow}^{\dag}~\cdots~\hat{c}_{N\uparrow}^{\dag}~\hat{c}_{1\downarrow}^{\dag}~\cdots~\hat{c}_{N\downarrow}^{\dag} )^{T}$, and the $4N \times 4N$ single-particle BdG Hamiltonian $H$ respects time-reversal symmetry
\begin{align}
    \left( \sigma_y \otimes \tau_x \otimes I_{N} \right) H^{*} \left( \sigma_y \otimes \tau_x \otimes I_{N} \right) = H
\end{align}
and chiral symmetry
\begin{align}
    \left( \sigma_z \otimes I_{2N} \right) H\left( \sigma_z \otimes I_{2N} \right) = - H.
\end{align}
Owing to chiral symmetry, the single-particle BdG Hamiltonian $H$ can be written as the off-diagonal form in Eq.~(\ref{seq: H - chiral - h}).
In this representation, time-reversal symmetry imposes Eq.~(\ref{seq: DIII-TRS}) on the $2N\times 2N$ matrix $h$, by which $h$ can be generally represented as Eq.~(\ref{seq: classifying space DIII}).
Then, the single-particle BdG Hamiltonian $H$ is diagonalized as
\begin{align}
    H = U \left( \sigma_z \otimes I_{2N} \right) U^{\dag},\quad
    U \coloneqq \frac{1}{\sqrt{2}} \begin{pmatrix}
        I_{2N} & I_{2N} \\
        h^{\dag} & -h^{\dag}
    \end{pmatrix},
\end{align}
leading to
\begin{align}
    \hat{H} &= \hat{\Gamma}^{\dag} U \left( \sigma_z \otimes I_{2N} \right) U^{\dag} \hat{\Gamma} \nonumber \\
    &= 2 \sum_{i=1}^{N} \left( \hat{\tilde{c}}_{i\uparrow}^{\dag} \hat{\tilde{c}}_{i\uparrow} + \hat{\tilde{c}}_{i\downarrow}^{\dag} \hat{\tilde{c}}_{i\downarrow} - 1\right)
\end{align}
with the Nambu spinor $\hat{\tilde{\Gamma}} \coloneqq ( \hat{\tilde{c}}_{1\uparrow}~\cdots~\hat{\tilde{c}}_{N\uparrow}~\hat{\tilde{c}}_{1\downarrow}~\cdots~\hat{\tilde{c}}_{N\downarrow}~\hat{\tilde{c}}_{1\uparrow}^{\dag}~\cdots~\hat{\tilde{c}}_{N\uparrow}^{\dag}~\hat{\tilde{c}}_{1\downarrow}^{\dag}~\cdots~\hat{\tilde{c}}_{N\downarrow}^{\dag} )^{T} \coloneqq U^{\dag} \hat{\Gamma}$.
Then, the ground state is obtained as the state annihilated by all $\hat{\tilde{c}}_{i\uparrow}$'s and $\hat{\tilde{c}}_{i\downarrow}$'s (i.e., $\forall\,i~~\hat{\tilde{c}}_{i\uparrow} \ket{\Omega} = \hat{\tilde{c}}_{i\downarrow} \ket{\Omega} = 0$).
For this ground state, the correlation matrix reads
\begin{align}
    C &= \braket{\Omega\,|\,\hat{\Gamma} \hat{\Gamma}^{\dag}\,|\,\Omega} \nonumber \\
    &= U \braket{\Omega\,|\,\hat{\tilde{\Gamma}} \hat{\tilde{\Gamma}}^{\dag}\,|\,\Omega} U^{\dag} \nonumber \\
    &= U \begin{pmatrix}  
        I_{2N} & 0 \\
        0 & 0
    \end{pmatrix} U^{\dag},
\end{align}
from which we calculate entanglement entropy by Eq.~(\ref{seq: entanglement sum - BdG}).
The eigenvalues of the correlation matrix $C$ are two-fold degenerate because of time-reversal symmetry and come in $\left( \lambda, 1-\lambda \right)$ pairs because of particle-hole symmetry.

\subsection{Class C}

Generic BdG Hamiltonians in class C read Eq.~(\ref{seq: BdG}), where the Nambu spinor is given as $\hat{\Gamma} \coloneqq ( \hat{c}_{1\uparrow}~\cdots~\hat{c}_{N\uparrow}~\hat{c}_{1\downarrow}^{\dag}~\cdots~\hat{c}_{N\downarrow}^{\dag} )^{T}$, and the $2N \times 2N$ single-particle BdG Hamiltonian $H$ respects
\begin{align}
    \left( \tau_y \otimes I_N \right) H^{T} \left( \tau_y \otimes I_N \right) = - H.
\end{align}
Similarly to Eq.~(\ref{seq: class C A-U}), the single-particle BdG Hamiltonian $H$ is diagonalized as
\begin{align}
    H = \ii U \begin{pmatrix}
        0 & I_N \\
        -I_N & 0
    \end{pmatrix} U^{\dag},\quad
    U \in \mathrm{Sp} \left( N \right),
\end{align}
where we assume the absence of zero-energy modes and flatten the single-particle spectrum.
Then, with the Nambu spinor
\begin{gather}
    \hat{\tilde{\Gamma}} \coloneqq
    \begin{pmatrix}
        \hat{\tilde{c}}_{1\uparrow} \\
        \vdots \\
        \hat{\tilde{c}}_{N\uparrow} \\
        \hat{\tilde{c}}_{1\downarrow}^{\dag} \\
        \vdots \\
        \hat{\tilde{c}}_{N\downarrow}^{\dag} \\
    \end{pmatrix} \coloneqq V^{\dag} U^{\dag}
    \begin{pmatrix}
        \hat{c}_{1\uparrow} \\
        \vdots \\
        \hat{c}_{N\uparrow} \\
        \hat{c}_{1\downarrow}^{\dag} \\
        \vdots \\
        \hat{c}_{N\downarrow}^{\dag} \\
    \end{pmatrix} = V^{\dag} U^{\dag} \hat{\Gamma}, \\ 
    V \coloneqq \frac{1}{\sqrt{2}} \begin{pmatrix}
        1 & 1 \\
        -\ii & \ii 
    \end{pmatrix} \otimes I_N,
\end{gather}
the BdG Hamiltonian reduces to
\begin{align}
    \hat{H} = \sum_{i=1}^{N} \left( \hat{\tilde{c}}_{i\uparrow}^{\dag} \hat{\tilde{c}}_{i\uparrow} + \hat{\tilde{c}}_{i\downarrow}^{\dag} \hat{\tilde{c}}_{i\downarrow} - 1 \right).
\end{align}
Thus, the ground state $\ket{\Omega}$ is obtained as the state annihilated by all $\hat{\tilde{c}}_{i\uparrow}$'s and $\hat{\tilde{c}}_{i\downarrow}$'s (i.e., $\forall\,i~~\hat{\tilde{c}}_{i\uparrow} \ket{\Omega} = \hat{\tilde{c}}_{i\downarrow} \ket{\Omega} = 0$).
For this ground state, the correlation matrix reads
\begin{align}
    C &= \braket{\Omega\,|\,\hat{\Gamma} \hat{\Gamma}^{\dag}\,|\,\Omega} \nonumber \\
    &= UV \braket{\Omega\,|\,\hat{\tilde{\Gamma}} \hat{\tilde{\Gamma}}^{\dag}\,|\,\Omega} V^{\dag} U^{\dag} \nonumber \\
    &= UV \begin{pmatrix}  
        I_N & 0 \\
        0 & 0
    \end{pmatrix} V^{\dag} U^{\dag},
\end{align}
from which we calculate entanglement entropy by Eq.~(\ref{seq: entanglement sum - BdG}).

\subsection{Class CI}

Generic BdG Hamiltonians in class CI read Eq.~(\ref{seq: BdG}), where the Nambu spinor is given as $\hat{\Gamma} \coloneqq ( \hat{c}_{1\uparrow}~\cdots~\hat{c}_{N\uparrow}~\hat{c}_{1\downarrow}^{\dag}~\cdots~\hat{c}_{N\downarrow}^{\dag} )^{T}$, and the $2N \times 2N$ single-particle BdG Hamiltonian $H$ respects time-reversal symmetry
\begin{align}
    \left( \tau_x \otimes I_N \right) H^{*} \left( \tau_x \otimes I_N \right) = H
\end{align}
and chiral symmetry
\begin{align}
    \left( \tau_z \otimes I_N \right) H\left( \tau_z \otimes I_N \right) = - H.
\end{align}
Owing to chiral symmetry, the single-particle BdG Hamiltonian $H$ can be written as the off-diagonal form in Eq.~(\ref{seq: H - chiral - h}).
In this representation, time-reversal symmetry imposes Eq.~(\ref{seq: CI-TRS}) on the $N\times N$ matrix $h$, by which $h$ can be generally represented as Eq.~(\ref{seq: classifying space CI}).
Then, the single-particle BdG Hamiltonian $H$ is diagonalized as
\begin{align}
    H = U \left( \tau_z \otimes I_{N} \right) U^{\dag},\quad
    U \coloneqq \frac{1}{\sqrt{2}} \begin{pmatrix}
        I_N & I_N \\
        h^{\dag} & -h^{\dag}
    \end{pmatrix},
\end{align}
leading to
\begin{align}
    \hat{H} &= \hat{\Gamma}^{\dag} U \left( \tau_z \otimes I_N \right) U^{\dag} \hat{\Gamma} \nonumber \\
    &= \sum_{i=1}^{N} \left( \hat{\tilde{c}}_{i\uparrow}^{\dag} \hat{\tilde{c}}_{i\uparrow} + \hat{\tilde{c}}_{i\downarrow}^{\dag} \hat{\tilde{c}}_{i\downarrow} - 1\right)
\end{align}
with the Nambu spinor $\hat{\tilde{\Gamma}} \coloneqq ( \hat{\tilde{c}}_{1\uparrow}~\cdots~\hat{\tilde{c}}_{N\uparrow}~\hat{\tilde{c}}_{1\downarrow}^{\dag}~\cdots~\hat{\tilde{c}}_{N\downarrow}^{\dag} )^{T} \coloneqq U^{\dag} \hat{\Gamma}$.
Then, the ground state is obtained as the state annihilated by all $\hat{\tilde{c}}_{i\uparrow}$'s and $\hat{\tilde{c}}_{i\downarrow}$'s (i.e., $\forall\,i~~\hat{\tilde{c}}_{i\uparrow} \ket{\Omega} = \hat{\tilde{c}}_{i\downarrow} \ket{\Omega} = 0$).
For this ground state, the correlation matrix reads
\begin{align}
    C &= \braket{\Omega\,|\,\hat{\Gamma} \hat{\Gamma}^{\dag}\,|\,\Omega} \nonumber \\
    &= U \braket{\Omega\,|\,\hat{\tilde{\Gamma}} \hat{\tilde{\Gamma}}^{\dag}\,|\,\Omega} U^{\dag} \nonumber \\
    &= U \begin{pmatrix}  
        I_N & 0 \\
        0 & 0
    \end{pmatrix} U^{\dag},
\end{align}
from which we calculate entanglement entropy by Eq.~(\ref{seq: entanglement sum - BdG}).

\section{Analytical calculation of typical quantum entanglement}
    \label{sec: Weingarten}

We analytically derive the typical entanglement entropy of free fermions in the tenfold AZ symmetry classes. 
While the calculations can be performed by various methods, we here adopt the resolvent method and Weingarten calculus. 
Using these methods, we analytically show that while the leading $O \left( N \right)$ term is unchanged even in the presence of AZ symmetries, the subleading $O \left( 1 \right)$ term receives different contributions depending on symmetry classes. 
These analytical calculations agree with our numerical results in Appendix~\ref{sec: numerics - AZ}.

\subsection{Resolvent method}
    \label{subsec: resolvent}

Let us first review the basic ingredients of the resolvent method, which is a powerful method to compute the spectral density from the moments.
As described in Appendix~\ref{sec: numerics - AZ}, in free fermions, entanglement entropy is computed by the truncated correlation matrix $C_A$ supported on a subregion $A$.
Let $D(\lambda) \coloneqq \sum_{i} \delta \left( \lambda - \lambda_i \right)$ be the spectral density of $C_A$ and $\langle D(\lambda)\rangle$ be its ensemble average.
The average entanglement entropy of the subregion $A$ is 
\begin{equation}
    \langle S\rangle=\int_0^1 d\lambda\left[-\lambda \log \lambda-(1-\lambda) \log (1-\lambda)\right]\langle D(\lambda)\rangle.
\end{equation}
To obtain the spectral density, we introduce the resolvent $R(z)$ of $C_A$ by
\begin{equation}
    R(z) \coloneqq \tr \left( \frac{I}{zI-C_A} \right)
\end{equation}
for $z \in \mathbb{C}$ and the identify matrix $I$.
The spectral density $D(\lambda)$ is obtained from the resolvent $R(z)$ by
\begin{align}
D(\lambda)
&\coloneqq \sum_i\delta(\lambda-\lambda_i) \nonumber \\
&= -\frac{1}{\pi}\mathrm{Im}\lim_{\varepsilon\ra 0^{+}} \tr\left(\frac{I}{(\lambda+\ii\varepsilon)I-C_A}\right) \nonumber \\
&=-\frac{1}{\pi}\mathrm{Im}\lim_{\varepsilon\ra 0^{+}}R(\lambda+\ii\varepsilon),
\end{align}
where we use $\lim_{\varepsilon\ra 0^{+}} 1/\left( x+\ii\varepsilon \right) = \mathcal{P} \left( 1/x \right) - \ii \pi \delta(x)$ for $x \in \mathbb{R}$.
To evaluate the resolvent, let us expand it at $z\ra\infty$ by
\begin{equation}
R(z)=\tr \left( \frac{I}{z} \right)+\sum_{n=1}^\infty \frac{\tr\,C_A^n}{z^{n+1}}.
\end{equation}
Thus, for random free fermions, the trace of moments of the truncated correlation matrix, $\tr\,C_A^n$, allows us to obtain the spectral density of $C_A$ and consequently the average entanglement entropy. 

\subsection{Standard class}
    \label{sec:standard-analy}

We apply the resolvent method to derive the typical entanglement entropy of the threefold standard (Wigner-Dyson) classes (i.e., classes A, AI, and AII), where Hamiltonians are concerned only with time-reversal symmetry. 
As discussed in Appendix~\ref{sec: numerics - AZ}, when the total system size is $N$ ($2N$ for class AII), a single-particle Hamiltonian is diagonalized by a matrix $U$ that belongs to the unitary group $\mathrm{U}(N)$, orthogonal group $\mathrm{O}(N)$, and symplectic group $\mathrm{Sp}(N)$, respectively. 
Consequently, the trace of moments, $\tr\,C_A^n$, is evaluated by the Weingarten calculus for $\mathrm{U}(N)$, $\mathrm{O}(N)$, and $\mathrm{Sp}(N)$, respectively~\cite{Matsumoto-13}. 
We show that the $O (N)$ term of average entanglement entropy is the same for all the symmetry classes while the $O (1)$ term depends on the symmetry classes. 
In the following, we consider the half filling and half bipartition for clarity, namely, $N_A=N/2$ and $M=N/2$. 

\onecolumngrid

\subsubsection{Class A} 
For class A, the trace of moments is expressed by the submatrix $V$ of the unitary matrix $U$ in Eq.~(\ref{seq: submatrix}):
\begin{align}
    \langle \tr C_A^n\rangle&=\langle\tr(V^\dg V)^n \rangle \nonumber \\
    &=\left(\sum_{i_1=1}^{N/2}\cdots \sum_{i_{2n}=1}^{N/2}\right)\int dU U^*_{i_2 i_1} U_{i_2 i_3} U^*_{i_4i_3} U_{i_4 i_5}\cdots U^*_{i_{2n}i_{2n-1}} U_{i_{2n}i_1} \nonumber \\
    &=\sum_{\{i_k,i_k',j_k,j_k'\}=1}^{N/2}\int dU U_{i_1 j_1}\cdots U_{i_n j_n}U^*_{i_1' j_1'}\cdots U^*_{i_n' j_n'}\delta_{i_1' i_1}\cdots \delta_{i_n' i_n}\delta_{j_2' j_1}\cdots \delta_{j_n' j_{n-1}}\delta_{j_1' j_n},
\end{align}
where the curly bracket $\lbrace i_k,i_k',j_k,j_k'\rbrace$ under the summation includes all $i_k,i_k',j_k,j_k'$ appearing in the integrand $U_{i_1 j_1}\cdots U_{i_n j_n}U^*_{i_1' j_1'}\cdots U^*_{i_n' j_n'}$.
The integral (average) over $N\times N$ unitary random matrices $U$ with the Haar measure (namely, the circular unitary ensemble) is evaluated by the Weingarten formula, summing over elements $\sigma,\tau$ in the permutation group $S_n$:
\begin{equation}
    \int dU U_{i_1 j_1}\cdots U_{i_n j_n} U^*_{i_1' j_1'}\cdots U^*_{i_n' j_n'} = \sum_{\sigma,\tau\in S_n} \delta_{i_1,i'_{\sigma(1)}}\cdots \delta_{i_n,i'_{\sigma(n)}}\delta_{j_1,j'_{\tau(1)}}\cdots \delta_{j_n,j'_{\tau(n)}} \Wg^{\mathrm{U}}(N,\sigma\tau^{-1}).
\end{equation}
Here, $\Wg^{\mathrm{U}}(N,\sigma)$ is the Weingarten function for the unitary group $\mathrm{U} \left( N \right)$ as a function of the size $N$ of the unitary matrix and the permutation group element $\sigma\in S_n$.
In our convention, the product of two permutations is performed from the right to the left; note that some software, such as Mathematica, uses different conventions.
From the Weingarten formula, the trace of moments reduces to
\begin{align}
\langle \tr C_A^n\rangle &=\sum_{\{i_k,i_k',j_k,j_k'\}=1}^{N/2}\delta_{i_1' i_1}\cdots \delta_{i_n' i_n}\delta_{j_2' j_1}\cdots \delta_{j_n' j_{n-1}}\delta_{j_1' j_n} \sum_{\sigma,\tau\in S_{n}} \delta_{i_1,i'_{\sigma(1)}}\cdots \delta_{i_n,i'_{\sigma(n)}}\delta_{j_1,j'_{\tau(1)}}\cdots \delta_{j_n,j'_{\tau(n)}} \Wg^{\mathrm{U}}(N,\sigma\tau^{-1}) \nonumber \\
&= \sum_{\sigma,\tau\in S_n}\Wg^{\mathrm{U}}(N,\sigma\tau^{-1}) \sum_{\lbrace i_k,j_k\rbrace=1}^{N/2} \delta_{i_1 i_{\sigma(1)}} \cdots \delta_{i_n i_{\sigma(n)}} \delta_{j_2 j_{\tau(1)}} \cdots \delta_{j_1 j_{\tau(n)}} \nonumber\\
& =  \sum_{\sigma,\tau\in S_n}\Wg^{\mathrm{U}}(N,\sigma\tau^{-1}) \left(\frac{N}{2}\right)^{C(\sigma)+C(\tau\eta^{-1})},
    \label{eqn:class-A-trace}
\end{align}
where $C(\sigma)$ is the number of cycles in the permutation $\sigma$, and $\eta$ denotes the shift permutation $\eta(i)=i+1$. 

Since we are interested in the large-$N$ limit, we need to expand the Weingarten function $\Wg^{\mathrm{U}}(N,\sigma)$ in terms of $N$. The permutation $\sigma$ can be written in terms of the product of cycles $C_i$ of length $|C_i|$, and the expansion of the Weingarten function takes the following form:
\begin{equation}
    \Wg^\mathrm{U}(N,\sigma)=N^{-n-|\sigma|}\prod_i (-1)^{|C_i|-1}c_{|C_i|-1}+O(N^{-n-|\sigma|-2}),
    \label{eq:WgU-largeN}
\end{equation}
where $|\sigma|$ denotes the number of transposition of $\sigma$, and $c_n \coloneqq \left( 2n \right)!/n! \left( n+1 \right)!$ is the Catalan number.
The Weingarten function for the unitary group is special, in the sense that the leading term and the subleading term in the large-$N$ expansion differ only by order $1/N^2$; 
there is no $O (1/N)$ contribution in $\Wg^\mathrm{U}(N,\sigma)$. 
Consequently, we see in the following that there is no $O (1)$ term in the typical entanglement entropy for class A.

Now, we plug the leading term of Eq.~\eqref{eq:WgU-largeN} into Eq.~\eqref{eqn:class-A-trace} and list the values of $ \langle \tr C_A^n \rangle$ for $n=1, 2, \cdots, 6$ up to the $O(1)$ term: 
\begin{equation}
        \begin{tabularx}{0.7\textwidth}{ 
   >{\centering\arraybackslash}X 
  | >{\centering\arraybackslash}X 
   >{\centering\arraybackslash}X 
  >{\centering\arraybackslash}X 
  >{\centering\arraybackslash}X 
  >{\centering\arraybackslash}X 
  >{\centering\arraybackslash}X }
  \hline
           $n$ & $1$ & $2$ & $3$ & $4$ & $5$ & $6$ \\
            \hline
           $\langle \tr C_A^n \rangle$  &  $\dfrac{1}{4}N$ & $\dfrac{3}{16}N$ & $\dfrac{5}{32}N$ & $\dfrac{35}{256}N$ & $\dfrac{63}{512}N$ & $\dfrac{231}{2048}N$\\
           \hline
    \end{tabularx}
\end{equation}
From these values, we conjecture 
\begin{equation}
    \langle \tr C_A^n \rangle=\frac{1}{2^{2n}} \binom{2n-1}{n} N + O(N^{-1})\qquad \left( n\geq 1 \right).
\end{equation}
Then, the average resolvent of $C_A$ is
\begin{equation}
    \langle R(z)\rangle=\frac{N}{2z}+\sum_{n=1}^\infty \frac{\langle \tr C_A^n\rangle}{z^{n+1}}
    =\frac{N}{2z}+\frac{N}{2z}\sum_{n=1}^\infty \frac{(2n-1)!}{2^n 2^{n-1}(n-1)!}\frac{1}{n!z^n} + O(N^{-1})
    =\frac{N}{2}\frac{1}{\sqrt{z(z-1)}}+O(N^{-1}).
\end{equation}
Taking the discontinuity across the real axis, we obtain the spectral density of $C_A$ as
\begin{equation}
    \langle D(\lambda)\rangle=-\frac{1}{\pi}\mathrm{Im}\lim_{\epsilon\ra 0} R(\lambda+i\epsilon)
    =\frac{N}{2\pi\sqrt{\lambda(1-\lambda)}}1_{[0,1]}+O(N^{-1}),
\end{equation}
where $1_{[0,1]}$ is defined to be $1$ in the interval $[0,1]$ and $0$ otherwise. We integrate it to find the average entanglement entropy
\begin{equation}
     \langle S\rangle=\int_0^1 d\lambda\left[-\lambda \log \lambda-(1-\lambda) \log (1-\lambda)\right]\langle D(\lambda)\rangle
     = \left( \log 2-\frac{1}{2} \right) N+O(N^{-1}),
\end{equation}
which agrees with our numerical calculations in Appendix~\ref{sec: numerics - AZ} and the previous results in Refs.~\cite{Liu-18, Bianchi-22} at half filling and half bipartition. 
As mentioned before, there is no $O (1)$ term in the average entanglement entropy for class A.

\subsubsection{Class AI}
The treatment for class AI runs parallel to that for class A, and the difference lies in the Weingarten formula for the orthogonal group. 
Recall that in class AI, Hamiltonians respect time-reversal symmetry with the sign $T^* T=+1$ and are diagonalized by matrices $O$ that belong to the orthogonal group $\rO(N)$. 
Therefore, the trace of moments is expressed by the submatrix $V$ of the orthogonal matrix $O$:
\begin{align}
    \langle \tr C_A^n\rangle&=\langle\tr(V^\dg V)^n \rangle \nonumber \\
    &=\sum_{i_k=1}^{N/2}\int dO O_{i_2 i_1} O_{i_2 i_3} O_{i_4i_3} O_{i_4 i_5}\cdots O_{i_{2n}i_{2n-1}} O_{i_{2n}i_1} \nonumber \\
    &=\sum_{\{i_k,i_k',j_k,j_k'\}=1}^{N/2}\int dO O_{i_1 j_1}\cdots O_{i_n j_n}O_{i_1' j_1'}\cdots O_{i_n' j_n'}\delta_{i_1' i_1}\cdots \delta_{i_n' i_n}\delta_{j_2' j_1}\cdots \delta_{j_n' j_{n-1}}\delta_{j_1' j_n} \nonumber \\
     &=\sum_{\{i_k,j_k\}=1}^{N/2}\int dO O_{i_1 j_1}\cdots O_{i_n j_n}O_{i_{n+1} j_{n+1}}\cdots O_{i_{2n} j_{2n}}\delta_{i_{n+1} i_1}\cdots \delta_{i_{2n} i_n}\delta_{j_{n+2} j_1}\cdots \delta_{j_{2n} j_{n-1}}\delta_{j_{n+1} j_n}.
\label{eqn:classAI-avg}
\end{align}
To perform the integral (average) over the $N\times N$ random orthogonal matrix $O$, the average is evaluated by the Weingarten formula for the orthogonal group~\cite{Matsumoto-13}. Let $M_{2n}$ be the set of all pair partitions on $\lbrace 1,2,\cdots 2n\rbrace$. 
Each pair partition $\sigma\in M_{2n}$ is uniquely expressed by
\begin{equation}
    \lbrace \lbrace \sigma(1),\sigma(2)\rbrace,\lbrace \sigma(3),\sigma(4)\rbrace,\cdots \lbrace \sigma(2n-1),\sigma(2n)\rbrace \rbrace,
\end{equation}
with $\sigma(2i-1)<\sigma(2i)$ for $1\leq i\leq n$ and with $\sigma(1)<\sigma(3)<\cdots <\sigma(2n-1)$. 
As a simple example for $n=2$, the set $M_4$ consists of three elements $\sigma_1=\lbrace\lbrace 1,2\rbrace, \lbrace 3,4\rbrace\rbrace$, $\sigma_2=\lbrace\lbrace 1,3\rbrace,\lbrace 2,4\rbrace\rbrace$, $\sigma_3=\lbrace\lbrace 1,4\rbrace,\lbrace 2,3\rbrace\rbrace$. 
Now, the Weingarten formula for the orthogonal group is expressed by the summation over elements $\sigma,\tau$ in the set of pair partitions $M_{2n}$:
\begin{equation}
    \int dO O_{i_1 j_1}\cdots O_{i_{2n}j_{2n}} = \sum_{\sigma,\tau\in M_{2n}}\Wg^{\mathrm{O}}(N;\sigma,\tau) \prod_{k=1}^n \delta_{i_{\sigma(2k-1)}i_{\sigma(2k)}} \delta_{j_{\tau(2k-1)}j_{\tau(2k)}}.
\end{equation}
Here, $\Wg^{\mathrm{O}}(N;\sigma,\tau)$ is the Weingarten function for the orthogonal group, which is an element of the Weingarten matrix.

By definition, the Weingarten matrix is the pseudo-inverse matrix of the Gram matrix, which is determined by the graph constructed from $\sigma$ and $\tau$. 
For example, for $n=2$, one can derive from the definition
\begin{equation}
\Wg^{\mathrm{O}}(N;\sigma_i,\sigma_j)=
\begin{cases}
    \cfrac{N+1}{N(N+2)(N-1)} & \left( i=j \right); \\
    - \cfrac{1}{N(N+2)(N-1)} & \left( i\neq j \right),
\end{cases}
\end{equation}
with $i,j=1,2,3$. An immediate observation is that for the large-$N$ expansion of $\Wg^{\mathrm{O}}(N;\sigma_i,\sigma_j)$, the leading term and the subleading term differ by $1/N$, rather than $1/N^2$ in the large-$N$ expansion of $\Wg^{\mathrm{U}}$. 
This suggests the appearance of $O (1)$ term in typical entanglement entropy for class AI, as shown below. 
In practice, the Weingarten function for the orthogonal group is evaluated by the zonal spherical functions of the Gelfand pair; 
see, for example, Ref.~\cite{Collins-09} for the explicit expressions up to $n=6$.

From the Weingarten formula and Eq.~\eqref{eqn:classAI-avg}, the trace of moments of $C_A$ reduces to
\begin{equation}
    \langle \tr C_A^n \rangle=\sum_{\sigma,\tau\in M_{2n}}\Wg^{\mathrm{O}}(N;\sigma,\tau) \left[\sum_{\lbrace i_k\rbrace=1}^{N/2}\prod_{k=1}^n\delta_{i_k i_{n+k} } \prod_{k=1}^n \delta_{i_{\sigma(2k-1)}i_{\sigma(2k)}}\right]    
\left[  \sum_{\lbrace j_k\rbrace=1}^{N/2}\prod_{k=1}^n \delta_{j_k j_{n+1+(k \bmod n)}} \prod_{k=1}^n  \delta_{j_{\tau(2k-1)}j_{\tau(2k)}}\right].
\end{equation}
The terms in the square brackets can be evaluated graphically. For the first square bracket, given $\sigma\in M_{2n}$, one can define a graph consisting of vertices $\lbrace 1,2,\cdots 2n\rbrace$ and the edge set consisting of $(k,n+k)$ and $(\sigma(2k-1),\sigma(2k))$. 
Namely, each delta function is regarded as an edge. Let us denote $L(\sigma,0)$ as the number of loops in this graph, and the first square bracket takes the value $(N/2)^{L(\sigma,0)}$. Similar treatment can be applied to the second square bracket, where the edge set consists of $(k,n+1+(k\bmod n))$ and $(\tau(2k-1),\tau(2k))$,  and we denote the number of loops as $L(\tau,1)$. These graphical evaluations lead to
\begin{equation}
    \langle \tr C_A^n\rangle=\sum_{\sigma,\tau\in M_{2n}}\Wg^{\rO}(N;\sigma,\tau) \left(\frac{N}{2}\right)^{L(\sigma,0)}\left(\frac{N}{2}\right)^{L(\tau,1)}.
\end{equation}
From the explicit forms of $\Wg^{\mathrm{O}}$, we obtain $\langle \tr C_A^n \rangle$;
we list $\langle \tr C_A^n \rangle$ below for $n=1, 2, \cdots, 5$ up to $O(1)$:
\begin{equation}
        \begin{tabularx}{0.9\textwidth}{ 
   >{\centering\arraybackslash}X 
  | >{\centering\arraybackslash}X 
   >{\centering\arraybackslash}X 
  >{\centering\arraybackslash}X 
  >{\centering\arraybackslash}X 
  >{\centering\arraybackslash}X }
  \hline
           $n$ & $1$ & $2$ & $3$ & $4$ & $5$ \\
            \hline
           $\langle \tr C_A^n \rangle$  &  $\dfrac{1}{4}N$ & $\dfrac{3}{16}N+\dfrac{1}{16}$ & $\dfrac{5}{32}N+\dfrac{3}{32}$ & $\dfrac{35}{256}N+\dfrac{29}{256}$ & $\dfrac{63}{512}N+\dfrac{65}{512}$ \\
           \hline
    \end{tabularx}
\end{equation}
Based of these values, we conjecture the closed form formula:
\begin{equation}
    \langle \tr C_A^n \rangle=\frac{1}{2^{2n}} \binom{2n-1}{n} N + \frac{1}{4}-\frac{1}{2^{2n}}\binom{2n-1}{n} + O(N^{-1})\qquad \left( n\geq 1 \right).
\end{equation}
Then, the average resolvent for $C_A$ in class AI is
\begin{equation}
    \langle R(z)\rangle=\frac{N}{2z}+\sum_{n=1}^\infty \frac{\langle \tr C_A^n\rangle}{z^{n+1}}=\frac{N}{2\sqrt{z(z-1)}}
    +\frac{1}{4} \left( \frac{1}{z} + \frac{1}{z-1} \right) -\frac{1}{2\sqrt{z(z-1)}}
    +O(N^{-1}),
\end{equation}
and the corresponding spectral density of $C_A$ is
\begin{equation}
    \langle D(\lambda)\rangle= \frac{N-1}{2\pi\sqrt{\lambda(1-\lambda)}}1_{[0,1]}+\frac{1}{4}\delta(\lambda)+\frac{1}{4}\delta(\lambda-1)+O(N^{-1}).
        \label{seq: DOS-AI}
\end{equation}
As a consistency check, one can verify $\int_0^1 d\lambda \langle D(\lambda)\rangle=N/2$, which is the number of filled particles. The two delta functions $\delta(\lambda),\delta(\lambda-1)$ do not contribute to the entanglement entropy but are needed to ensure $\int_0^1 d\lambda \langle D(\lambda)\rangle=N/2$. Finally, the average entanglement entropy is
\begin{equation}
    \langle S\rangle=\int_0^1 d\lambda\left[-\lambda\log \lambda-(1-\lambda)\log(1-\lambda)\right]\langle D(\lambda)\rangle=\left( \log 2-\frac{1}{2} \right) (N-1) +O(N^{-1}).
\end{equation}
Indeed, the Weingarten function for the orthogonal group leads to the $O (1)$ term in the average entanglement entropy, which effectively removes one fermion. 

\subsubsection{Class AII}
The treatment for class AII is also similar to classes A and AI. In
class AII, single-particle Hamiltonians respect time-reversal symmetry with the sign $T^* T=-1$ and are diagonalized by matrices $U$ that belong to the symplectic group $\rSp(N)$.
Therefore, the trace of moments can be expressed by a submatrix $V$ of a $2N\times 2N$ unitary symplectic matrix. Note that the size of the matrix is twice larger than classes AI and AII because of the Kramers degeneracy. 
For a fair comparison of entanglement, we divide the final typical entanglement entropy by two. 
Specifically, a unitary symplectic matrix $U$ satisfies 
\begin{equation}
    U^\dg = (I_N\otimes \sigma_y)U^T (I_N\otimes \sigma_y),
    \label{eqn:symp}
\end{equation}
where the time-reversal operator is chosen to be $T = I_N\otimes \sigma_y$ with the Pauli matrix $\sigma_y$.
Let us define the exchange function $p$ and parity function $s$ by
\begin{equation}
    p(i) \coloneqq
    \begin{cases}
        i+1 & \left( \text{$i$ is odd} \right);\\
        i-1 & \left( \text{$i$ is even} \right),
    \end{cases} \qquad s(i) \coloneqq
    \begin{cases}
        1 & \left( \text{$i$ is odd} \right);\\
        -1 & \left( \text{$i$ is even} \right).
    \end{cases}
    \label{eqn:ps}
\end{equation}
Using them, we write Eq.~\eqref{eqn:symp} in the component form as
\begin{equation}
    U^\dg_{ij}=s(i)s(j) U_{p(j)p(i)}.
\end{equation}
Then, the trace of moments of $C_A$ is 
\begin{align}
    \langle \tr C_A^n\rangle &= \langle \tr (V^\dg V)^n\rangle \nonumber \\
    &= \sum_{\lbrace i_k\rbrace=1}^{N}\int dU U^\dg_{i_1 i_2} U_{i_2 i_3} U^\dg_{i_3 i_4} U_{i_4 i_5}\cdots U^\dg_{i_{2n-1} i_{2n}} U_{i_{2n} i_1} \nonumber \\
    &=\sum_{\lbrace i_k\rbrace =1}^N\int dU \left[ \prod_{k=1}^{2n} s(i_k) \right] U_{p(i_2)p(i_1)} U_{i_2 i_3}U_{p(i_4)p(i_3)} U_{i_4 i_5}\cdots U_{p(i_{2n})p(i_{2n-1})} U_{i_{2n} i_1} \nonumber \\
    &=\sum_{\lbrace i_k,j_k\rbrace =1}^N\int dU \left[ \prod_{k=1}^{2n} s(i_k) \right] U_{i_1 j_1}U_{i_2 j_2}\cdots U_{i_n j_n}U_{i_{n+1} j_{n+1}}\cdots U_{i_{2n} j_{2n}} \nonumber \\
    &\qquad \qquad \qquad \qquad \qquad \times \delta_{i_1,p(i_{n+1})}\cdots \delta_{i_n,p(i_{2n})}\delta_{j_1,p(j_{n+2})}\cdots \delta_{j_{n-1},p(j_{2n})}\delta_{j_n,p(j_{n+1})}.
\label{eqn:trCA-AII}
\end{align}
To perform the integral (average) over the $2N\times 2N$ random unitary symplectic matrix $U$, we use the following Weingarten formula for the symplectic group:
\begin{equation}
    \int dU U_{i_1 j_1} U_{i_2 j_2}\cdots U_{i_{2n},j_{2n}}=\sum_{\sigma,\tau\in M_{2n}}\prod_{k=1}^n\langle i_{\sigma(2k-1)},i_{\sigma(2k)}\rangle \prod_{k=1}^n\langle j_{\tau(2k-1)},j_{\tau(2k)}\rangle \Wg^{\rSp}(N;\sigma,\tau),
\end{equation}
where $\Wg^{\rSp}(N;\sigma,\tau)$ is the Weingarten function for the symplectic group, and $\langle i,j \rangle$ is defined by
\begin{equation}
    \langle i,j\rangle
    \coloneqq s(i)\delta_{i,p(j)}
    =\begin{cases}
        \delta_{i, j-1} & \left( \text{$i$ is odd} \right); \\
        -\delta_{i, j+1} & \left( \text{$i$ is even} \right). \\
    \end{cases}
\end{equation}
From the Weingarten formula and Eq.~\eqref{eqn:trCA-AII}, the trace of moments of $C_A$ is simplified to
\begin{equation}
\begin{aligned}
&\langle \tr C_A^n \rangle=\\
&\sum_{\sigma,\tau\in M_{2n}}\Wg^{\mathrm{\rSp}}(N;\sigma,\tau) \left[\sum_{\lbrace i_k\rbrace=1}^{N}\prod_{k=1}^n \langle i_k, i_{n+k} \rangle \prod_{k=1}^n \langle i_{\sigma(2k-1)},i_{\sigma(2k)}\rangle\right]    
\left[  \sum_{\lbrace j_k\rbrace=1}^{N}\prod_{k=1}^n \langle j_k,j_{n+1+(k \bmod n)}\rangle \prod_{k=1}^n  \langle j_{\tau(2k-1)},j_{\tau(2k)}\rangle\right].
\end{aligned}
\end{equation}
Similar to class AI, the terms in the square brackets can be evaluated graphically. The graph and edges are defined in the same manner as class AI, with a possible minus sign from $\langle , \rangle$ that we denote as $S(\sigma,0)$ and $S(\tau,1)$ for the first and second square brackets. 
More specifically, given a graph with directed edges (putting arrows on the edges), we count the number of reversed arrow $A(l)$ of every loop $l$ in the graph and the length of loop $|l|$. The sign function is given as $S(\sigma,0)=\prod_l (-1)^{A(l)}(-1)^{|l|/2}$, and same for $S(\tau,1)$. 
Thus, the first square bracket takes the value $S(\sigma,0)N^{L(\sigma,0)}$, and the second bracket takes the value $S(\tau,1)N^{L(\tau,1)}$.
These graphical evaluations lead to
\begin{equation}
    \langle \tr C_A^n\rangle = \sum_{\sigma,\tau\in M_{2n}}\Wg^{\rSp}(N;\sigma,\tau) S(\sigma,0) S(\tau,1) N^{L(\sigma,0)}N^{L(\tau,1)}.
\end{equation}
The Weingarten function for the symplectic group, $\Wg^{\rSp}(N;\sigma,\tau)$, is computed by the relation to $\Wg^{\rO}(N;\sigma,\tau)$,
\begin{equation}
    \Wg^{\rSp}(N;\sigma,\tau)=(-1)^n \epsilon(\sigma^{-1}\tau)\Wg^{\rO}(-2N;\sigma,\tau),
    \label{eqn:WgSp}
\end{equation}
where $\epsilon(\sigma)$ is the signature of the permutation $\sigma$.
We embed the set $M_{2n}$ in the permutation group $S_{2n}$ so that $\sigma^{-1}\tau$ will be well defined. 
Based on the values of $\langle \tr C_A^n\rangle$ for $n=1, 2, \cdots, 5$ up to the $O(1)$ contribution,
\begin{equation}
        \begin{tabularx}{0.9\textwidth}{ 
   >{\centering\arraybackslash}X 
  | >{\centering\arraybackslash}X 
   >{\centering\arraybackslash}X 
  >{\centering\arraybackslash}X 
  >{\centering\arraybackslash}X 
  >{\centering\arraybackslash}X }
  \hline
           $n$ & $1$ & $2$ & $3$ & $4$ & $5$ \\
            \hline
           $\langle \tr C_A^n \rangle$  &  $\dfrac{1}{2}N$ & $\dfrac{3}{8}N-\dfrac{1}{16}$ & $\dfrac{5}{16}N-\dfrac{3}{32}$ & $\dfrac{35}{128}N-\dfrac{29}{256}$ & $\dfrac{63}{256}N-\dfrac{65}{512}$ \\
           \hline
    \end{tabularx}
\end{equation}
we conjecture the closed form formula
\begin{equation}
    \langle \tr C_A^n \rangle=\frac{1}{2^{2n-1}} \binom{2n-1}{n} N - \frac{1}{4}+\frac{1}{2^{2n}}\binom{2n-1}{n} + O(N^{-1})\qquad \left( n\geq 1 \right).
\end{equation}
From the resolvent method, the average entanglement entropy is obtained as 
\begin{equation}
    \langle S\rangle= \left(\log 2-\frac{1}{2} \right) \left( N+\frac{1}{2} \right) +O(N^{-1}).
\end{equation}
Here, the single-particle entanglement spectrum exhibits the Kramers degeneracy due to time-reversal symmetry, and we calculate entanglement entropy only from half of the entanglement spectrum.
We see that time-reversal symmetry in class AII effectively increases half of a fermion, which also agrees with our numerical results in Appendix~\ref{sec: numerics - AZ}. 
This completes the proof of Eq.~\eqref{eqn:ent-standard} in the standard classes. 

\subsection{Chiral class}
    \label{sec:chiral-analy}

\subsubsection{Volume-law term of entanglement entropy}

Let us now move on to the typical entanglement entropy in the chiral classes (classes AIII, BDI, and CII). 
Hamiltonians in classes AIII and BDI (class CII) with $N$ unit cells can be diagonalized by the $2N\times 2N$ ($4N\times 4N$) unitary matrix $U$ in Eq.~\eqref{eqn:U-chiral}, 
where $h$ belongs to $\rU(N)$, $\rO(N)$, and $\rSp(N)$ for classes AIII, BDI, and CII, respectively. For later convenience, we write the $N\times N$ ($2N\times 2N$) matrix $-h^\dg$ as
\begin{equation}
    -h^\dg \eqqcolon \left(
    \begin{array}{cc}
        A & B \\
        C & D
    \end{array}
    \right),
    \label{eqn:h-decomp}
\end{equation}
where $A$, $B$, $C$, $D$ are $N/2\times N/2$ ($N\times N$) matrices.
Now, for the case of half filling and half bipartition, namely, $N_A=M=N$ for classes AIII and BDI or $N_A=M=2N$ for class CII, the $N_A\times M$ submatrix $V$ takes the form of
\begin{equation}
    V = \frac{1}{\sqrt{2}}\left(
    \begin{array}{cc}
        I_{N/2(N)} & 0 \\
       A & B
    \end{array}
    \right)^{T},
    \label{eqn:chiral-V}
\end{equation}
from which the truncated correlation matrix is constructed as
\begin{equation}
    C_{A,\Ch} = V^\dg V =\frac{1}{2}
    \left(
    \begin{array}{cc}
        I_{N/2(N)} & A^T \\
        A^* & A^* A^T+B^* B^T
    \end{array}
    \right)=\frac{1}{2} \left(
    \begin{array}{cc}
        I_{N/2(N)} & A^T \\
        A^* & I_{N/2(N)}
    \end{array}
    \right).
\end{equation}
Hence, the trace of moments of $C_{A,\Ch}$ can be expressed by $A^*$ and $A^T$. 
Specifically, $\tr C_{A,\Ch}^n$ is the summation of $\tr(A^T A^*)^m$, which reduces to $\langle \tr C_A^m\rangle$ for classes A, AI, and AII, respectively. 
Since the leading $O (N)$ term of $\langle \tr C_A^m\rangle$ is the same for the three standard classes, the leading $O (N)$ term of $\langle \tr C_{A,\Ch}^n\rangle$ should also be the same for the three chiral classes.
In the following, we list the first several $\langle \tr C_{A,\Ch}^n\rangle $: 
\begin{equation}
    \begin{aligned}
    \langle \tr C_{A,\Ch}\rangle& = \langle \tr I\rangle = \frac{N}{2}, \\
 \langle\tr C_{A,\Ch}^2 \rangle &= \frac{1}{2}\langle\tr( I+A^T A^*)\rangle=\frac{3}{8}N, \\
 \langle\tr C_{A,\Ch}^3 \rangle& = \frac{1}{4}\langle\tr( I+ 3A^T A^*)\rangle=\frac{5}{16}N, \\
\langle \tr C_{A,\Ch}^4\rangle & = \frac{1}{8} \langle\tr( I+6 A^T A^*+(A^T A^*)^2)\rangle=\frac{35}{128}N + O(1), \\
 \langle\tr C_{A,\Ch}^5 \rangle & = \frac{1}{16} \langle\tr(
 I+10 A^T A^*+5(A^T A^*)^2
 )\rangle=\frac{63}{256}N + O(1), \\
 \langle\tr C_{A,\Ch}^6 \rangle&=\frac{1}{32}\langle\tr(I+15(A^T A^*)+15(A^T A^*)^2+(A^T A^*)^3)\rangle=\frac{231}{1024}N + O(1).
 \end{aligned}
 \label{eqn:tr-CACh}
 \end{equation}
 We conjecture that the $O (N)$ term in $\langle \tr C_{A,\Ch}^n\rangle$ for the chiral classes is
 \begin{equation}
     \langle \tr C_{A,\Ch}^n\rangle = \frac{1}{2^{2n-1}}\binom{2n-1}{n} N + O(1).
 \end{equation}
From the resolvent method, the $O (N)$ term of the average entanglement entropy is
\begin{equation}
    \langle S\rangle= (2\log 2 -1) N +O(1),
\end{equation}
which agrees with our numerical calculations in Appendix~\ref{sec: numerics - AZ}. 

\subsubsection{Constant terms of entanglement entropy}
We next focus on the $O (1)$ contribution to the typical entanglement entropy. 
From Eq.~\eqref{eqn:tr-CACh}, an immediate consequence is that the $O (1)$ terms of $\langle \tr C_{A,\Ch}^n\rangle$ and $\braket{S}$ are zero for class AIII because the $O (1)$ term in $\langle \tr C_{A}^m\rangle$ is zero for class A.
This again agrees with our numerical calculations in Appendix~\ref{sec: numerics - AZ}.  
Another direct consequence is that the $O (1)$ contribution to $\langle S \rangle$ in class CII equals $(-1/2)$ times that in class BDI owing to the relationship between classes AI and AII. Thus, we only need to evaluate the $O (1)$ term in $\langle S \rangle$ for class BDI.

Let us evaluate the $O (1)$ term in $ \langle \tr C_{A,\Ch}^n\rangle$ for class BDI. Formally, Eq.~\eqref{eqn:tr-CACh} can be written as 
\begin{equation}
    \tr C_{A,\Ch}^n = \frac{1}{2^{n-1}} \sum_{m=0}^{\infty} g_{m}^{(n)} \tr (A^T A^*)^m
\end{equation}
with $\sum_{m=0}^{\infty} g_{m}^{(n)} = 2^{n-1}$ and $g_0^{(n)} = 1$.
From the previous results of $\tr C_A^m$ in class AI, we have $\langle\tr (A^T A^*)^m\rangle=\langle\tr C_A^m\rangle = f_m N+(\frac{1}{4}-f_m) +O(N^{-1})$ with $f_m \coloneqq \frac{1}{2^{2m}} \binom{2m-1}{m}$ for $m\geq 1$. 
We also introduce $f_0 = \frac{1}{2}$, satisfying $\langle\tr (A^T A^*)^0\rangle=\langle\tr C_A^0\rangle = f_0 N$.
Then, from the $O (N)$ calculations above, we have $\sum_{m=0}^\infty f_m g_m^{(n)}=\frac{1}{2^n} \binom{2n-1}{n}$.  
From these relations, the $O (1)$ term in $ \langle \tr C_{A,\Ch}^n\rangle$ for class BDI is obtained as
\begin{align}
     \langle \tr C_{A,\Ch,O(1)}^n\rangle&=\frac{1}{2^{n-1}}\sum_{m=1}^{\infty} g_m^{(n)} \left( \frac{1}{4}-f_m \right) 
    = \frac{1}{4}+\frac{1}{2^{n+1}}- \frac{1}{2^{2n-1}} \binom{2n-1}{n}.
\end{align}
Including back the $O (N)$ term, we have
\begin{equation}
    \langle \tr (C_{A,\Ch}^n)\rangle=\frac{1}{2^{2n-1}}\binom{2n-1}{n} \left( N-1 \right) + \frac{1}{4}+\frac{1}{2^{n+1}} + O(N^{-1}).
\end{equation}
Then, the average resolvent for $C_{A,\Ch}$ is
\begin{align}
    \langle R(z)\rangle&= \frac{N}{z}+\sum_{n=1}^\infty \frac{\langle \tr C_{A,\Ch}^n\rangle}{z^{n+1}} 
    = \frac{N-1}{\sqrt{z(z-1)}}+ \frac{1}{4} \left(  \frac{1}{z} + \frac{1}{z-1} \right) + \frac{1}{2z-1} +O(N^{-1}),
\end{align}
leading to the spectral density
\begin{equation}
    \langle D(\lambda)\rangle = \frac{N-1}{\pi\sqrt{\lambda(1-\lambda)}}1_{[0,1]}+\frac{1}{4}\delta \left( \lambda-1 \right)+\frac{1}{4}\delta \left( \lambda \right)+\frac{1}{2}\delta \left( \lambda-\frac{1}{2} \right)+O(N^{-1}). 
\end{equation}
One can verify $\int_0^1 d\lambda \langle D(\lambda)\rangle=N$, as expected from $N_A = N$. 
It is notable that $\langle D(\lambda)\rangle$ contains the delta function $\left( 1/2 \right) \delta(\lambda-1/2 )$ at $\lambda = 1/2$, which corresponds to the ``zero mode'' in the single-particle entanglement spectrum. 
This term contributes to the entanglement entropy by $\left( 1/2 \right) \log 2$. 
After integration, the average entanglement entropy is
\begin{equation}
    \langle S\rangle
    =(2\log 2 -1) (N-1) + \frac{1}{2}\log 2 + O(N^{-1})
    = (2\log 2-1) N - \left( \frac{3}{2}\log 2 - 1 \right)+O(N^{-1}),
\end{equation}
which agrees with the numerical results in Appendix~\ref{sec: numerics - AZ}. This completes the proof of Eq.~\eqref{eqn:ent-chiral} in the chiral classes. 

\subsection{Bogoliubov-de Gennes (BdG) class}
    \label{sec:BdG-analy}

Finally, we consider the BdG classes (classes D, C, DIII, and CI), where Hamiltonians are subject to particle-hole symmetry.
In class D, the average entanglement entropy was analytically obtained for particle-number-nonconserving BdG Hamiltonians~\cite{Bianchi-21}.
In addition, while the classifying space for class D is $\mathrm{O} \left( 2N \right)/\mathrm{U} \left( N \right)$, that for class C is $\mathrm{Sp} \left( N \right)/\mathrm{U} \left( N \right)$.
Thus, the $O \left( 1 \right)$ constant term of the average entanglement entropy for class C should have the opposite sign to that for class D, similar to the standard and chiral classes.
Based on these facts, the average density of the single-particle entanglement spectrum should be
\begin{align}
    \braket{D \left( \lambda \right)} = \frac{N - \left( 1-\alpha \right)/2}{\pi \sqrt{\lambda \left( 1-\lambda \right)}} + \frac{1-\alpha}{2} \delta \left( \lambda - \frac{1}{2} \right) + O \left( N^{-1} \right)
        \label{seq: density-D&C}
\end{align}
with $\alpha = 0$ (class D) and $\alpha = 2$ (class C), which leads to the average entanglement entropy in Eq.~(\ref{eqn:ent-D&C}).
Below, we also confirm Eq.~(\ref{seq: density-D&C}) by the Weingarten calculus.
In classes DIII and CI, on the other hand, the simultaneous presence of time-reversal symmetry leads to chiral symmetry, and we derive the typical entanglement entropy by the formalism in the chiral classes in Appendix~\ref{sec:chiral-analy}.

\subsubsection{Classes D and C}

In class D, $2N\times 2N$ single-particle Hamiltonians respect particle-hole symmetry $H^*=-H$ and can be diagonalized by unitary matrices $U$ in Eq.~\eqref{eqn:U-class-D}.
We write the $2N \times 2N$ orthogonal matrix $O$ in Eq.~\eqref{eqn:U-class-D} as 
\begin{equation}
    O \eqqcolon \left(
    \begin{array}{cc}
        A & B \\
        C & D
    \end{array}
    \right),
\end{equation}
where $A$, $B$, $C$, $D$ are $N \times N$ matrices.
For the case of half filling, the correlation matrix is
\begin{equation}
    C_{\mathrm{D}} = \frac{1}{2}O 
    \begin{pmatrix}
        I_N & \ii\times I_N\\
        -\ii\times I_N & I_N
    \end{pmatrix}
    O^T=\frac{1}{2}\left[I_{2N}+
     O\begin{pmatrix}
        0 & \ii\times I_N\\
        -\ii\times I_N & 0
    \end{pmatrix}O^T
    \right].
\end{equation}
When choosing the subsystem as the first $N$ fermions, the truncated correlation matrix is
\begin{equation}
    C_{A,\mathrm{D}} = \frac{1}{2}\left[ I_N + \ii \left( AB^T - BA^T \right) \right] \eqqcolon \frac{1}{2}\left[ I_{2N}+J_A \right],
\end{equation}
where we introduce $J_A \coloneqq \ii \left( AB^T - BA^T \right)$. 
To obtain the moments $\tr\,C_{A,\mathrm{D}}^n$, we need to compute $\tr\,J_A^m$. 
One can see $\tr\,J_A^m = 0$ for odd $m$ since $J_A$ is an antisymmetric matrix. 
In the following, we describe the procedure for computing $\tr\,J_A^m$ for even $m$. 

As an example, we calculate $\tr\,J_A^2$ for $m=2$;
the generalization to larger even $m \geq 4$ is straightforward. 
For $m=2$, the average trace of moments of $J_A$ is
\begin{align}
    \braket{ \tr\,J_A^2 } &= -\braket{ \tr \left( AB^T AB^T - AB^T BA^T - BA^T A B^T + BA^T BA^T \right) } 
    = - 2 \braket{ \tr \left( AB^T AB^T - AB^T BA^T \right) }.
\end{align}
From the Weingarten formula for the orthogonal group, the first term $\langle\tr\,AB^T A B^T \rangle$ is calculated as 
\begin{align}
    &\langle\tr\,AB^T A B^T \rangle 
    =\sum_{i_1,i_2,i_3,i_4=1}^N \int dO  O_{i_1 i_2}O_{i_3 i_2+N} O_{i_3 i_4}O_{i_1 i_4+N} \nonumber \\
    &\qquad = \sum_{\lbrace i_k,j_k\rbrace=1}^N\int dO O_{i_1 j_1}O_{i_2,j_2+N} O_{i_3 j_3} O_{i_4, j_4+N}\delta_{i_1 i_4}\delta_{i_2 i_3}\delta_{j_1 j_2}\delta_{j_3 j_4} \nonumber \\
    &\qquad =\sum_{\sigma\in M_4,\tau\in M'_4}\Wg^{\mathrm{O}}(2N;\sigma,\tau)\left[\sum_{\lbrace i_k\rbrace=1}^N\delta_{i_{\sigma(1)} i_{\sigma(2)}} \delta_{i_{\sigma(3)} i_{\sigma(4)}} \delta_{i_1 i_4}\delta_{i_2 i_3}\right]  \left[\sum_{\lbrace j_k\rbrace=1}^N\delta_{j_{\tau(1)} j_{\tau(2)}} \delta_{j_{\tau(3)} j_{\tau(4)}} \delta_{j_1 j_2}\delta_{j_3 j_4}\right].
\end{align}
Here, $\sigma$ takes values from the set of pair partitions $M_4$, and $\tau$ is only allowed to take values from a subset $M_4' \coloneqq \lbrace\lbrace1,3\rbrace,\lbrace2,4\rbrace\rbrace \in M_4$. 
Then, we evaluate $\langle \tr\,AB^T AB^T\rangle$ graphically by counting the number of loops in each square bracket, similar to the treatment in class AI. The second term $\langle \tr\,AB^T BA^T\rangle$ can also be evaluated in a similar manner by taking $M_4'$ as $M_4'=\lbrace\lbrace1,4\rbrace,\lbrace2,3\rbrace\rbrace$. 

Assembling the above results, we obtain $\langle\tr J_A^2\rangle=\frac{N}{2}-\frac{1}{4}+O(N^{-1})$. 
Similarly, for $m=4$, we obtain $\langle \tr J_A^4\rangle=\frac{3N}{8}-\frac{3}{16}+O(N^{-1})$. 
From $\langle \tr J_A^m\rangle$, we compute $\langle \tr\,C_{A,\mathrm{D}}^n\rangle$ and show the values up to $n=5$ and $O(N^{-1})$:

\begin{equation}
        \begin{tabularx}{0.8\textwidth}{ 
   >{\centering\arraybackslash}X 
  | >{\centering\arraybackslash}X 
   >{\centering\arraybackslash}X 
  >{\centering\arraybackslash}X 
  >{\centering\arraybackslash}X 
  >{\centering\arraybackslash}X
 }
  \hline
           $n$ & $1$ & $2$ & $3$ & $4$ & $5$  \\
            \hline
          $\langle \tr\,C_{A,\mathrm{D}}^n\rangle$  & $\dfrac{N}{2}$ & $\dfrac{3N}{8}-\dfrac{1}{16}$ & $\dfrac{5N}{16}-\dfrac{3}{32}$ & $\dfrac{35N}{128}-\dfrac{27}{256}$ & $\dfrac{63N}{256}-\dfrac{55}{512}$\\
           \hline
    \end{tabularx}
\end{equation}
Based on these results, we conjecture
\begin{equation}
    \langle \tr\,C_{A,\mathrm{D}}^n\rangle=\frac{1}{2^{2n}}\binom{2n-1}{n}(2N-1)+\frac{1}{2^{n+1}}+O(N^{-1}).
\end{equation}
From the resolvent method, the corresponding spectral density of $C_{A,\mathrm{D}}$ is obtained as Eq.~\eqref{seq: density-D&C} with $\alpha = 0$.
The treatment for class C runs parallel to class D by replacing the orthogonal matrix with the unitary symplectic matrix, and the average density of the single-particle entanglement spectrum is obtained as Eq.~\eqref{seq: density-D&C} with $\alpha = 2$.

\subsubsection{Class CI}
In class CI, we choose chiral symmetry as $S=\sigma_z\otimes I_N$ and time-reversal symmetry as $T=\sigma_x\otimes I_N$, the combination of which leads to particle-hole symmetry $C=\ii \sigma_y\otimes I_N$ with $C^*C=-1$. 
This is to be distinguished from class BDI, where particle-hole symmetry satisfies $C^* C = +1$.
Because of chiral symmetry, a $2N \times 2N$ Hamiltonian can be diagonalized by a $2N\times 2N$ unitary matrix $U$ in Eq.~\eqref{eqn:U-chiral}.
As discussed in Eq.~\eqref{seq: CI-TRS}, time-reversal symmetry imposes $h^T=h$, and $h$ is drawn from the circular orthogonal ensemble. 
Then, $h$ can be expressed by
\begin{equation}
    -h^\dg = f^T f,\quad f\in \mathrm{U} (N),
\end{equation}
which indicates that we may use the Weingarten calculus for the unitary group to calculate typical entanglement entropy. 

As in Eq.~\eqref{eqn:h-decomp}, we decompose $-h^\dg$ into four blocks $A,B,C,D$. Let us denote the corresponding truncated correlation matrix in class CI as $C_{A,\mathrm{CI}}=V^\dg V$, where $V$ is defined as Eq.~\eqref{eqn:chiral-V}. 
From the discussions in the chiral classes, $\langle \tr C_{A,\mathrm{CI}}^n\rangle$ is obtained from $\langle \tr(A^T A^*)^m\rangle$. 
To obtain $\langle \tr(A^T A^*)^m\rangle$, we write the $N\times N$ matrix $f$ as
\begin{equation}
    f = \left(
    \begin{array}{cc}
        a & b \\
        c & d
    \end{array}
    \right),\quad f\in \mathrm{U} (N),
\end{equation}
where $a$, $b$, $c$, $d$ are $N/2 \times N/2$ matrices.
In particular, the components of $a$ and $c$ are related to those of $f$ via $a_{ij}=f_{ij}$ and $c_{ij}=f_{i+N/2,j}$. 
Using these submatrices of $f$, we express the matrix $A$ and its component as
\begin{align}
    A = a^T a + c^T c, \quad    A_{ij}=\sum_{k=1}^{N/2}f_{ki}f_{kj}+f_{k+N/2,i}f_{k+N/2,j}.
\end{align}
Applying the Weingarten formula for the unitary group to the components of $f$, we compute $\langle \tr(A^T A^*)^m\rangle$. To shed light on the calculations, we below outline the steps explicitly for $m=3$, which can be readily generalized to arbitrary $m$. 

For $m=3$, we compute
\begin{equation}
    \langle \tr(A^T A^*)^3\rangle=\langle \tr[( a^T a + c^T c)( a^\dg a^* + c^\dg c^*)( a^T a + c^T c)( a^\dg a^* + c^\dg c^*)( a^T a + c^T c)( a^\dg a^* + c^\dg c^*)] \rangle.
\end{equation}
While this includes 64 terms, not all of them contribute to the average entanglement entropy due to the delta symbols in the Weingarten formula. To have a nonzero contribution, the number of $c^T c$ should match the number of $c^\dg c^*$. In the case of $m=3$, the number of $\lbrace c^T c, c^\dg c^*\rbrace$ pairs can be 0,1,2,3, and the number of terms of each type is 1,9,9,1, respectively. Only 20 terms out of the 64 terms have nonzero contributions. Due to the symmetry between $a$ and $c$, we only need to consider the first two cases where there is no $\lbrace c^T c, c^\dg c^*\rbrace$ pair and one $\lbrace c^T c, c^\dg c^*\rbrace$ pair, and multiply the final result by two. 
Let us first focus on the case with no $\lbrace c^T c, c^\dg c^*\rbrace$ pair:
\begin{align}
    &\langle \tr(a^T a a^\dg a^* a^T a a^\dg a^* a^T a a^\dg a^*)\rangle \nonumber \\
    =& \sum_{\lbrace i_k,j_k,i_k',j_k'\rbrace=1}^{N/2}\delta_{i_1 i_2} \delta_{i_3 i_4} \delta_{i_5 i_6} \delta_{i'_1 i'_2} \delta_{i'_3 i'_4} \delta_{i'_5 i'_6} \left(\delta_{j_2 j_1'}\delta_{j_3 j_2'}\cdots \delta_{j_6 j_5'}\delta_{j_1 j_6'}\right) \left[\int df f_{i_1 j_1}\cdots f_{i_6 j_6}f^*_{i'_1 j'_1}\cdots f^*_{i'_6 j'_6}\right] \nonumber \\
    =&  \sum_{\lbrace i_k,j_k,i_k',j_k'\rbrace=1}^{N/2}\left(\prod_{k=1}^m\delta_{i_{2k-1} i_{2k}}  \delta_{i'_{2k-1} i'_{2k}} \right)\left(\prod_{k=1}^{2m}\delta_{j_k j'_{k-1}}\right) \left[\sum_{\sigma,\tau\in S_{2m}}\left(\prod_{k=1}^{2m}\delta_{i_k,i'_{\sigma(k)}} \delta_{j_k,j'_{\tau(k)}}\right)\Wg^{\rU}(N,\sigma\tau^{-1})\right] \nonumber \\
    =& \sum_{\sigma,\tau\in S_{2m}}\Wg^{\rU}(N,\sigma\tau^{-1})\left[\sum_{\lbrace i_k,i'_k\rbrace=1}^{N/2}\left(\prod_{k=1}^m\delta_{i_{2k-1} i_{2k}}  \delta_{i'_{2k-1} i'_{2k}} \right)\left(\prod_{k=1}^{2m}\delta_{i_k,i'_{\sigma(k)}}\right)\right]\left[
    \sum_{\lbrace j_k,j'_k\rbrace=1}^{N/2} \left(\prod_{k=1}^{2m}\delta_{j_k j'_{k-1}}\right)\left(\prod_{k=1}^{2m}\delta_{j_k,j'_{\tau(k)}}\right)
    \right].
    \label{eqn:CI-analy}
\end{align}
From the second to the third line, we apply the Weingarten formula for the unitary group to the terms in the square bracket. The above result is applicable to arbitrary $m$. 
Similar to the previous cases, the terms in the square brackets of the last line can be evaluated graphically by counting the number of loops in the corresponding graph. 
Next, let us consider the case with one $\lbrace c^T c, c^\dg c^*\rbrace$ pair. 
There are such 9 terms; we write down one of them explicitly as
\begin{align}
    &\langle \tr(a^T a c^\dg c^* c^T c a^\dg a^* a^T a a^\dg a^*)\rangle 
    = \sum_{\lbrace i_k,j_k,i_k',j_k'\rbrace=1}^{N/2}\delta_{i_1 i_2} \delta_{i_3 i_4} \delta_{i_5 i_6} \delta_{i'_1 i'_2} \delta_{i'_3 i'_4} \delta_{i'_5 i'_6} \left(\delta_{j_2 j_1'}\delta_{j_3 j_2'}\cdots \delta_{j_6 j_5'}\delta_{j_1 j_6'}\right) \nonumber \\
    &\qquad\qquad\qquad \times\left[\int df \left(f_{i_1 j_1}f_{i_2 j_2} f_{i_3+N/2,j_3} f_{i_4+N/2,j_4} f_{i_5,j_5} f_{i_6 j_6}\right)\left(f^*_{i'_1+N/2, j'_1}f^*_{i'_2+N/2, j'_2} f^*_{i'_3 j'_4} f^*_{i'_4 j'_4} f^*_{i'_5 j'_5} f^*_{i'_6 j'_6}\right)\right].
\end{align}
When applying the Weingarten formula, $\tau$ can still take all permutations in $S_6$ while valid $\sigma$ is restricted. In particular, to obtain a nonzero contribution, $\sigma$ needs to be in the subgroup: $\mathrm{Perm}[3,4]\,\mathrm{Perm}[1,2,5,6]\in S_6$, i.e., $\sigma(1),\sigma(2)\in\lbrace 3,4\rbrace$, and $\sigma(3),\sigma(4),\sigma(5),\sigma(6)\in \lbrace 1,2,5,6\rbrace$. For example, $\lbrace 3,4,1,5,2,6\rbrace$ is an element in this subgroup.
Thus, the last line of Eq.~\eqref{eqn:CI-analy} is still applicable as long as we replace the summation $\sum_{\sigma\in S_{2m}}$ by the summation of elements in the subgroup. Each one of the 9 terms corresponds to a summation of a different subgroup. Assembling everything, we obtain $\langle \tr(A^T A^*)^3\rangle$. 

Using the above formalism, we calculate $\langle \tr(A^T A^*)^m\rangle$. 
The first three values for $m=1,2,3$ are $\frac{N}{4}+\frac{1}{4}+O(N^{-1})$, $\frac{3N}{16}+\frac{1}{4}+O(N^{-1})$, and $\frac{5N}{32}+\frac{1}{4}+O(N^{-1})$, respectively. 
From these values and Eq.~\eqref{eqn:tr-CACh}, the first seven values of $\langle \tr C_{A,\mathrm{CI}}^n\rangle$ are up to $O(1)$
\begin{equation}
        \begin{tabularx}{0.9\textwidth}{ 
   >{\centering\arraybackslash}X 
  | >{\centering\arraybackslash}X 
   >{\centering\arraybackslash}X 
  >{\centering\arraybackslash}X 
  >{\centering\arraybackslash}X 
  >{\centering\arraybackslash}X
  >{\centering\arraybackslash}X
  >{\centering\arraybackslash}X}
  \hline
           $n$ & $1$ & $2$ & $3$ & $4$ & $5$ & $6$ & $7$ \\
            \hline
          $\langle \tr C_{A,\mathrm{CI}}^n\rangle$  &  $\dfrac{1}{2}N$ & $\dfrac{3}{8}N+\dfrac{1}{8}$ & $\dfrac{5}{16}N+\dfrac{3}{16}$ & $\dfrac{35}{128}N+\dfrac{7}{32}$ & $\dfrac{63}{256}N+\dfrac{15}{64}$ & $\dfrac{231N}{1024}+\dfrac{31}{128}$ & $\dfrac{429N}{2048}+\dfrac{63}{256}$ \\
           \hline
    \end{tabularx}
\end{equation}
Based of these values, we conjecture the closed form formula,
\begin{equation}
    \langle \tr C_{A,\mathrm{CI}}^n \rangle=\frac{1}{2^{2n-1}} \binom{2n-1}{n} N +\frac{1}{4}-\frac{1}{2^{n+1}} + O(N^{-1}) \qquad \left( n\geq 1 \right).
\end{equation}
Then, the resolvent is
\begin{align}
    \langle R(z)\rangle&=\frac{N}{z}+\sum_{n=1}^\infty \frac{\langle \tr C_{A,\mathrm{CI}}^n\rangle}{z^{n+1}}
    =\frac{N}{\sqrt{z(z-1)}}+ \frac{1}{4} \left( \frac{1}{z} + \frac{1}{z-1} \right) - \frac{1}{2z-1} +O(N^{-1}),
\end{align}
and the corresponding spectral density is
\begin{equation}
    \langle D(\lambda)\rangle = \frac{N}{\pi\sqrt{\lambda(1-\lambda)}}1_{[0,1]}+\frac{1}{4}\delta \left( \lambda-1 \right)+\frac{1}{4}\delta \left( \lambda \right)-\frac{1}{2}\delta \left( \lambda-\frac{1}{2} \right)+O(N^{-1}). 
\end{equation}
One can verify $\int_0^1 d\lambda \langle D(\lambda)\rangle=N$, as expected. 
The singular term $\left( 1/2 \right) \delta(\lambda-1/2)$ contributes to the entanglement entropy by $\left( 1/2 \right) \log 2$, and the average entanglement entropy is thus
\begin{equation}
    \langle S\rangle= (2\log 2 -1) N - \frac{1}{2}\log 2+O(N^{-1}),
\end{equation}
which agrees with the numerical results in Appendix~\ref{sec: numerics - AZ}. 

\subsubsection{Class DIII}
The treatment for class DIII runs parallel to that for class CI with the subtlety that some permutations might contribute a minus sign, as explained in the following. 
Specifically, we choose chiral symmetry as $S=\sigma_z\otimes I_{2N}$ and time-reversal symmetry as $T=\sigma_x\otimes I_N\otimes \ii\sigma_y$, 
the combination of which leads to particle-hole symmetry $C=\sigma_y\otimes I_N\otimes \sigma_y$ with $C^* C = +1$. 
This is to be distinguished from class CII, where particle-hole symmetry satisfies $C^* C=-1$. 
The presence of chiral symmetry allows us to diagonalize a Hamiltonian by a $4N\times 4N$ unitary matrix $U$ in Eq.~\eqref{eqn:U-chiral}. 
As discussed in Eq.~\eqref{seq: DIII-TRS}, time-reversal symmetry imposes $(I_N\otimes \sigma_y)h^T(I_N\otimes \sigma_y)=h$, and $h$ is drawn from the circular symplectic ensemble. Then, $h$ can be expressed by
\begin{equation}
    -h^\dg = (I_N\otimes \sigma_y) f^T (I_N\otimes \sigma_y)f,\quad f\in \mathrm{U}(2N).
    \label{eqn:DIII-decomp}
\end{equation}
Similar to class CI, we use the Weingarten calculus for the unitary group to compute typical entanglement entropy for class DIII. 

We again decompose $-h^\dg$ into four blocks $A,B,C,D$ and denote the corresponding truncated correlation matrix in class DIII as $C_{A,\mathrm{DIII}}=V^\dg V$, where $V$ is defined as Eq.~\eqref{eqn:chiral-V}. 
From the discussions in the chiral classes, $\langle \tr C_{A,\mathrm{DIII}}^n\rangle$ can be obtained from $\langle \tr(A^T A^*)^m\rangle$. 
To obtain $\langle \tr(A^T A^*)^m\rangle$, we decompose the $2N\times 2N$ matrix $f$ as
\begin{equation}
    f = \left(
    \begin{array}{cc}
        a & b \\
        c & d
    \end{array}
    \right),\quad f\in \mathrm{U} (2N),
\end{equation}
where $a$, $b$, $c$, $d$ are $N \times N$ matrices.
From Eq.~\eqref{eqn:DIII-decomp}, the matrix $A$ is expressed as
\begin{equation}
    A = (I_{N/2}\otimes\sigma_y) a^T (I_{N/2}\otimes\sigma_y) a + (I_{N/2}\otimes\sigma_y) c^T (I_{N/2}\otimes\sigma_y) c.  
\end{equation}
For the sake of clarity, we assume that $N$ is an even number. 
The matrix component of $A$ is explicitly given as
\begin{equation}
    A_{ij}=\sum_{k=1}^N s(i) s(k) \left( f_{p(k),p(i)}f_{k,j}+f_{p(k)+N,p(i)}f_{k+N,j} \right),  
\end{equation}
where the exchange function $p$ and parity function $s$ are defined in Eq.~\eqref{eqn:ps}, as discussed in class AIII. 
Since $\langle \tr(A^T A^*)^m\rangle$ can now be expressed by components of $f$, which is drawn from the random unitary ensemble, it can be readily computed by the Weingarten formula for the unitary group. 
The treatment is similar to class CI, with a minor difference that the presence of $s(i)$ might contribute a minus sign.

To illustrate this difference, let us examine the case of $m=3$:
\begin{align}
    \langle \tr (A^T A^*)^3\rangle=&\langle \tr \left(a^T(I_{N/2}\otimes \sigma_y)aa^\dg(I_{N/2}\otimes \sigma_y)a^* +a^T(I_{N/2}\otimes \sigma_y)ac^\dg(I_{N/2}\otimes \sigma_y)c^*\right. \nonumber \\
    &\qquad\qquad \left.+c^T(I_{N/2}\otimes \sigma_y)ca^\dg(I_{N/2}\otimes \sigma_y)a^*+c^T(I_{N/2}\otimes \sigma_y)cc^\dg(I_{N/2}\otimes \sigma_y)c^*\right)^3\rangle.
\end{align}
As in class CI, there are 64 terms in this expansion, and 20 of them have nonzero contributions, where the number of $\lbrace c^T(I_{N/2}\otimes \sigma_y)c, c^\dg (I_{N/2}\otimes \sigma_y)c^*\rbrace$ pairs can be 0,1,2,3. For the sake of simplicity, let us focus on the first term, which is the case where with no $\lbrace c^T(I_{N/2}\otimes \sigma_y)c, c^\dg (I_{N/2}\otimes \sigma_y)c^*\rbrace$ pair; 
the generalization to the other cases is straightforward. 
The first term is given as
\begin{align}
    &\langle \tr(a^T (I_{N/2}\otimes \sigma_y)a a^\dg (I_{N/2}\otimes \sigma_y)a^* a^T (I_{N/2}\otimes \sigma_y)a a^\dg(I_{N/2}\otimes \sigma_y) a^* a^T(I_{N/2}\otimes \sigma_y) a a^\dg (I_{N/2}\otimes \sigma_y)a^*)\rangle \nonumber \\
    =& \sum_{\lbrace i_k,j_k,i_k',j_k'\rbrace=1}^{N}
    s(i_1)s(i_3) s(i_5)s(i_2')s(i_4')s(i_6')
    \delta_{i_1 p(i_2)} \delta_{i_3 p(i_4)} \delta_{i_5 p(i_6)} \delta_{i'_1 p(i'_2)} \delta_{i'_3 p(i'_4)} \delta_{i'_5 p(i'_6)} \left(\delta_{j_2 j_1'}\delta_{j_3 j_2'}\cdots \delta_{j_6 j_5'}\delta_{j_1 j_6'}\right) \nonumber \\
    &\qquad\qquad\qquad\qquad\qquad\qquad\qquad\qquad\qquad\qquad\qquad\qquad\qquad\qquad\qquad\qquad\qquad\times\left[\int df f_{i_1 j_1}\cdots f_{i_6 j_6}f^*_{i'_1 j'_1}\cdots f^*_{i'_6 j'_6}\right] \nonumber \\
    =&  \sum_{\lbrace i_k,j_k,i_k',j_k'\rbrace=1}^{N}\left(\prod_{k=1}^m s(i_{2k-1})s(i'_{2k})\delta_{i_{2k-1} p(i_{2k})}  \delta_{i'_{2k-1} p(i'_{2k})} \right)\left(\prod_{k=1}^{2m}\delta_{j_k j'_{k-1}}\right) \left[\sum_{\sigma,\tau\in S_{2m}}\left(\prod_{k=1}^{2m}\delta_{i_k,i'_{\sigma(k)}} \delta_{j_k,j'_{\tau(k)}}\right)\Wg^{\rU}(N,\sigma\tau^{-1})\right] \nonumber \\
    =& \sum_{\sigma,\tau\in S_{2m}}\Wg^{\rU}(N,\sigma\tau^{-1}) \nonumber \\
    &\times\left[\sum_{\lbrace i_k,i'_k\rbrace=1}^{N}\left(\prod_{k=1}^m s(i_{2k-1})s(i'_{2k})\delta_{i_{2k-1} p(i_{2k})}  \delta_{i'_{2k-1} p(i'_{2k})} \right)\left(\prod_{k=1}^{2m}\delta_{i_k,i'_{\sigma(k)}}\right)\right]\left[
    \sum_{\lbrace j_k,j'_k\rbrace=1}^{N} \left(\prod_{k=1}^{2m}\delta_{j_k j'_{k-1}}\right)\left(\prod_{k=1}^{2m}\delta_{j_k,j'_{\tau(k)}}\right)
    \right].
\end{align}
We can see that the only difference from class CI is in the first square bracket, where the product of $s(i)$ might bring a minus sign. As in the previous cases, the terms in the square bracket can be evaluated graphically, and the extra minus sign can be taken care of by introducing directed edges. 

Using the above formalism, we calculate the values of $\langle \tr(A^T A^*)^m\rangle$. 
The first three values for $m=1,2,3$ are $\frac{N}{2}-\frac{1}{4}+O(N^{-1})$, $\frac{3N}{8}-\frac{1}{4}$, and $\frac{5N}{16}-\frac{1}{4}$, respectively. 
From these values and Eq.~\eqref{eqn:tr-CACh}, the first seven values of $\langle \tr C_{A,\mathrm{DIII}}^n\rangle$ are up to $O(1)$
\begin{equation}
        \begin{tabularx}{0.9\textwidth}{ 
   >{\centering\arraybackslash}X 
  | >{\centering\arraybackslash}X 
   >{\centering\arraybackslash}X 
  >{\centering\arraybackslash}X 
  >{\centering\arraybackslash}X 
  >{\centering\arraybackslash}X
  >{\centering\arraybackslash}X
  >{\centering\arraybackslash}X}
  \hline
           $n$ & $1$ & $2$ & $3$ & $4$ & $5$ & $6$ & $7$ \\
            \hline
          $\langle \tr C_{A,\mathrm{DIII}}^n\rangle$  &  $N$ & $\dfrac{3}{4}N-\dfrac{1}{8}$ & $\dfrac{5}{8}N-\dfrac{3}{16}$ & $\dfrac{35}{64}N-\dfrac{7}{32}$ & $\dfrac{63}{128}N-\dfrac{15}{64}$ & $\dfrac{231N}{512}-\dfrac{31}{128}$ & $\dfrac{429N}{1024}-\dfrac{63}{256}$ \\
           \hline
    \end{tabularx}
\end{equation}
Based of these values, we conjecture the closed form formula,
\begin{equation}
    \langle \tr C_{A,\mathrm{DIII}}^n \rangle=\frac{1}{2^{2n-2}} \binom{2n-1}{n} N -\frac{1}{4}+\frac{1}{2^{n+1}} + O(N^{-1}) \qquad \left( n\geq 1 \right).
\end{equation}
Following the resolvent method, we obtain the average entanglement entropy in class DIII as
\begin{equation}
    \langle S\rangle=(2\log 2-1)N+\frac{1}{4}\log 2+O(N^{-1}).
\end{equation}
We see that the $O (1)$ contribution in class DIII is related to the $O (1)$ contribution in class CI by a factor of $-1/2$. 
We have already seen the same pattern between classes AI and AII, as well as classes BDI and CII. This finishes the proof of Eq.~\eqref{eqn:ent-DIII&CI}. 

\clearpage

\twocolumngrid
\section{Wigner surmise of typical quantum entanglement}
    \label{sec: Wigner surmise}

We analytically and numerically calculate the average and variance of entanglement entropy for small systems $N=2$ in the ten AZ symmetry classes (Table~\ref{stab: Wigner surmise}), similar to the Wigner surmise.
While these results quantitatively deviate from the large-$N$ results, they are qualitatively similar to the large-$N$ results.
In BdG Hamiltonians that break the conservation of the particle number, the average is one half and the variance is one quarter in comparison with the results for particle-number-conserving free fermions (Table~\ref{stab: Wigner surmise - BdG}).
Notably, the typical entanglement entropy for $N=2$ cannot be well described by the random-matrix indices $\left( \alpha, \beta \right)$ in contrast to the large-$N$ results;
the universal values of the typical entanglement entropy for large $N$ should originate from the many-level effect.

\begin{table}[b]
	\centering
	\caption{Wigner surmise of typical quantum entanglement for particle-number-nonconserving Bogoliubov-de Gennes (BdG) Hamiltonians in classes D, DIII, C, and CI.
    The Altland-Zirnbauer (AZ) symmetry classes consist of time-reversal symmetry (TRS), particle-hole symmetry (PHS), and chiral symmetry (CS). 
    For TRS and PHS, the entries ``$\pm 1$" mean the presence of symmetry and its sign, and the entries ``$0$" mean the absence of symmetry.
    For CS, the entries ``$1$" and ``$0$" mean the presence and absence of symmetry, respectively.
    The average and variance of entanglement entropy are calculated numerically for $N=2$.
    In the numerical calculations, each datum is averaged over $10^8$ ensembles.
    All the results of entanglement entropy are calculated for particle-number-nonconserving BdG Hamiltonians with the half bipartition.
    }
     \begin{tabular}{cccccc} \hline \hline
     ~AZ class~ & ~TRS~ & ~PHS~ & ~CS~ & ~$\braket{S}_{\rm numerical}$~ & ~$\braket{\left( \Delta S\right)^2}_{\rm numerical}$~  \\ \hline
     D & $0$ & $+1$ & $0$ & ~~$0.5000$~~ & ~~$0.0350$~~ \\
     DIII & $-1$ & $+1$ & $1$ & $0.5068$ & $0.0205$ \\
     C & $0$ & $-1$ & $0$ & ~~$0.3333$~~ & ~~$0.0321$~~ \\
     CI & $+1$ & $-1$ & $1$ & $0.2772$ & $0.0445$ \\ \hline \hline
     \end{tabular}
  	\label{stab: Wigner surmise - BdG}
\end{table}

\subsection{Standard class}

In class A, the $2\times 2$ unitary matrix $U \in \mathrm{U} \left( 2 \right)$ in Eq.~(\ref{seq: U-group}) can be parameterized as
\begin{align}
    U = e^{\ii \gamma/2} \begin{pmatrix}
        e^{\ii \phi_1} \cos \theta & e^{\ii \phi_2} \sin \theta \\
        - e^{-\ii \phi_2} \sin \theta & e^{-\ii \phi_1} \cos \theta
    \end{pmatrix}
        \label{seq: U(2)}
\end{align}
with the Haar measure $dU = \sin \left( 2\theta \right) d\theta d\gamma d\phi_1 d\phi_2/( 8\pi^3 )$ ($\gamma, \phi_1, \phi_2 \in \left[ 0, 2\pi \right]$, $\theta \in \left[0, \pi/2 \right]$).
Then, the truncated correlation matrix in Eq.~(\ref{seq: CA}) is $C_A = \cos^2 \theta$, and hence the entanglement entropy for given $\theta$ is obtained as
\begin{align}
    S = - \left( \cos^2 \theta \right) \log \left( \cos^2 \theta \right) - \left( \sin^2 \theta \right) \log \left( \sin^2 \theta \right).
        \label{seq: Wigner-EE-A}
\end{align}
The average entanglement entropy is given as
\begin{align}
    \braket{S} = \int_{0}^{\pi/2} d\theta \left( \sin 2\theta \right) S = \frac{1}{2},
\end{align}
and the variance of entanglement entropy is given as
\begin{align}
    \braket{\left( \Delta S \right)^2} 
    &= \int_{0}^{\pi/2} d\theta \left( \sin 2\theta \right) \left( S - \braket{S} \right)^2 \nonumber \\
    &= \frac{21-2\pi^2}{36} = 0.035022 \cdots.
\end{align}
In addition, the average density of the single-particle entanglement spectrum is 
\begin{align}
    \braket{D \left( \lambda \right)} 
    &= \braket{\delta \left( \lambda - \cos^2 \theta \right)} \nonumber \\
    &= \int_{0}^{\pi/2} d\theta \left( \sin 2\theta \right) \delta \left( \lambda - \cos^2 \theta \right) \nonumber \\
    &= 1.
\end{align}

In class AI, the $2\times 2$ orthogonal matrix $O \in \mathrm{O} \left( 2 \right)$ can be parameterized as
\begin{align}
    O = \begin{pmatrix}
        \cos \theta & - \sin \theta \\
        \sin \theta & \cos \theta
    \end{pmatrix}
        \label{seq: O(2)}
\end{align}
with the Haar measure $dO = d\theta/2\pi$ ($\theta \in \left[ 0, 2\pi \right]$).
Then, the entanglement entropy for given $\theta$ is obtained as Eq.~(\ref{seq: Wigner-EE-A}).
The average entanglement entropy is
\begin{align}
    \braket{S} = \int_{0}^{2\pi} \frac{d\theta}{2\pi} S = 2\log 2 - 1 = 0.386294 \cdots,
\end{align}
and the variance of entanglement entropy is
\begin{align}
    \braket{\left( \Delta S \right)^2} 
    &= \int_{0}^{2\pi} \frac{d\theta}{2\pi} \left( S - \braket{S} \right)^2 \nonumber \\
    &= \frac{5\pi^2}{24} - 2 = 0.0561676 \cdots.
        \label{seq: Wigner-AI-variance}
\end{align}
The different classifying space and the concomitant different Haar measure directly lead to the different typical entanglement entropy.
The average density of the single-particle entanglement spectrum is 
\begin{align}
    \braket{D \left( \lambda \right)} 
    &= \braket{\delta \left( \lambda - \cos^2 \theta \right)} \nonumber \\
    &= \int_{0}^{2\pi} \frac{d\theta}{2\pi}\,\delta \left( \lambda - \cos^2 \theta \right) \nonumber \\
    &= \frac{1}{\pi\sqrt{\lambda \left( 1-\lambda \right)}},
\end{align}
which coincides with the large-$N$ results of $\braket{D \left( \lambda \right)}$ in Eq.~(\ref{seq: DOS-AI}) for $N=3$.

In class AII, the $4\times 4$ symplectic matrix $U \in \mathrm{Sp} \left( 2 \right)$ can be parameterized as~\cite{Haake-textbook}
\begin{align}
    U = \begin{pmatrix}
        e^{\ii \phi\,(\bm{n} \cdot \bm{\sigma})} \cos \theta &  - \sin \theta \\
        \sin \theta & e^{- \ii \phi\,(\bm{n} \cdot \bm{\sigma})} \cos \theta
    \end{pmatrix}
        \label{seq: Sp(2)}
\end{align}
with the unit vector $\bm{n}$ on the sphere $S^2$ and the Haar measure $dU = \left( 3/2 \right) \sin^3 \left( 2\theta \right) d\theta \left( 2/\pi \right) \sin^2 \phi\,d\phi d^2n$ ($\phi \in \left[ 0, \pi \right]$, $\theta \in \left[ 0, \pi/2 \right]$).
While this parametrization does not give the most general  $4\times 4$ symplectic matrix, it suits the present purpose of obtaining the typical quantum entanglement entropy.
Then, the entanglement entropy for given $\theta$ is obtained as Eq.~(\ref{seq: Wigner-EE-A}).
The average entanglement entropy is
\begin{align}
    \braket{S} = \int_{0}^{\pi/2} d\theta \left( \frac{3}{2} \sin^3  2\theta \right) S = \frac{7}{12} = 0.583333 \cdots,
\end{align}
and the variance of entanglement entropy is
\begin{align}
    \braket{\left( \Delta S \right)^2} 
    &= \int_{0}^{\pi/2} d\theta \left( \frac{3}{2} \sin^3  2\theta \right) \left( S - \braket{S} \right)^2 \nonumber \\
    &= \frac{97}{144} - \frac{\pi^2}{15} 
    = 0.0156375 \cdots.
\end{align}
In addition, the average density of the single-particle entanglement spectrum is 
\begin{align}
    \braket{D \left( \lambda \right)} 
    &= \braket{\delta \left( \lambda - \cos^2 \theta \right)} \nonumber \\
    &= \int_{0}^{\pi/2} d\theta \left( \frac{3}{2} \sin^3  2\theta \right) \delta \left( \lambda - \cos^2 \theta \right) \nonumber \\
    &= 6\lambda \left( 1-\lambda \right),
\end{align}
which vanishes at $\lambda = 0, 1$ in contrast to $\braket{D \left( \lambda \right)}$ in class AI.

\subsection{Chiral class}

In class AIII, the $2 \times 2$ unitary matrix $h$ in Eq.~(\ref{seq: h-group}) can be parameterized as Eq.~(\ref{seq: U(2)}).
Then, the truncated correlation matrix $C_A$ in Eq.~(\ref{seq: CA}) is obtained as
\begin{align}
    C_A = \frac{1}{2} \begin{pmatrix}
        1 & -e^{-\ii\,(\gamma/2 + \phi_1)} \cos \theta \\
        -e^{\ii\,(\gamma/2 + \phi_1)} \cos \theta & 1
    \end{pmatrix},
\end{align}
whose eigenvalues are $\{ \left( 1 \pm \cos \theta \right)/2 \} = \{ \cos^2 \left( \theta/2 \right), \sin^2 \left( \theta/2 \right) \}$.
Hence, the entanglement entropy for given $\theta$ is obtained as
\begin{align}
    S &= - 2 \left[ \left( \cos^2 \left( \theta/2 \right) \right) \log \left( \cos^2 \left( \theta/2 \right) \right) \right. \nonumber \\
    &\qquad\qquad\qquad\left. + \left( \sin^2 \left( \theta/2 \right) \right) \log \left( \sin^2 \left( \theta/2 \right) \right) \right].
        \label{seq: Wigner-EE-AIII}
\end{align}
The average entanglement entropy is
\begin{align}
    \braket{S} = \int_{0}^{\pi/2} d\theta \left( \sin 2\theta \right) S = \frac{2\log 2 + 1}{3} = 0.795431 \cdots,
\end{align}
and the variance of entanglement entropy is
\begin{align}
    \braket{\left( \Delta S \right)^2} 
    &= \int_{0}^{\pi/2} d\theta \left( \sin 2\theta \right) \left( S - \braket{S} \right)^2 \nonumber \\
    &= \frac{3- 8 \log 2 + 8 \left( \log 2\right)^2}{9} = 0.144272 \cdots.
\end{align}
In addition, the average density of the single-particle entanglement spectrum is 
\begin{align}
    \braket{D \left( \lambda \right)} 
    &= \int_{0}^{\pi/2} d\theta \left( \sin 2\theta \right) \left[ \delta \left( \lambda - \cos^2 \left( \theta/2 \right) \right) \right. \nonumber \\
    &\qquad\qquad\qquad\qquad\quad \left. + \delta \left( \lambda - \sin^2 \left( \theta/2 \right) \right) \right] \nonumber \\
    &= 4 \left| 2\lambda - 1\right|,
\end{align}
which vanishes linearly toward the chiral-symmetric point $\lambda = 1/2$.

In class BDI, the $2 \times 2$ orthogonal matrix $h$ in Eq.~(\ref{seq: h-group}) can be parameterized as Eq.~(\ref{seq: O(2)}).
Then, the eigenvalues of the truncated correlation matrix $C_A$ are given as $\{ \cos^2 \left( \theta/2 \right), \sin^2 \left( \theta/2 \right) \}$, and hence the entanglement entropy for given $\theta$ is obtained as Eq.~(\ref{seq: Wigner-EE-AIII}).
The average entanglement entropy is
\begin{align}
    \braket{S} = \int_{0}^{2\pi} \frac{d\theta}{2\pi} S = 2 \left( 2\log 2 - 1 \right) = 0.772589 \cdots,
\end{align}
and the variance of entanglement entropy is
\begin{align}
    \braket{\left( \Delta S \right)^2} 
    &= \int_{0}^{2\pi} \frac{d\theta}{2\pi} \left( S - \braket{S} \right)^2 \nonumber \\
    &= \frac{5\pi^2}{6} - 8 = 0.22467 \cdots.
\end{align}
Notably, in comparison with class AI, the average $\braket{S}$ is twice larger, and the variance $\braket{\left( \Delta S \right)^2}$ is four times larger.
The average density of the single-particle entanglement spectrum is 
\begin{align}
    \braket{D \left( \lambda \right)} 
    &= \int_{0}^{2\pi} \frac{d\theta}{2\pi} \left[ \delta \left( \lambda - \cos^2 \left( \theta/2 \right) \right) + \delta \left( \lambda - \sin^2 \left( \theta/2 \right) \right) \right] \nonumber \\
    &= \frac{2}{\pi \sqrt{\lambda \left( 1-\lambda \right)}},
\end{align}
which does not vanish even at the chiral-symmetric point $\lambda = 1/2$.

In class CII, the $4 \times 4$ symplectic matrix $h$ in Eq.~(\ref{seq: h-group}) can be parameterized as Eq.~(\ref{seq: Sp(2)}).
Then, the eigenvalues of the truncated correlation matrix $C_A$ are given as $\{ \cos^2 \left( \theta/2 \right), \cos^2 \left( \theta/2 \right), \sin^2 \left( \theta/2 \right), \sin^2 \left( \theta/2 \right) \}$, and hence the entanglement entropy for given $\theta$ is obtained as Eq.~(\ref{seq: Wigner-EE-AIII}).
The average entanglement entropy is
\begin{align}
    \braket{S} 
    &= \int_{0}^{\pi/2} d\theta \left( \frac{3}{2} \sin^3  2\theta \right) S \nonumber \\
    &= \frac{79 + 132 \log 2}{210} = 0.811883 \cdots,
\end{align}
and the variance of entanglement entropy is
\begin{align}
    \braket{\left( \Delta S \right)^2} 
    &= \int_{0}^{\pi/2} d\theta \left( \frac{3}{2} \sin^3  2\theta \right) \left( S - \braket{S} \right)^2 \nonumber \\
    &= \frac{16021 - 44160 \log 2 + 38016 \left(  \log 2\right)^2}{44100} \nonumber \\ 
    &= 0.0833679 \cdots.
\end{align}
In addition, the average density of the single-particle entanglement spectrum is 
\begin{align}
    \braket{D \left( \lambda \right)} 
    &= \int_{0}^{\pi/2} d\theta \left( \frac{3}{2} \sin^3  2\theta \right) \left[ \delta \left( \lambda - \cos^2 \left( \theta/2 \right) \right) \right. \nonumber \\
    &\qquad\qquad\qquad\qquad\left. + \delta \left( \lambda - \sin^2 \left( \theta/2 \right) \right) \right] \nonumber \\
    &= 96 \lambda \left( 1- \lambda \right) \left| 2\lambda - 1\right|^3,
\end{align}
which vanishes cubically toward the chiral-symmetric point $\lambda = 1/2$.

\onecolumngrid
\subsection{Bogoliubov-de Gennes (BdG) class}

In class D, the $4 \times 4$ orthogonal matrix $O \in \mathrm{O} \left( 4 \right)$ in Eq.~(\ref{seq: classifying space D}) can be parameterized as
\begin{align}
    O = \begin{pmatrix}
        \cos \theta_{13} 
        & - \sin \theta_{13} \cos \theta_{23} 
        & \sin \theta_{13} \sin \theta_{23} \cos \theta_{33} 
        & - \sin \theta_{13} \sin \theta_{23} \sin \theta_{33} \\
        \sin \theta_{13} 
        & \cos \theta_{13} \cos \theta_{23} 
        & - \cos \theta_{13} \sin \theta_{23} \cos \theta_{33} 
        & \cos \theta_{13} \sin \theta_{23} \sin \theta_{33} \\
        0
        & \sin \theta_{23}
        & \cos \theta_{23} \cos \theta_{33}
        & - \cos \theta_{23} \sin \theta_{33} \\
        0 
        & 0
        & \sin \theta_{33}
        & \cos \theta_{33}
    \end{pmatrix} \begin{pmatrix}
        O_3 & 0 \\
        0 & 1
    \end{pmatrix}
\end{align}
with 
\begin{align}
    O_3 \coloneqq \begin{pmatrix}
        \cos \theta_{12} 
        & - \sin \theta_{12} \cos \theta_{22}
        & \sin \theta_{12} \sin \theta_{22} \\
        \sin \theta_{12}
        & \cos \theta_{12} \cos \theta_{12}
        & - \cos \theta_{12} \sin \theta_{12} \\
        0
        & \sin \theta_{22}
        & \cos \theta_{22}
    \end{pmatrix} \begin{pmatrix}
        \cos \theta_{11} & - \sin \theta_{11} & 0 \\
        \sin \theta_{11} & \cos \theta_{11} & 0 \\
        0 & 0 & 1
    \end{pmatrix} \in \mathrm{O} \left( 3 \right).
\end{align}
Here, the six parameters take $\theta_{11}$, $\theta_{12}$, $\theta_{13} \in \left[ 0, 2\pi \right]$ and $\theta_{22}$, $\theta_{23}$, $\theta_{33} \in \left[ 0, \pi\right]$, and the Haar measure is given as $dO = \left( \sin \theta_{22} \right) \left( \sin \theta_{23} \right) \left( \sin^2 \theta_{33} \right) d\theta_{11} d\theta_{12} d\theta_{22} d\theta_{13} d\theta_{23} d\theta_{33}/ ( 16\pi^4 )$.
From this orthogonal matrix $O \in \mathrm{O} \left( 4 \right)$, the truncated correlation matrix in Eq.~(\ref{seq: CA}) is obtained as $C_A = \left( I_2 + \delta \sigma_y \right)/2$ with
\begin{align}
    \delta &\coloneqq - \sin \theta_{11} \left[ \cos \left( \theta_{12} - \theta_{33} \right) \sin \theta_{13} + \cos \theta_{13} \cos \theta_{23} \sin \left( \theta_{12} - \theta_{33} \right) \right] \nonumber \\
    &\qquad\quad+ \cos \theta_{11} \left[ - \cos \theta_{13} \sin \theta_{22} \sin \theta_{23} + \cos \theta_{22} \sin \theta_{12} \left( - \cos \theta_{33} \sin \theta_{13} + \cos \theta_{13} \cos \theta_{23} \sin \theta_{33}\right) \right. \nonumber \\    
    &\qquad\qquad\qquad\qquad\qquad\qquad \left. + \cos \theta_{12} \cos \theta_{22} \left( \cos \theta_{13} \cos \theta_{23} \cos \theta_{33} + \sin \theta_{13} \sin \theta_{33}\right)\right].
\end{align}
\twocolumngrid
\noindent
Then, the average entanglement entropy is given as
\begin{align}
    \braket{S} = \int dO~S = 1,
\end{align}
and the variance of entanglement entropy is given as
\begin{align}
    \braket{\left( \Delta S \right)^2} = \int dO \left( S - \braket{S} \right)^2 = \frac{21 - 2\pi^2}{9} = 0.140088 \cdots.
\end{align}

In class DIII, the $4\times 4$ unitary matrix $h$ in Eq.~(\ref{seq: classifying space DIII}) belongs to the circular symplectic ensemble and hence is parameterized as
\begin{align}
    h = e^{-\ii a\,(x_1 \tau_y + x_2 \tau_z + x_3 \sigma_x \tau_y + x_4 \sigma_y \tau_y + x_5 \sigma_z \tau_y)}
\end{align}
with the unit vector $\bm{x} = \left( x_1, x_2, x_3, x_4, x_5 \right)$ on the four-dimensional sphere $S^4$, and the Haar measure is given as $\left( 8/3\pi \right) \left( \sin^4 a \right) da d^4 x$ ($a \in \left[ 0, \pi\right]$).
Then, the truncated correlation matrix is obtained as
\begin{align}
    C_A = \frac{1}{2} \left( I_2 + \sigma_x \cos a - \sigma_y x_2 \sin a \right) \otimes I_2.
\end{align}
The average entanglement entropy is
\begin{align}
    \braket{S} = \int d^4 x \int_0^{\pi} da \left( \frac{8}{3\pi} \sin^4 a \right) S  = 1.013604 \cdots,
        \label{seq: Wigner-DIII-average}
\end{align}
and the variance of entanglement entropy is
\begin{align}
    \braket{\left( \Delta S \right)^2} 
    &= \int d^4 x \int_0^{\pi} da \left( \frac{8}{3\pi} \sin^4 a \right) \left( S - \braket{S} \right)^2 \nonumber \\
    &= 0.0818924 \cdots.
        \label{seq: Wigner-DIII-variance}
\end{align}

In class C, the $4 \times 4$ symplectic matrix $U \in \mathrm{Sp} \left( 2 \right)$ in Eq.~(\ref{seq: classifying space C}) can be parameterized as Eq.~(\ref{seq: Sp(2)}).
Then, the truncated correlation matrix is obtained as 
\begin{align}
    C_A = \frac{1}{2} \left( I_2 + \sin \left( 2\theta \right) \sin \phi \left( \bm{n} \cdot \bm{\sigma}^T \right) \right).
\end{align}
The average entanglement entropy is
\begin{align}
    \braket{S} = \int_{0}^{\pi/2} d\theta \left( \frac{3}{2} \sin^3  2\theta \right) \int_0^{\pi} d\phi \left( \frac{2}{\pi} \sin^2 \phi \right) S = \frac{2}{3},
\end{align}
and the variance of entanglement entropy is
\begin{align}
    \braket{\left( \Delta S \right)^2} 
    &= \int_{0}^{\pi/2} d\theta \left( \frac{3}{2} \sin^3  2\theta \right) \nonumber \\
    &\qquad\quad \int_0^{\pi} d\phi \left( \frac{2}{\pi} \sin^2 \phi \right) \left( S - \braket{S} \right)^2 \nonumber \\
    &= 0.128497 \cdots.
        \label{seq: Wigner-C-variance}
\end{align}

In class CI, the $2\times 2$ unitary matrix $h$ in Eq.~(\ref{seq: classifying space CI}) belongs to the circular orthogonal ensemble and hence is parameterized as
\begin{align}
    h = e^{-\ii a\,(\cos \theta \sigma_x + \sin \theta \sigma_z)}
\end{align}
with the Haar measure $\left( 1/2 \right) \sin a da \left( d\theta/2\pi \right)$ ($a \in \left[ 0, \pi\right]$, $\theta \in \left[ 0, 2\pi \right]$).
Then, the truncated correlation matrix is obtained as
\begin{align}
    C_A = \frac{1}{2} \left( I_2 + \sigma_x \cos a + \sigma_z \sin a \sin \theta \right).
\end{align}
The average entanglement entropy is
\begin{align}
    \braket{S} = \int_{0}^{2\pi} \frac{d\theta}{2\pi} \int_0^{\pi} da \left( \frac{\sin a}{2} \right) S  = 0.554363 \cdots,
        \label{seq: Wigner-CI-average}
\end{align}
and the variance of entanglement entropy is
\begin{align}
    \braket{\left( \Delta S \right)^2} 
    &= \int_{0}^{2\pi} \frac{d\theta}{2\pi} \int_0^{\pi} da \left( \frac{\sin a}{2} \right) \left( S - \braket{S} \right)^2 \nonumber \\
    &= 0.177858 \cdots.
        \label{seq: Wigner-CI-variance}
\end{align}

\bibliography{Gaussian_Page}

\end{document}